\documentclass[useAMS,usenatbib]{mn2e}
\usepackage{graphicx,amssym}
\newcommand{\bc}{\begin{center}}
\newcommand{\ec}{\end{center}}

\title[From dwarf spheroidals to cDs: simulating the galaxy population in a $\Lambda$CDM cosmology]
      {From dwarf spheroidals to cDs: Simulating the galaxy population in a $\Lambda$CDM cosmology}
\author[Qi Guo et al.]
       {\parbox{18cm}{Qi Guo$^{1,2}$ \thanks{Email:guoqi@durham.ac.uk}, Simon White$^1$, Michael Boylan-Kolchin$^1$, Gabriella De Lucia$^3$,  
Guinevere Kauffmann$^1$, Gerard Lemson$^1$, Cheng Li$^{1}$, Volker Springel$^{1,4}$, 
Simone Weinmann$^1$}
       \\     
       \\
       $^{1}$ Max Planck Institut f\"ur					       
         Astrophysik, Karl-Schwarzschild-Str. 1, 85741 Garching, Germany\\  
	$^2$ Institute for Computational Cosmology, Department of Physics, University of Durham, South Road, Durham, DH1 3LE, UK \\ 
       $^3$  INAF - Astronomical Observatory of Trieste, via G.B. Tiepolo 11, I-34143 Trieste, Italy   \\ 
        $^{4}$ Heidelberg Institute for Theoretical Studies, Schloss-Wolfsbrunnenweg 35, 69118 Heidelberg, Germany}
\begin{document}

\date{Accepted  ???? ??. 2010 ???? ??}

\pagerange{\pageref{firstpage}--\pageref{lastpage}} 
\pubyear{2010}

\maketitle

\label{firstpage}

\begin{abstract}
We have updated and extended our semi-analytic galaxy formation
modelling capabilities and applied them simultaneously to the stored
halo/subhalo merger trees of the Millennium and Millennium-II
simulations. These differ by a factor of 125 in mass resolution,
allowing explicit testing of resolution effects on predicted galaxy
properties. We have revised the treatments of the transition between
the rapid infall and cooling flow regimes of gas accretion, of the
sizes of bulges and of gaseous and stellar disks, of supernova
feedback, of the transition between central and satellite status as
galaxies fall into larger systems, and of gas and star stripping once
they become satellites.  Plausible values of efficiency and scaling
parameters yield an excellent fit not only to the observed abundance
of low-redshift galaxies over 5 orders of magnitude in stellar mass
and 9 magnitudes in luminosity, but also to the observed abundance of
Milky Way satellites. This suggests that reionisation effects may not
be needed to solve the ``missing satellite'' problem except, perhaps, for the
faintest objects.  The same model matches the observed large-scale
clustering of galaxies as a function of stellar mass and colour. The
fit remains excellent down to $\sim 30$~kpc for massive galaxies. For
$M_* < 6\times 10^{10}M_\odot$, however, the model overpredicts
clustering at scales below $\sim 1$~Mpc, suggesting that the assumed
fluctuation amplitude, $\sigma_8=0.9$, is too high.  The observed
difference in clustering between active and passive galaxies is
matched quite well for all masses. Galaxy distributions within rich clusters agree between the
simulations and match those observed, but only if galaxies without
dark matter subhalos (so-called orphans) are included. Even at MS-II
resolution, schemes which assign galaxies only to resolved dark matter
subhalos cannot match observed clusters. Our model predicts a larger
passive fraction among low-mass galaxies than is observed, as well as
an overabundance of $\sim10^{10}M_\odot$ galaxies beyond $z\sim
0.6$. (The abundance of $\sim10^{11}M_\odot$ galaxies is matched out
to $z\sim 3$.) These discrepancies appear to reflect deficiencies in
the way star-formation rates are modelled.
\end{abstract}

\section{Introduction}
\label{sec:SAMintro}

The $\Lambda$CDM model has been successful in interpreting a wide
variety of observations. These include the cosmic microwave
background fluctuations at $z\sim 1000$ \citep[e.g.][]{Dunkley2009},
the large-scale clustering of galaxies in the low-redshift universe
\citep[e.g.][]{Percival2010}, the cosmic shear field measured by weak
gravitational lensing \citep[e.g.][]{Fu2008}, the
high-redshift power spectrum probed by the Lyman $\alpha$ forest
\citep[e.g.][]{Viel2009,Paschos2009}, and the
abundance \citep[e.g][]{Vikhlinin2009} and baryon
fractions \citep[e.g][]{Allen2008} of galaxy clusters. Current N-body
simulations can follow the growth of representative samples of dark
matter halos at high resolution and in their full cosmological context
on scales ranging from those of rich clusters to those of dwarf
galaxies. The formation of galaxies does not, however, trace that of their dark
matter halos in a simple manner, and the exponentially growing body of
high-quality galaxy data coming from large surveys cannot be properly
compared to the $\Lambda$CDM model without a careful treatment of
baryonic processes. Such detailed comparison is the most promising
route to clarifying the complex astrophysics underlying galaxy
formation, and it may also uncover problems with the
$\Lambda$CDM model which are not evident on larger scales.

In the standard scenario of galaxy formation, originally proposed
by \cite{White1978}, gas cools and condenses at the centres of a population of
hierarchically merging dark matter haloes.  These and earlier
ideas \citep[e.g.][]{Rees1977} were adapted by \cite{Blumenthal1984} to
the specific initial conditions predicted in a CDM dominated universe, showing
that the dichotomy between rapid and slow cooling regimes provides a natural
explanation for the dichotomy between individual galaxies and larger systems
like groups and clusters. Within such models, and in particular within the
current standard $\Lambda$CDM model, dark matter halos grow through accretion
and merging to produce a present-day halo mass function which has a very
different shape from the observed luminosity function of
galaxies \citep[e.g.][]{Benson2003a}. If one nevertheless matches the two,
assuming bigger galaxies to live in bigger halos, the ratio of halo mass to
central galaxy light is found to minimize for galaxies similar to the Milky
Way and to increase rapidly for both more massive and less massive
systems. The maximal efficiency for converting available baryons into stars is
about 20\%, and much lower efficiencies are found for the halos of rich
clusters or dwarf galaxies \citep{Moster2010, Guo2010}.

A popular explanation for the low efficiency of galaxy formation in
massive halos is that a supermassive black hole releases
vast amounts of energy when it accretes gas from its surroundings, and
that this suppresses cooling onto (and hence star formation in) the
host galaxy \citep{Silk1998, Birzan2004, Croton2006, Bower2006}. For
low-mass halos, \cite{White1978} argued that the supernova-driven
winds invoked by \cite{Larson1974} to explain the metallicities of
dwarf galaxies might sufficiently reduce the efficiency of galaxy
formation. Although this has been the preferred theoretical
explanation ever since, there is still no convincing observational
evidence that dwarf galaxy winds have the properties needed to
reproduce the faint end of the galaxy luminosity function in a
$\Lambda$CDM universe. Current simulations of dwarf galaxy formation,
while predicting substantial winds, neverthless suppress star
formation much less effectively than is required \citep{Sawala2010}.  The
UV and X-ray backgrounds heat the intergalactic medium and are also
thought to affect galaxy formation in small
halos \citep{Doroshkevich1967,Couchman1986,Efstathiou1992,Gnedin2000,Benson2002,Hoeft2006,Okamoto2008,Hambrick2009}

A related issue is the {\it missing satellite} problem. According to
the $\Lambda$CDM model, the halo of the Milky Way accreted many lower
mass halos as it grew, many of which should have contained small
galaxies.  Just as low-mass isolated halos produce too many dwarf
field galaxies unless their galaxy formation efficiency is extremely
low, so these accreted halos overpredict the number of dwarf
satellites around the Milky Way unless their star formation is
similarly suppressed \citep{Kauffmann1993}. Simulations of the growth
of such halos revealed correspondingly large numbers of surviving dark
matter subhalos as soon as their resolution was high
enough \citep{Klypin1999, Moore1999}, and increasing resolution has
predicted ever larger numbers of ever smaller
objects \citep{Diemand2007, Springel2008, Boylan2009}.  The number of
known satellites of the Milky Way has also increased in
recent years, through effective use of the Sloan Digital Sky Survey
(SDSS) to detect extremely low luminosity
systems \citep{Willman2005,Zucker2006,Belokurov2007,Koposov2008},
but the apparent discrepancy between the predicted and observed
numbers has steadily grown.  Environmental effects undoubtedly play a
role in determining satellite properties: after a galaxy falls into a
larger system, its gas may be stripped, leading to a rapid decline in
star formation, dimming its light and reddening its colour.  Tidal
stripping may remove stars or even destroy the satellite
altogether, contributing gas to the disk of the central galaxy and
stars to its stellar halo. Nevertheless, since the subhalos survive in
the simulations, such disruption cannot explain the apparent
discrepancy. Many low-mass subhalos must be dark if the $\Lambda$CDM
model is correct.

An entirely different resolution of these problems could lie in a
modification of the $\Lambda$CDM model itself. A number of authors
have suggested that the suppression of small-scale structure
expected in a warm dark matter model might reduce the
abundance of low-mass halos enough to alleviate the tension
\citep[e.g.][]{Bode2001,Zavala2009,Maccio2010}. The strongest 
constraint on this low-mass cut-off currently comes from observations
of small-scale structure in the high-redshift intergalactic medium, as
observed through the Ly~$\alpha$ forest in quasar spectra. These place
an upper limit on the cut-off wavelength and the corresponding halo
mass, which implies a lower limit on the mass of the dark matter
particle \citep{Viel2008,Boyarsky2009a,Boyarsky2009b}. Taken at face value,
this recent work appears to exclude significant warm dark matter
effects on any but the very faintest galaxies.

In recent years, the completion of SDSS has allowed a determination of
the galaxy stellar mass function down to a stellar mass of
$10^{7}M_{\odot}$. Above about $10^{8}M_{\odot}$ these mass functions
are robust against incompleteness and cosmic variance and have very
small uncertainties, other than an overall systematic coming from the
poorly known stellar initial mass function \citep{Baldry2008,Li2009}.
The large sample size makes it possible to retain small mass function
errors for subsamples split according to additional galaxy properties
such as colour and environment \citep{Peng2010}. This results in a
considerable sharpening of the constraints on galaxy formation within
the $\Lambda$CDM model \citep{Guo2010, Sawala2010}, making it timely
to reassess the viability of current models, in particular their
ability to reproduce the faint end of the galaxy luminosity (or
stellar mass) function and the faint satellite abundance around the
Milky Way.

In this paper we use the combination of the Millennium Simulation 
\citep[MS,][]{Springel2005} and the 
Millennium-II Simulation \citep[MS-II,][]{Boylan2009} to address this
issue. The latter has 125 times better mass resolution and 5 times better
force resolution than the MS, but follows evolution within a box of 5 times
smaller side. We update our earlier MS-based galaxy formation
models \citep[][hereafter collectively referred to as
DLB07]{Springel2005,Croton2006,DeLucia2007} to include a better treatment of a
number of physical processes, and we apply the improved model to both
simulations simultaneously. This allows us to test explicitly how limited
resolution affects our results. We demonstrate that together, the two
simulations enable study of the formation, evolution and clustering of
galaxies ranging from the faint dwarf satellites of the Milky Way to the most
massive cD galaxies. Uncertain astrophysical processes are strongly
constrained by the precise, low-redshift abundance and clustering data
provided by the SDSS.  Models consistent with these data can be tested against
other observational data, notably the satellite abundance around the Milky
Way, but also, for example, the Tully-Fisher relations of giant and dwarf
galaxies or the properties of high-redshift galaxy populations.

\begin{figure}
\bc
\hspace{-0.6cm}
\resizebox{8.5cm}{!}{\includegraphics{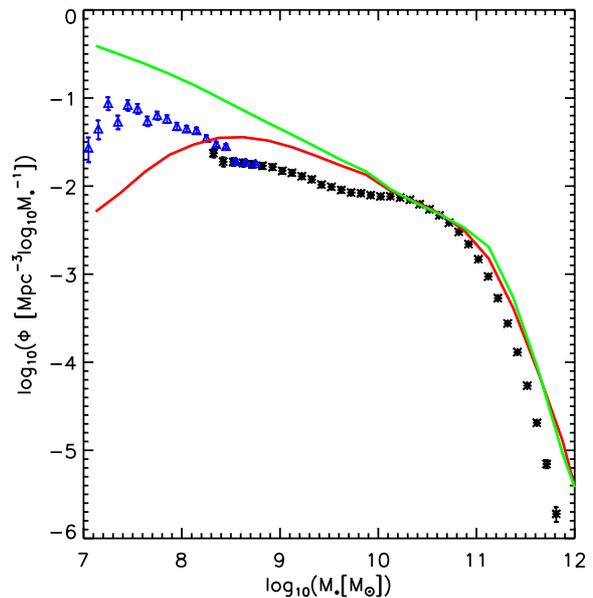}}\\%
\caption{Stellar mass functions predicted by the galaxy formation
model of DLB07. The green curve is the prediction for the MS-II and
the red curve is that for the MS. Results for the two simulations
agree well above $10^{9.5}M_\odot$, but resolution effects cause an
underprediction at lower masses in the MS. Black stars show the
observed function for SDSS/DR7 with error bars including both counting
and cosmic variance uncertainties \citep{Li2009,Guo2010}. Blue triangles
are results for a low-redshift sample (0.0033$<z<$0.05) from SDSS/DR4
taken from \cite{Baldry2008}; these are corrected for
surface-brightness incompleteness, but the error bars do not include
cosmic variance uncertainties. Clearly the model substantially
overpredicts the abundance of low-mass galaxies and slightly
overpredicts the masses of high-mass galaxies. }
\label{fig:delucia}
\ec
\end{figure}

Previous generations of semi-analytic galaxy formation models have
been able to reproduce the properties of observed galaxy populations
in ever increasing detail \citep{White1991, Kauffmann1993,Cole1994,
Kauffmann1999,Somerville1999,Cole2000,Springel2001,Hatton2003,
Kang2005,Baugh2005,Croton2006,
Bower2006,DeLucia2007,Somerville2008,Font2008,Guo2009,Weinmann2009}.
The DLB07 model was built for the MS simulation and has been
extensively compared to the abundance, intrinsic properties and
clustering of galaxies, both in the local universe and at high
redshift. These comparisons have generally been limited to galaxies
with stellar masses of at least $10^{9}M_{\odot}$, corresponding
approximately to the resolution limit of the MS. When the same model
is applied to the MS-II, it significantly overpredicts the observed
abundance of galaxies near this limit and it substantially
overpredicts the abundance at lower masses (see
Fig.~\ref{fig:delucia}). The high-mass cut-off is also at slightly
larger mass than in the new SDSS data, although it was consistent with
earlier datasets \citep{Croton2006}.  Clearly, galaxy formation
efficiencies are substantially too high at low halo mass in the DLB07
model, and slightly too high at high halo mass \citep[see also, for
example,][]{Fontanot2009}

In the following section, we revisit the DLB07 model, improving the treatment
of a number of physical processes and retuning the uncertain efficiency
parameters to obtain a better fit to the new SDSS data on abundance and
clustering. In particular, we change the treatments of supernova feedback, of
the reincorporation of ejected gas, of the sizes of galaxies, of the
distinction between satellite and central galaxies, and of environmental
effects on galaxies.  Our paper is organised as follows. In Sec.~\ref{sec:sim}
we briefly describe the two N-body simulations on which we implement our
galaxy formation model. A detailed description of the semi-analytic model
itself is presented in Sec.~\ref{sec:model}. In Sec.~\ref{sec:results} we
compare both the abundance and the clustering of galaxies as a function of
stellar mass, luminosity and colour to low-redshift data from the SDSS.  We
also compare model predictions to the observed abundance of satellite galaxies
around the Milky Way, to the Tully-Fisher relation of isolated galaxies, and
to the galaxy number density profiles, stellar mass functions, and
intergalactic light fractions of clusters. A final subsection focusses on a
few model predictions at high redshift. Sec.~\ref{sec:conclusion} presents a
concluding discussion of our results.

\section{N-body Simulations}
\label{sec:sim}

We build virtual catalogues of the galaxy population by implementing galaxy
formation models on the stored output of two very large cosmological N-body
simulations, the Millennium Simulation \citep[MS,][]{Springel2005} and the
Millennium-II Simulation \citep[MS-II][]{Boylan2009}. Both simulations assume
a $\Lambda$CDM cosmology with parameters based on a combined analysis of the
2dFGRS \citep{Colless2001} and the first-year WMAP data \citep{Spergel2003}.
The parameters are $\Omega_{\rm m}=0.25$, $\Omega_{\rm b}=0.045$,
$\Omega_\Lambda=0.75$, $n=1$, $\sigma_8=0.9$ and $H_0= 73~{\rm
  km~s^{-1}Mpc^{-1}}$.  These cosmological parameters are not consistent
  with more recent analyses of the CMB data \citep[e.g.][]{Komatsu2010} but the relatively small off-sets are not significant for most of the issues addressed
in this paper, with the important exception of the small-scale clustering
analysis of section~\ref{sec:correl}.) The parameter which deviates most
  from recent estimates is $\sigma_8$ which is quoted as 0.809$\pm$0.024 for
  WMAP7 by \cite{Komatsu2010}, disgreeing with the simulation value by almost
  $4\sigma$.  As shown by \cite{Angulo2010} simulations based on the MS
  cosmology can be scaled to represent neighboring cosmologies such as WMAP7
  to the precision needed for making galaxy catalogues. In future work we will
  use this to show how predictions for galaxy properties are affected by the
  small parameter shifts to WMAP7 and other currently allowed cosmologies.

Both MS and the MS-II trace $2160^3$ particles from redshift 127 to the
present day. The MS was carried out in a periodic box of side 685~Mpc
and the MS-II in a box of side 137~Mpc. The corresponding particle
masses are $1.18\times 10^{9}M_\odot$ and $9.45\times 10^{6}M_\odot$,
respectively. The smallest halos/subhalos we consider contain 20 bound
particles, and it will turn out that the MS-II has just
sufficient resolution to study dwarf galaxies as faint as those
seen around the Milky Way. On the other hand, the large volume of the
MS makes it possible to study rare objects like rich clusters and
bright quasar hosts. In addition, a comparison of the two simulations
where both have good statistics allows us to study how the limited
resolution of the MS affects its model galaxy populations.

The particle data were stored at 64 and 68 times for the MS and the
MS-II, respectively, with the last 60 being identical in the two
simulations. At each output time, the post-processing pipeline produced a
friends-of-friends (FOF) catalogue by linking particles with
separation less than 0.2 of the mean value \citep{Davis1985}. The
SUBFIND algorithm \citep{Springel2001} was then applied to each FOF
group to identify all its bound substructures (subhalos). The merger
trees which are the basis for our galaxy formation modelling are then
constructed by linking each subhalo found in a given output to one
and only one descendent at the following output
\citep{Springel2005,DeLucia2007,Boylan2009}. Note that all our
galaxy formation models are thus based on the growth and merging of
the population of {\it subhalos}, not on the growth and merging of the
population of {\it halos}.  This is an important distinction which
allows us to build much more realistic models for the galaxy
population, in particular for its merging rates and its clustering,
than would otherwise be the case. We refer readers
to \cite{Springel2005} and \cite{Boylan2009} for full descriptions of
the two simulations.

The most massive self-bound subhalo in a FOF group is referred to as
its main subhalo (sometimes the main halo) and usually contains most
of its mass. Other subhalos of the FOF group are referred to as
satellite subhalos.  After implementation of the galaxy formation
model, each FOF group hosts a ``central galaxy'', which sits at the
potential minimum of the main subhalo. Other associated galaxies may
sit at the potential minima of smaller subhalos, or may no longer
correspond to a resolved dark matter substructure (``orphans''). These
galaxies are collectively referred to as satellites, although we note
that in this paper we break with our previous practice by assuming
that the physical processes affecting satellite galaxies only begin to
differ from those affecting central galaxies when a satellite first
enters the virial radius of the larger system. This is to account for
the fact that FOF groups quite often link two essentially independent
dark matter clumps, and the two central galaxies are expected to keep
evolving quasi-independently while this is the case.

We define the centre of a FOF group to be its potential minimum and
its virial radius to be the radius of the largest sphere with this
centre and a mean overdensity exceeding 200 times the critical
value. The total mass within the virial radius is defined as the
virial mass of the group. Virial radius and virial mass are then
related by
\begin{equation}
R_{\rm vir}= \left(\frac{G}{100}\frac{M_{\rm vir}}{H^2(z)}\right)^{1/3}.
\end{equation}
The virial radius usually lies almost entirely within the boundary of
the FOF group and, as a result, the virial mass is typically somewhat
smaller than the FOF mass (and also typically somewhat larger than the
mass of the main subhalo).

\section{Galaxy Formation Models}
\label{sec:model}
Galaxies form at the centres of dark matter halos and gain stars by formation
from their interstellar medium (ISM) and by accretion of satellite
galaxies. We assume the ISM to form a disk and to be replenished both by
diffuse infall from the surroundings and by gas from accreted satellite
galaxies. Diffuse infall can occur directly from the intergalactic medium (in
a so-called ``cold flow'') or through cooling of a hot halo surrounding the
galaxy. The interaction of these processes with each other and with flows
driven by supernovae and by active galactic nuclei is responsible for the
overall evolution of each galaxy, which thus cannot be followed realistically
without superposing a complex network of baryonic astrophysics on the assembly
history of its dark matter component.  Physical understanding of most of these
baryonic processes is quite incomplete and is based largely on simplified
numerical simulations and on the phenomenology of observed
systems. Descriptions of the processes are thus necessarily both approximate
and uncertain, and models of the kind we build here may offer the best means
to constrain them empirically using observational data.

Here we implement simplified galaxy formation recipes onto the subhalo
merger trees built from the MS and the MS-II. Treating baryonic
evolution by post-processing cosmological N-body simulations in this
way makes it possible to explore a wide model and parameter space in a
relatively short amount of time. In general, our models build on those
developed in \cite{Springel2005}, \cite{Croton2006}, and \cite{DeLucia2007} hereafter collectively referred to as
DLB07. The baryonic content of galaxies is split into five components,
a stellar bulge, a stellar disk, a gas disk, a hot gas halo, and an
ejecta reservoir. These components exchange material through a variety
of processes and their total mass grows through accretion from the
intergalactic medium. As noted above the main modifications here concern the
definition of satellite galaxies, a mass-dependent model for supernova
feedback, the gradual stripping and disruption of satellite galaxies,
more realistic treatments of the growth of gaseous and stellar disks,
a model to calculate bulge end elliptical galaxy sizes, and an updated
reionization model. We determine the free parameters of these models
using the observed abundance, structure and clustering of low redshift
galaxies as a function of stellar mass, luminosity and colour.
In the following we describe our new galaxy formation model in detail.  
For a more general review of semi-analytic models, we refer the
reader to \cite{Baugh2006, BensonBower2010,Bensonarxiv2010}

\subsection{Reionization}
\label{sec:reionize}
It is now well established that the global baryon to dark matter
mass ratio is 15-20\%. In galaxy clusters, the observed baryon
fraction is close to but somewhat below this value, and is mostly in
the form of hot gas. In halos like that of the Milky Way, on the other
hand, at most about 20\% of the expected baryons are detected and
these are primarily in the form of stars; the detected baryon fraction
is apparently even lower in the halos of dwarf
galaxies \citep[e.g.][]{Guo2010}. One mechanism which may contribute to the
low efficiency of dwarf galaxy formation is photo-heating of
pregalactic gas by the UV background. This inhibits gas condensation
within dark matter halos if the thermal energy exceeds the halo
potential well depth. This effect was first pointed out
by \cite{Doroshkevich1967} and was later investigated in the context
of CDM models by \cite{Couchman1986} and \cite{Efstathiou1992}. 

In recent years a number of simulations of this process have been carried
out. Here we use a fitting function of the form originally proposed
by \cite{Gnedin2000} to describe how the baryon fraction in a halo depends on
mass and redshift:
\begin{equation}
f_b(z,M_{\rm vir}) = f_b^{\rm cos}\left(
1+\left(2^{\alpha/3}-1\right)\left[\frac{M_{\rm vir}}{M_F(z)}\right]^{-\alpha} 
\right)^{-3/\alpha}.
\end{equation}
In this formula, $f_b^{\rm cos}$=17\% is the universal baryon fraction as given
by first-year WMAP estimates \citep{Spergel2003}, and $\alpha = 2$ is a fit to
the simulations in \cite{Okamoto2008}.  $M_F$ is the characteristic halo mass
of this ``filter''. In halos with $M_{\rm vir}\gg M_F(z)$ the baryon fraction is
set to the universal value, while in halos with $M_{\rm vir}\ll M_F(z)$ it drops
as $(M_{\rm vir}/M_F)^3$. The redshift dependence of $M_F$ is determined by the
details of how the reionization process occurred. Here we use a table of
$M_F(z)$ kindly provided by \cite{Okamoto2008} from their simulations; $M_F$
varies from $\sim 6.5\times 10^9M_{\odot}$ at $z=0$ to $\sim
10^7M_{\odot}$ just after reionization at $z\sim 8$. These results are
consistent with those found earlier by \cite{Hoeft2006}. In DLB07, a lower
value of $\alpha$ and a different $M_F(z)$ were adopted \citep[following
][]{Kravtsov2004} leading to the substantially higher value $M_F\sim 3 \times
10^{10}M_{\odot}$ at $z=0$. These differences in simulation results appear to
reflect differences in resolution and in the treatment of radiative transfer.
Although we adopt the more recent results as ``standard'', we will rediscuss
how these issues affect dwarf galaxy formation in
Sec.~\ref{sec:resultreionization}, showing that reionization does not appear
to be a major factor except, possibly, for the faintest Milky Way satellite
systems.

\subsection{Cooling}
\label{sec:cooling}

The pressure of the intergalactic medium has little effect on the growth of
more massive halos. A fraction $\sim f_b^{\rm cos}$ of the infalling material
is expected to be diffuse gas, and thus must shock as it joins the halo. At
early times and in low-mass halos, post-shock cooling is rapid and the
accretion shock is very close to the central
object, which thus gains new material at
essentially the free-fall rate; at late times and in massive halos, post-shock
cooling times exceed halo sound crossing times, the accretion shock moves away
from the galaxy, and infalling gas forms a quasi-static hot atmosphere from
which it condenses onto the central galaxy through a cooling flow \citep{Forcada-Miro1997,Birnboim2003}. The
critical mass separating these two regimes is around $10^{12}M_{\odot}$ and is
weakly dependent on redshift but strongly dependent on $f_b^{\rm cos}$ and on
the metallicity of the infalling gas \citep{Rees1977,White1991}. These rapid
infall and quasi-static cooling flow regimes have featured in almost all
galaxy formation models of the last two decades \citep[e.g.][and also, of
course DLB07]{Kauffmann1993, Cole1994, Kauffmann1999, Somerville1999,Cole2000,
Springel2001,Hatton2003, Kang2005,Somerville2008} as well as (by construction)
in all direct simulations of the galaxy formation
process \citep[e.g.][]{Navarro1997,Steinmetz1999,Springel2003}. The simple
criterion of \cite{White1991} is used to separate the two regimes in most
semi-analytic models, and tests with both
one-dimensional \citep{Forcada-Miro1997} and three-dimensional
\citep{Benson2001,Yoshida2002,Cattaneo2007} simulations have shown 
it to provide an adequate description. More recent numerical work has focussed
on the fact that the two regimes can effectively coexist near the transition,
with cold gas falling in narrow streams through a hot gas atmosphere or even a
galactic wind (see, for example, Fig.2 of \cite{Springel2003}
or \cite{Dekel2009}). A recent reanalysis by \cite{Benson2010}
emphasised that the details of how such ``cold flows'' are treated has
little effect on predicted galaxy properties once the necessary strong
feedback is included.

Here we use the simple model of \cite{Springel2001} to estimate the
gas cooling rate in the hot halo regime. We assume that infalling gas
is shock-heated to the virial temperature of the host halo at an
accretion shock, and that its distribution interior to this shock is a
quasistatic isothermal sphere with density falling as the inverse
square of radius. The cooling time at each radius can then be
calculated as
\begin{equation}
t_{\rm cool}(r)=\frac{3\mu m_{\rm H}kT_{\rm
vir}}{2\rho_{\rm hot}(r)\Lambda(T_{\rm hot},Z_{\rm hot})},
\label{coolingtime}
\end{equation}
where $\mu m_{\rm H}$ is the mean particle mass, $k$ is the Boltzmann
constant, $\rho_{\rm hot}(r)$ is the hot gas density at radius $r$, $\Lambda
(T_{\rm hot},Z_{\rm hot})$ is the temperature- and metallicity-dependent cooling
function \citep{Sutherland1993}, and $Z_{\rm hot}$ is the metallicity of the hot halo
gas. $T_{\rm hot}=35.9(V_{\rm vir}/{\rm km~s}^{-1})^2\rm K$ is the assumed virial temperature of the
host halo. For main subhalos, the gas temperature is updated according to the
current circular velocity at the virial radius at each snapshot, while for
satellite subhalos, we assume the gas temperature to be constant at the value
it had when the subhalo was accreted onto its main halo.

The cooling radius is then estimated through
\begin{equation}
r_{\rm cool}=\left[\frac{t_{\rm dyn,h}m_{\rm hot}\Lambda(T_{\rm hot},Z_{\rm hot})}{6\pi\mu m_{\rm
H}kT_{\rm vir}R_{\rm vir}}\right]^{\frac{1}{2}}.
\end{equation}
The definition of the halo dynamical time $t_{\rm dyn,h}$ involves an
arbitrary constant. Here we adopt the convention $t_{\rm dyn,h}\equiv R_{\rm
vir}/V_{\rm vir}=0.1H(z)^{-1}$ as in \cite{DeLucia2004}. Readers are refereed
to \cite{Croton2006} and \cite{Somerville2008} for the discussion of other
possible choices of $t_{\rm dyn,h}$ when defining $r_{\rm cool}$. When $r_{\rm
cool}<R_{\rm vir}$, we assume that we are indeed in the cooling flow regime, and that the cooling rate onto the central galaxy
is\footnote{Note that the coefficient on the {\it rhs} of this equation differs
  from that in the corresponding equation (equ.~6) of \cite{Croton2006}. By
  checking the original code, we have verified that a factor of 0.5 was
  erroneously omitted when programming equ.~6 in this paper, and that this
  error then propagated through all the later DLB07 papers, with the result
  that the equations which actually correspond to the models of
  \cite{Croton2006} and DLB07 are those given here. For consistency in
  comparing to the earlier work, we have kept these assumptions in our new
  model.}

\begin{equation}
\dot M_{\rm cool}=m_{\rm hot}\frac{r_{\rm cool}}{R_{\rm vir}}\frac{1}{t_{\rm dyn,h}}.
\label{eq:mcool}
\end{equation}

When $r_{\rm cool}>R_{\rm vir}$, on the other hand, we assume that we are in
the rapid infall regime and gas accretes onto the central object in
free-fall, thus on the halo dynamical time:
\begin{equation}
\dot M_{\rm cool}= \frac{m_{\rm hot}}{t_{\rm dyn,h}}.
\label{eq:mcool2}
\end{equation}
Note that condensation is smoother in time in this model (which is essentially
identical to that of \cite{DeLucia2004}) than in the model of DLB07, who
assumed the hot gas to fall onto the central object instantaneously as soon as
it satisfies $r_{\rm cool}>R_{\rm vir}$.  In a situation of steady accretion
onto a low-mass halo, the DLB07 model results in non-steady behaviour; after a
cooling ``event'' empties the halo, its hot gas atmosphere is replenished by
infall until it again reaches the rapid cooling threshold, triggering another
cooling event. Although the time-average condensation rate is equal to the
infall rate onto the halo, condensation occurs in bursts which induce
(possibly) unrealistic star formation bursts in the central object. The model
of Equ.~(\ref{eq:mcool2}) eliminates this behaviour. For a low-mass halo
experiencing steady infall, condensation onto the central object is now also
steady, and the hot gas atmosphere has constant mass equal to the gas
infall rate times the halo dynamical time. The coefficient of unity in
Equ.~(\ref{eq:mcool2}) ensures continuity of the condensation rate as a halo
transitions between the rapid infall and hot halo regimes. With these changes,
condensation rates onto galaxies fluctuate strongly only in response to
variations in the accretion onto their halos, not as a consequence of
discontinuities in the way we treat the various regimes.

Another substantive difference in the way we treat cooling with
respect to the model of DLB07 is that we now allow satellite galaxies
to have their own hot gas halos which can be removed dynamically by tidal and
ram-pressure effects.  This hot gas can continue to cool onto the
(satellite) galaxy, adding to its interstellar medium and providing
additional fuel for star formation. We discuss this in more
detail in sect.~\ref{sec:stripping} below.

\subsection{Disk Sizes}
\label{sec:modeldisk}
Disk sizes are not only interesting in their own right, but are also
important because they determine the surface density of gas in disks,
which in turn determines the star formation rate. DLB07 followed the
simple model of \cite{Mo1998} which assumes that the specific angular
momentum of a galaxy disk is the same as that of the dark halo in
which it is embedded. This results in the characteristic size of a
disk scaling as the product of the virial radius and the spin
parameter of its host halo. \cite{Mo1998} intended this as a simple
model for a population of isolated disk galaxies at a single time, and
several difficulties arise when it is applied to individual objects as
they grow in time. For example, halo spin parameters often change
discontinuously by quite large amounts as new material is accreted,
but it is not plausible that this should result in instantaneous
changes in size of the central disk. Here we introduce a new and more
realistic disk model which distinguishes between gas and stellar disks
and allows each of them to grow continuously in mass and angular
momentum in a physically plausible way.

We assume that the change in the total vector angular momentum of the 
gas disk during a timestep can be expressed as
\begin{equation}
\Delta{\vec J}_{\rm gas}=\delta {\vec J}_{\rm gas,cooling} + \delta
{\vec J}_{\rm gas,acc} + \delta {\vec J}_{\rm gas,SF},
\end{equation}
where $\delta {\vec J}_{\rm gas,cooling}$, $\delta {\vec J}_{\rm gas,acc}$ and
$\delta {\vec J}_{\rm SF}$ are respectively the angular momentum changes
due to addition of gas by cooling, to accretion from minor mergers,
and to gas removal through star formation.

When new gas condenses onto the central galaxy, we assume it to carry 
specific angular momentum which matches the current value for the dark
matter halo ${\vec J}_{\rm DM}/M_{\rm DM}$. The angular momentum change due to
this gas can thus be expressed as
\begin{equation}
\delta {\vec J}_{\rm gas,cooling}=\dot {M}_{\rm cool}\frac{{\vec
J}_{\rm DM}}{M_{\rm DM}}\delta t ,
\end{equation}
where $\dot M_{\rm cool}$ is the condensation rate from Equ.~(\ref{eq:mcool}) or
Equ.~(\ref{eq:mcool2}), and $\delta t$ is the timestep.  When a minor merger
happens (which we define as the smaller galaxy having a baryonic mass less
than a third that of the larger) we assume the cold gas from the smaller
object to be added to the disk of the larger (see Sec.~\ref{sec:merge}),
carrying specific angular momentum equal to the current value for the dark
matter halo of the larger object.  The corresponding angular momentum change
in the gas disk is thus
\begin{equation}
\delta {\vec J}_{\rm gas,acc}={M}_{\rm sat,gas}\frac{{\vec J}_{\rm DM}}{M_{\rm DM}} ,
\end{equation}
where ${M}_{\rm sat,gas}$ is the cold gas mass in the satellite disk. 

When some gas from the cold disk is converted into stars, we assume it to have
the average specific angular momentum of the gas disk, ${\vec
J}_{\rm gas}/M_{\rm gas}$, so the change in angular momentum of the gas and stellar
disks is
\begin{equation}
\delta {\vec J}_{\rm gas,SF}= -\dot{M}_{*}\frac{{\vec
J}_{\rm gas}}{M_{\rm gas}}\delta t= -\delta {\vec J}_{*,\rm SF} ,
\end{equation}
where $\dot{M}_{*}$ is the star formation rate. 

When the cold gas in disks is reheated by SN feedback, we assume that the
outflowing material also carries away its ``fair share'' of the angular
momentum. As a result, the specific angular momentum of the gas disk is not
changed by the SN feedback process.

For the stellar disk, we assume the {\it total} change in (vector) angular
momentum over the timestep to be $\delta {\vec J}_{*,\rm SF}$.  Thus we are
assuming that the angular momentum of the disk is changed {\it only} by star
formation. In particular, bulge formation through disk instability (see
Sect.~\ref{sec:bulge}) is assumed to remove negligible angular momentum from
the disk.

We assume both the gas disk and the stellar disk to be thin, to be in
centrifugal equilibrium and to have exponential density profiles
\begin{equation}
\Sigma(R_{\rm gas})=\Sigma_{\rm gas0}\exp(-R_{\rm gas}/R_{\rm gas,d}),
\end{equation}
and
\begin{equation}
\Sigma(R_{*})=\Sigma_{*\rm 0}\exp(-R_{*}/R_{*,\rm d}),
\end{equation}
where $R_{\rm gas,d}$ and $R_{*,\rm d}$ are the scale-lengths of the gas
and stellar disks, and $\Sigma_{\rm gas0}$ and $\Sigma_{*\rm 0}$ are
the corresponding central surface densities. Assuming a flat
circular velocity curve, as would hold for a galaxy with negligible
self-gravity in an isothermal dark matter halo, the scale-lengths can 
be calculated as
\begin{equation}
R_{\rm gas,d}=\frac{J_{\rm gas}/M_{\rm gas}}{2V_{\rm max}},
\end{equation}
and
\begin{equation}
R_{*,\rm d} = \frac{J_{*}/M_{*,\rm d}}{2V_{\rm max}},
\end{equation}
where $M_{\rm gas}$ and $M_{*,\rm d}$ are the total masses of the two
disks. A significant issue here is the dark matter response to the
  gravity of the baryons as they condense within the halo. The simplest model
  for this effect assumes adiabatic contraction within a spherical halo
  \citep[e.g.][]{Barnes1984, Blumenthal1986} and has often been adopted in
  galaxy formation models \citep[e.g.][]{Mo1998, Cole2000}. However, recent
  simulations suggest that this simple scheme overestimates the effect
  \citep[e.g.] [] {Gnedin2004, Abadi2010}. Most recently, \cite{Tissera2010}
  found disk maximum rotation velocities very similar to the maximum circular
  velocities found for dark matter only haloes formed from identical
  $\Lambda$CDM initial conditions. Here, for simplicity, we adopt $V_{\rm
  max}$, the maximum circular velocity of the surrounding dark matter halo, as
the typical rotation velocity for both stellar and HI disks. Note that we keep
the rotation velocities of satellite galaxies fixed after infall. This is
because the inner regions of the dark matter subhalo, which determine the
rotation velocity of the disks, change rather little until the subhalo is
about to be destroyed \citep[e.g.][]{Hayashi2003,Kazantzidis2004}.

\begin{figure}
\bc
\hspace{-0.6cm}
\resizebox{8.5cm}{!}{\includegraphics{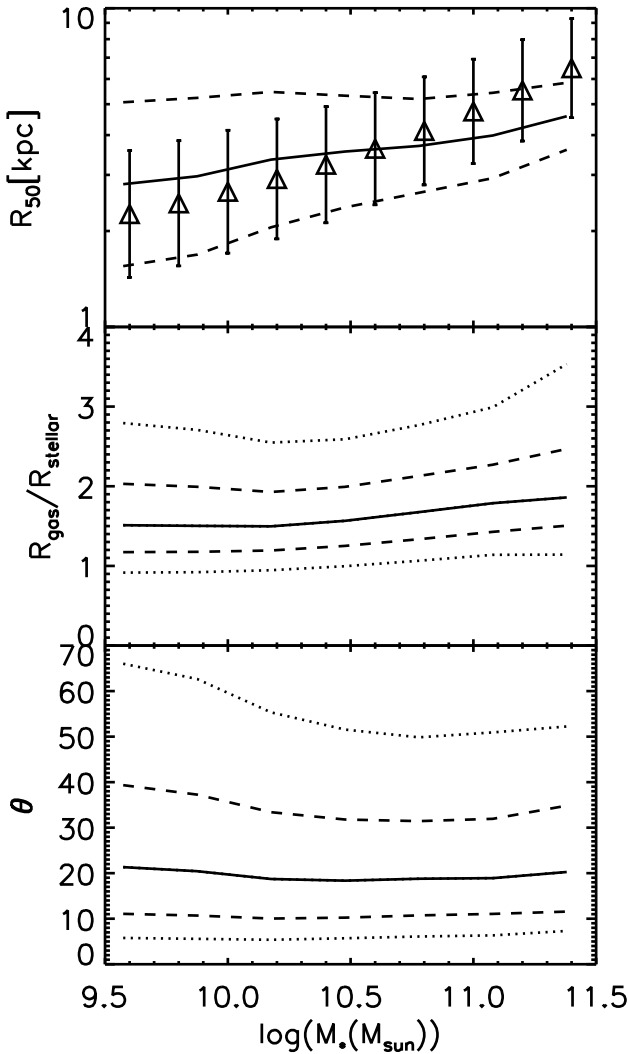}}\\%
\caption{The top panel gives the distribution of the radius containing 
half the stellar mass as a function of stellar mass for local late-type
galaxies. These are defined as having SDSS concentration parameter $c<2.86$
(see details in the text). The solid curve is the median half-mass radius
predicted by our model applied to the MS-II, while dashed curves indicate the
{\it rms} scatter in $\log R$ at each stellar mass. Symbols are the observed
median and scatter from the SDSS study by \cite{Shen2003}. The central panel
gives the 10, 25, 50, 75 and 90\% points of the distribution of the ratio of
sizes of the gaseous and stellar disks in our model, also as a function of
total stellar mass, while the bottom panel shows the same percentile points of
the distribution of the relative inclination of the two disks. }
\label{fig:disksize}
\ec
\end{figure}

Fig.~\ref{fig:disksize} shows a few results for this simple model implemented
on the MS-II to demonstrate that it gives results in fair agreement with
observation. The first panel compares model predictions for the distribution
of stellar half-mass radius for disk galaxies as a function of their stellar
mass to observational results from the Sloan Digital Sky
Survey \citep[SDSS,][]{Shen2003}. This SDSS study defined ``late-type''
galaxies according to the concentration parameter $c=R_{\rm 90}/R_{\rm 50}$, where
$R_{\rm 50}$ and $R_{\rm 90}$ are the radii which enclose 50\% and 90\% of the
projected stellar light. Galaxies with $c<2.86$ were taken to be late-type,
primarily spiral galaxies. To calculate $R_{\rm 50}$ and $R_{\rm 90}$ for our model
galaxies, we assume the above exponential model for the disk and a Jaffe
profile for the bulge (the modelling of bulge size will be described in
Sec.~\ref{sec:bulge}). In practice, $c<2.86$ corresponds roughly to galaxies
with $M_{*,\rm d}/M_{*,\rm tot} > 0.80$ in our model. The solid curves in the figure
show the median and $\pm 1\sigma$ range of the model distribution as a
function of stellar mass, while the symbols show the SDSS data. The amplitude,
slope and scatter of the observations are all fairly well reproduced, although
the predicted slope is somewhat shallower than observed. 

The second panel of Fig.~\ref{fig:disksize} shows the distribution of
the ratio of the sizes of the gas and stellar disks. It gives the
median, the upper and lower quartiles and the upper and lower 10\%
points as a function of stellar mass
for the same model galaxies plotted in the first panel of the figure. Gas
disks are typically larger than stellar disks by about a factor of
1.6 but the scatter in the ratio is large.  This agrees quite well
with the observational situation.  For the WHISP sample
of \cite{Noordermeer2005} the average ratio of disk sizes is 1.72 and values
within their sample of 49 galaxies range from 0.6 to 4.1.

The third panel of Fig.~\ref{fig:disksize} shows the distribution of the
misalignment angle $\theta = \arccos\left({\vec J}_{\rm gas}\cdot {\vec
J}_*/|{\vec J}_{\rm gas}| |{\vec J}_*|\right)$ between the two disks for several
ranges of stellar mass, again for the same galaxies. The distribution of
misalignment angles is quite broad and seems to depend very little on stellar
mass.  Warps are quite often seen in the outer parts of spiral galaxies,
particularly when the outer HI distribution is compared to the inner stellar
disk. The structure and evolution of the two components is quite strongly
coupled (e.g. \cite{Binney2008} section 6.6), but our simple model nevertheless gives
some indication of the extent to which misalignments might reflect changes
with time in the orientation of accreted angular momentum.

\subsection{Star Formation}
\label{sec:sfr}
In this paper we assume stars to form from cold gas in the disk according to
a simplified version of the empirical relation
which \cite{Kennicutt1998} found to give a good description of
galaxy-scale star formation in the bulk of low-redshift star-forming
galaxies.  Stars form efficiently only in regions where the surface
mass density exceeds a critical value which is plausibly related to
the \cite{Toomre1964} threshold for local instability of a
rotationally supported disk. Toomre's criterion is a function of local
velocity dispersion, of the surface densities of stars and gas, and of
the local rotation curve. We adopt a simple model which assumes a flat
rotation curve and a gas velocity dispersion which is everywhere 6
{\rm km/s}, leading to the critical surface density suggested
in \cite{Kauffmann1996} and \cite{Croton2006},
\begin{equation}
\Sigma_{\rm crit}(R)=12\times\left(\frac{V_{\rm max}}{200{\rm
km/s}}\right)\left(\frac{R}{10{\rm kpc}}\right)^{-1}
M_{\odot}{\rm pc}^{- 2}.
\end{equation}
We integrate this out to three exponential scale radii $R_{\rm gas,d}$ and then
divide by a factor of 2 to obtain a critical gas mass which is required for
any stars to form
\begin{equation}
\label{eq:mcrit} M_{\rm crit}=11.5\times
10^9\left(\frac{V_{\rm max}}{200{\rm km/s}}\right)\left(\frac{R_{\rm gas,d}}{10{\rm kpc}}\right)M_{\odot}.
\end{equation}
The final reduction by a factor of 2 is introduced to agree with the
assumptions of \cite{Croton2006} who took the cold gas surface
density to be constant with radius in disks at threshold.

The amount of cold disk gas that is converted into long-lived stars
per unit time is assumed to be
\begin{equation}
\dot{M}_*=\alpha(M_{\rm gas}-M_{\rm crit})/t_{\rm dyn}
\end{equation}
where $t_{\rm dyn} = 3R_{\rm gas,d}/V_{\rm max}$ is the characteristic
timescale at the edge of the star-forming disk, and $\alpha$ is an
adjustable efficiency parameter. We will adopt $\alpha=0.02$, which
results in a few percent of the gas in a disk being converted into
stars each rotation period. The star formation rates implied by this
model are, in the mean, quite similar to those in DLB07, but our
revised treatments of cooling and of disk size lead to considerably
smoother evolution than before, with less ``bursty'' star formation
histories in the bulk of the galaxy population.

\subsection{Supernova Feedback}
\label{sec:SNback}

During their short lives, massive stars emit large amounts of radiation
through optical and UV emission, and large amounts of mechanical energy
through their winds. As they die, comparable amounts of radiative and
mechanical energy are liberated by the final supernova (SN) explosion.  This
can dramatically reshape the surrounding interstellar medium, ionising and
heating it, and in many cases driving galactic-scale winds, Such effects are
generically referred to as SN feedback. As \cite{Larson1974} showed, they can
have a major impact on the evolution of low-mass galaxies, determining, for
example, their metallicities. \cite{White1978} argued that such SN
  feedback may induce the strong dependence of galaxy formation efficiency on
  halo mass required to explain why most stars live in galaxies with stellar
  mass close to the upper limit imposed by cooling constraints, and why the
  overall conversion of baryons into stars is relatively inefficient.  These
  ideas have subsequently been explored by many authors, particularly in the
  context of understanding the shape of the galaxy luminosity function at low
  luminosities \citep[e.g.][]{Benson2003a}. Here we assume that SN feedback
injects gas from the cold disk into the hot halo and, in addition, can
transfer halo gas to the ejecta reservoir.

We estimate the amount of cold disk gas that is reheated by SN
feedback and injected into the hot halo component as
\begin{equation}
\delta M_{\rm reheat} = \epsilon_{\rm disk} \times \delta M_* .
\end{equation}
where $\delta M_*$ is the mass of newly formed long-lived stars. DLB07 took
$\epsilon_{\rm disk}$ to be a constant, based on some observational indications
that mass outflow rates are typically a few times the star formation rate in
actively star-forming galaxies. We find that this scaling does not suppress
star formation in low-mass galaxies enough to reproduce the shallow slope of
the observed stellar mass function, so we have extended it to allow higher
ejection efficiencies in dwarf galaxies, taking
\begin{equation}
\epsilon_{\rm disk}=\epsilon\times\left[0.5+\left(\frac{V_{\rm
        max}}{70{\rm km/s}}\right)^{-\beta_1}\right],
\end{equation}
where $\epsilon$ and $\beta_1$ are free parameters describing the ratio of
reheated mass to new stellar mass in massive galaxies, and the scaling of this
ratio with $V_{\rm max}$ in dwarfs. The circular velocity dependence is motivated
by the fact that less energy is needed to heat a solar mass of gas to the halo
virial temperature and to eject it from the disk in lower mass galaxies. A
naive argument leads to the expectation $\beta_1\sim 2$, but a variety of
factors could lead to a different dependence, so we adjust both $\beta_1$ and
$\epsilon$ when fitting to observations, in particular to the observed stellar
mass function. Below we will take $\epsilon =6.5$ and $\beta_1=3.5$ in our
standard model.

We parametrise the total amount of energy effectively injected by
massive stars into disk and halo gas as:
\begin{equation}
\label{eq:SNeject}
\delta E_{\rm SN} =\epsilon_{\rm halo} \times\frac{1}{2}\delta M_* V_{\rm SN}^2.
\end{equation}
where $0.5V_{\rm SN}^2$ is the mean kinetic energy of supernova ejecta per unit
mass of stars formed, and, following \cite{Croton2006}, we take $V_{\rm SN}= 630 {\rm km/s}$, based on standard assumptions for the stellar initial mass function
and the energetics of SN explosions. In this case also, DLB07 assumed the
efficiency $\epsilon_{\rm halo}$ to be a constant. However, since dwarf galaxies
have stronger winds, lower metallicities and less dust than galaxies like our
own, it is plausible that radiative losses during the thermalisation of ejecta
energy are substantially smaller in dwarfs than in giants. We have therefore
allowed for this possibility explicitly by setting
\begin{equation}
\epsilon_{\rm halo}=\eta\times\left[0.5+\left(\frac{V_{\rm
        max}}{70{\rm km/s}}\right)^{-\beta_2}\right],
\end{equation}
where $\eta$ is a free parameter which encodes possible variations about our
IMF and SN assumptions, possible energy input from the winds and UV radiation
of massive stars, and the radiative losses during ejecta thermalisation, while
$\beta_2$ describes the dependence of this last factor on $V_{\rm max}$. Again, we
adjust these parameters when fitting to observations of the stellar mass
function and gas-to-star ratios of galaxies.  Our standard model below adopts
$\eta = 0.32$ and $\beta_2=3.5$.  We expect that $\epsilon_{\rm halo} < 1$ and,
unlike DLB07 or \cite{Bower2006}, we enforce this constraint in
all our models.

Given this energy input into disk and halo gas, the total amount of material
that can ejected from a halo/subhalo can be estimated as
\begin{equation}
\label{eq:Meject}
\delta
M_{\rm ejec}=\frac{\delta E_{\rm SN}-\frac{1}{2}\delta
M_{\rm reheat}V^2_{\rm vir}}{\frac{1}{2}V^2_{\rm vir}}. 
\end{equation}
If this equation gives $\delta M_{\rm ejec} <0$, we assume that the mass of
reheated gas saturates at $\delta M_{\rm reheat} = 
\delta{E_{\rm SN}}/(\frac{1}{2}V^2_{\rm vir})$ and that no gas is ejected from the
halo/subhalo.  

The reheating and ejection efficiencies, $\epsilon_{\rm disk}$ and
$\epsilon_{\rm halo}$, decline with increasing halo circular velocity,
saturating at $0.5\epsilon$ and $0.5\eta$, respectively. This
dependence is controlled by the values of $\beta_1$ and $\beta_2$
which are chosen to fit the abundance of low-mass galaxies.  $\beta_1$
primarily affects the low-mass slope of the stellar mass function,
while $\beta_2$ affects its amplitude. Our default model has a very
strong $V_{\rm max}$-dependence, $\beta_1=\beta_2=3.5$, but because of
saturation effects the results are not very sensitive to this
choice. For example, the adoption of $\beta_1=\beta_2=1.5$ predicts a
stellar mass function only slightly steeper than our default model and
overpredicts the abundance of galaxies of stellar mass $10^8M_{\odot}$
by 0.1 dex compared to the default model. This dependence of SN
feedback on $V_{\rm max}$ also affects the metallicities of low-mass
galaxies (see details in Sec.~\ref{sec:metals}). Compared to DLB07,
our model gives stronger feedback at low circular velocities. This is
the primary reason that it produces fewer dwarf galaxies and that
these have lower metallicities than in the earlier model.

The gas mass $M_{\rm ejec}$ thrown out of a system by these effects is stored in
an ejecta ``reservoir'' associated with the halo/subhalo, whence it may later be
reincorporated into the hot halo gas and so again become  available for 
cooling onto the central galaxy. In low-mass halos, hot winds are likely
to leave at a substantially higher velocity relative to the escape velocity
and so their gas is likely to be more difficult to reaccrete. To allow for
this possibility, we introduce a dependence of the reaccretion rate on
halo/subhalo virial velocity,
\begin{equation}
\dot M_{\rm ejec}=-\gamma\left(\frac{V_{\rm vir}}{220{\rm km/s}}\right)\left(\frac{M_{\rm ejec}}
{t_{\rm dyn,h}}\right),
\end{equation}
where $\gamma$ is a free parameter which we take to be 0.3. With these choices,
ejected gas is returned to the hot halo component in a few dynamical times for
galaxies like the Milky Way, but takes substantially longer in dwarf systems.

The association of hot gaseous halos and ejecta reservoirs with satellite
subhalos is a substantial change in our model with respect to DLB07. As
detailed in the next subsection we explicitly model the stripping of these
components by tidal and ram-pressure effects.

\subsection{Satellite Galaxies in Groups and Clusters}
\label{sec:stripping}

In the following, we classify galaxies into three types according to their
relationship to the dark matter distribution. Type 0 galaxies are the central
galaxies of main subhalos and so can be considered as the principal galaxies
of their FOF groups. Type 1 (satellite) galaxies lie at the centre of non-dominant
subhalos, while type 2 (satellite) galaxies are those which no longer have an
associated dark matter subhalo which is resolved by the simulation. The latter
are often referred to as ``orphan galaxies''.  All galaxies are born as
type 0. They usually become type 1 when they fall into a group or cluster, and
they may later become type 2. Type 2's may later merge into the central galaxy
of their halo.

FOF halos often link together two (or more) essentially disjoint dark matter
structures, joining them with low-density ``bridges'' of particles.  In such a
situation, one would expect the central galaxies of the various objects to
evolve independently until the smaller ones actually fall into the main body
of the system. To represent this we have changed the treatment of type 1
galaxies from that in DLB07.  When a type 0 galaxy first becomes type 1
(i.e. its FOF halo is first linked to a more massive FOF halo) we continue to
treat it as a type 0 galaxy (i.e. in the same manner as a galaxy at the centre
of a main subhalo) until it falls within $R_{\rm vir}$ of the centre of its new
FOF halo. At this point we switch on tidal and ram-pressure stripping
processes which can remove gas from the galaxy or even disrupt it
completely. In our model such processes {\it only} occur within $R_{\rm vir}$ so
that if a satellite passes outside $R_{\rm vir}$ again it is once more treated in
the same way as a type 0 galaxy.\footnote{Since only the main subhalo of a FOF
halo has an associated $R_{\rm vir}$, this quantity is not available for
``independent'' type 1 galaxies outside the $R_{\rm vir}$ of their new FOF
halo. For such objects we use the values of $R_{\rm vir}$ and $M_{\rm vir}$ recorded
at the last time they were type 0's when values of these quantities are
required in our cooling recipe.} This change primarily affects
galaxies between $R_{\rm vir}$ and the boundary of the FOF group. It leads to a
reduction in the number of ``true'' satellite galaxies (e.g. galaxies whose
evolution is influenced by being a non-central object within a larger
system). We discuss the number of galaxies affected by this change, as well as
the overall number of satellites and of orphans in our MS and MS-II models, in
Appendix A.

\subsubsection{Gas Stripping}

In most semi-analytic models, hot gas associated with a halo is assumed to be
stripped immediately after accretion onto a larger system, leading to a rapid
decline in star formation and a reddening in colour \citep[e.g.][]{Wang2007,
Weinmann2006,Baldry2006,Font2008}. In the real Universe \citep{Sun2007,
Jeltema2008} and in hydrodynamic simulations, however, the hot atmosphere of
massive satellite galaxies may survive for some considerable time after
accretion. \cite{McCarthy2008} found that for satellite galaxies with typical
structural and orbital parameters, up to 30\% of the initial hot halo gas can
remain in place for up to 10 Gyr. \cite{Weinmann2009} and \cite{Font2008}
constructed MS-based models for incremental, rather than instantaneous removal
of material through tidal stripping and ram-pressure stripping. In the
following we describe how we include both mechanisms in our own models, which
are similar to but different in detail from those of \cite{Weinmann2009}
and \cite{Font2008}.

We assume the hot gas in a subhalo to have a distribution that exactly
parallels that of the dark matter. Since tidal acceleration acts identically
on hot gas and dark matter at each location, we take the hot gas mass to be
reduced by tidal stripping in direct proportion to the subhalo's dark matter
mass. The latter is, of course, followed explicitly in a dynamically
consistent way by the original simulation. Thus we assume
\begin{equation}
\frac{M_{\rm hot}(R_{\rm tidal})}{M_{\rm hot,infall}}= \frac{M_{\rm DM}}{M_{\rm DM,infall}},
\end{equation}
where $M_{\rm DM,infall}$ and $M_{\rm hot,infall}$ were the DM mass of
the subhalo and the mass of its associated hot gas when its central
galaxy was last a type 0, and $M_{\rm DM}$ and $M_{\rm hot}$ are the
current masses of these two components. Recall that we assume
$\rho\propto r^{-2}$ for the hot gas distribution, thus $M_{\rm
hot}(r) \propto r$.  The tidal radius beyond which the hot gas is
stripped can be thus expressed as
\begin{equation}
R_{\rm tidal}=\left(\frac{M_{\rm DM}}{M_{\rm DM,infall}}\right)R_{\rm DM,infall}
\end{equation}
where $R_{\rm DM,infall}$ was the virial radius of the subhalo just before it
became a satellite.

In addition to tidal forces, the hot gas around satellite galaxies experiences
ram-pressure forces due to satellite's motion through the intracluster
medium (ICM). At a certain distance, $R_{\rm r.p.}$, from the centre of the
satellite, self-gravity is approximately balanced by this ram pressure:
\begin{equation}
\rho_{\rm sat}(R_{\rm r.p.})V_{\rm sat}^2=\rho_{\rm par}(R)V_{\rm orbit}^2,
\end{equation}
where $\rho_{\rm sat}(R_{\rm r.p.})$ is the hot gas density of the satellite at radius
$R_{\rm r.p.}$, $V_{\rm sat}$ is the virial velocity of the subhalo at infall (which
we assume to be constant as the subhalo orbits around the main halo),
$\rho_{\rm par}(R)$ is the hot gas density of the parent dark matter halo at
distance $R$ from the centre of its potential well, and $V_{\rm orbit}$ is the
orbital velocity of the satellite, which we estimate as the virial circular
velocity of the main halo. The ram-pressure dominates over gravity beyond
$R_{\rm r.p.}$ and hot gas at these radii is stripped.

We compare the two radii $R_{\rm tidal}$ and $R_{\rm r.p.}$ and define the minimum of
the two as the stripping radius
\begin{equation}
R_{\rm strip}=min(R_{\rm tidal},R_{\rm r.p.}).
\end{equation}
Beyond $R_{\rm strip}$, we assume all the hot gas to be removed without modifying
the gas profile within $R_{\rm strip}$. Thus the cooling rate onto the centre is
not affected, to lowest order, by this stripping. We assume gas in the
``ejecta reservoir'' of a subhalo to be stripped in proportion to the hot gas.
It is unclear where this reservoir should be located and whether or not it
will be affected by ram-pressure effects (e.g. whether it is primarily diffuse
or in relatively dense clouds). Thus we adopt the simple approach of stripping
it in proportion to the hot gas. Stripped material from each of these
components is added to the corresponding component associated with the central
(type 0) galaxy of the main subhalo, and so can never fall back onto its
original subhalo.

In addition to stripping, at least two other processes affect the hot gas
halos of satellites. One is cooling. The hot gas in satellite galaxies can
cool onto the central cold star-forming disk. We assume that the temperature
of the hot gas atmosphere is not changed by stripping and cooling processes,
remaining pegged to its value at infall. The cooling rate is calculated just
as in Sec.~\ref{sec:cooling}, which ensures continuity in cooling rates as
central galaxies turn into satellite galaxies. As cooling depletes the hot
atmosphere, we assume its density to drop everywhere, but its profile shape
and bounding radius to remain the same.

SN feedback also modifies the hot atmospheres and ejecta reservoirs of
satellite galaxies. As in central galaxies, star formation in
satellites releases large amounts of energy, reheating both cold ISM
gas and the hot gas atmosphere. \cite{Font2008} presented a model in
which both the hot and the reheated gas of satellites is stripped
primarily in the initial infall event.  They found that the satellite
galaxy properties are very sensitive to the way the secondary reheated
gas (which is only reheated after the galaxy has become a satellite)
is stripped. If it is stripped as in the initial infall event,
satellite galaxies lose all gas and become red very rapidly. In order
to retain gas and keep satellites blue for longer, they adopted a
stripping efficiency for this secondary reheated gas which is only
10\% of that at infall, and they do not strip any of the hot gas after
the initial stripping event. In our model, we have adopted continuous
stripping, in which hot and ejected components are stripped equally at
each timestep as long as the galaxy is a satellite. We assume the
reheated gas to extend to a radius equal to the virial radius of the
subhalo at infall. Taking into account the stripping mechanisms
discussed above, only reheated gas within $R_{\rm strip}$ (i.e. a
fraction $R_{\rm strip}/R_{\rm DM,inf}$ of the total reheated gas)
remains in the subhalo; the rest is added to the hot atmosphere of the
main halo. If SN is strong enough to predict that material should be
ejected from the subhalo altogether, then we use the formulae given
above for central galaxies (Equations~(\ref{eq:SNeject}) and
~(\ref{eq:Meject})), and distribute the ejected material between the
ejecta reservoirs of the satellite and central galaxies in the same
proportions as the reheated gas.  In general, the stripping of gas in
our model is more efficient than in \cite{Font2008}, but considerably
less efficient than in DLB07.

Our current model differs from that of DLB07 both because galaxies effectively
become satellites later (when they cross $R_{\rm vir}$ rather than when they
become part of a larger FOF group) and because satellites retain their hot gas
components and ejecta reservoirs until these are removed by stripping (rather
than losing them as soon as they become satellites).  Satellites thus retain
more fuel for star formation and can be expected to stay blue longer. Note
that ram-pressure stripping does not affect the cold gas component of
galaxies in our models. This is unrealistic for galaxies in the inner regions
of rich clusters and results in passive S0 galaxies retaining significant
gas and dust which should probably have been removed (see Fig.~\ref{fig:color}).

\begin{figure}
\bc
\hspace{-0.6cm}
\resizebox{8.5cm}{!}{\includegraphics{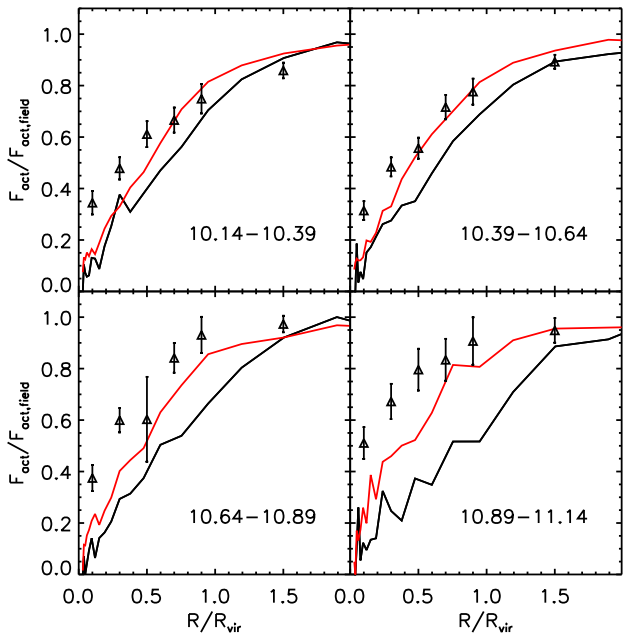}}\\%
\caption{The reduction in the fraction of  actively star-forming galaxies ($\dot{M}_*/M_* >
10^{-11} {\rm yr}^{-1}$) as a function of projected distance from
cluster centre in units of $R_{\rm vir}$. The four panels refer to
different ranges of $\log M_*/M_\odot$ as indicated by the
labels. Predictions from the preferred model of this paper applied to
the MS are shown in red, those from the model of DLB07 in
black. Symbols with error bars are SDSS data for a large sample of
nearby clusters taken from \cite{Weinmann2009}. For each curve the fraction of actively
star-forming galaxies is normalised by its ``field'' value, taken to be the
value at 20 $R_{\rm vir}$. This emphasises the effect of
cluster environment on star formation activity.}
\label{fig:satcolor}
\ec
\end{figure}

We illustrate the effect of our modification of stripping recipes in
Fig.~\ref{fig:satcolor}. We select 1000 galaxy clusters from the MS with
$M_{\rm vir}>2\times10^{14}M_{\odot}$. For each, we calculate the fraction of
actively star-forming galaxies as a function of projected distance from the
centre in units of $R_{\rm vir}$, and we average over all clusters.
``Actively star-forming'' here means having a specific star formation rate
(SSFR, the ratio of star formation rate to stellar mass) above
$10^{-11}$yr$^{-1}$.  We consider galaxies with velocity relative to cluster
centre (peculiar + $H_0\times$ line-of-sight distance difference) less than 3
$\times$ V$_{\rm vir}$, dividing them into four stellar mass bins as indicated
by the $\log M_*$ ranges given in the bottom right corner of each panel. To
emphasize the environmental effects which concern us here, each active galaxy
fraction is normalized to its ``field'' value, estimated at 20$R_{\rm vir}$.
Symbols with error bars are observational data from \cite{Weinmann2009} based
on the SDSS cluster sample of \cite{vonderLinden2007}. Predictions from our
model are in red, those from the model of DLB07 in black. Clearly, within
$R_{\rm vir}$, the changes we have introduced do slow the decline of star
formation in satellite galaxies, although the differences are not large. This is because satellites start to be stripped later in our new recipes, and
even thereafter they retain some hot gas which can fuel star formation and
keep them blue, whereas in DLB07, hot gas is stripped immediately once
galaxies become attached to a larger FOF group and star formation ceases once
their cold ISM gas is used up. Note, however, that the fraction of active
galaxies in the field differs between our model and DLB07, with our model
predicting somewhat lower active fractions in general, thereby worsening the
overall agreement with observation. This is a result of the enhanced feedback
we have introduced in order to match the observed stellar mass function (see
Sec~\ref{sec:bh}).

\subsubsection{Disruption}
\label{sec:disrup}

The stellar component in subhalos can also be stripped in the presence of very
strong tidal forces.  Usually, the galaxy is harder to disrupt than its dark
halo because it is more compact and denser.  We thus assume that the stellar
component of a satellite galaxy is affected by tidal forces only after its
subhalo has been entirely disrupted, i.e. it has become a type 2 galaxy. The
position of such a galaxy is identified with that of the most bound particle
of its subhalo at the last time the subhalo could be identified. To estimate
when stripping of stars is important we assume the satellite orbits in a
singular isothermal potential,
\begin{equation}
\phi(R) =V_{\rm vir}^2\ln R .
\end{equation}
Assuming conservation of energy and angular momentum along the orbit, its
pericentric distance can be estimated from:
\begin{equation}
\left(\frac{R}{R_{\rm peri}}\right)^2=\frac{\ln R/R_{\rm peri}+\frac{1}{2}
\left(V/V_{\rm vir}\right)^2}{\frac{1}{2}\left(V_t/V_{\rm vir}\right)^2},
\end{equation}
where $R$ is the current distance of the satellite from halo centre, and
$V$ and $V_t$ are the velocity of the satellite galaxy with respect to halo
centre and its tangential part, respectively.

We compare the main halo density at pericentre with the average baryon mass
(cold gas mass + stellar mass) density of satellite within its half mass
radius. If
\begin{equation}
\label{eq:disrupt}
\frac{M_{\rm DM,halo}(R_{\rm peri})}{R_{\rm peri}^3}\equiv\rho_{\rm DM,halo} > \rho_{\rm sat} \equiv 
\frac{M_{\rm sat}}{R_{\rm sat,half}^3},
\end{equation}
we assume the satellite galaxy is disrupted entirely. Its stars are
then assigned to a population of intracluster stars (ICS) and its cold
gas and the associated metals are added to the hot gas atmosphere of
the halo central galaxy. Note that this calculation does not fully
account for dynamical friction effects on the satellite orbit, which
are underestimated by the simulation once the remaining mass of a
subhalo drops below the stellar mass of its associated galaxy. Note
also that we do not model continuous disruption. Rather, once
Equ.~(\ref{eq:disrupt}) is satisfied, satellite galaxies are disrupted
completely. When a central type 0 galaxy merges in to a larger system
and becomes a type 1 satellite, it carries its ICS with it until it
becomes a type 2 galaxy.  At this point, its current central galaxy
acquires its ICS.
\subsection{Mergers}
\label{sec:merge}
Mergers can occur between a central galaxy and a satellite galaxy, and between
two satellite galaxies. In the MS, the minimum resolved subhalo has a mass of
$2.3 \times 10^{10}M_{\odot}$. The stellar mass of the galaxy within a given
subhalo is thus smaller than the subhalo mass, except in the case of very
massive satellites. In the MS-II, however, the minimum subhalo mass is
$1.9 \times 10^{8}M_{\odot}$, and the stellar mass of a galaxy often becomes
larger than the mass of its host subhalo well before we lose the track of the
subhalo. In this situation the simulation no longer correctly follows the
expected decay of the satellite orbit through dynamical friction. In this
paper we therefore modify the DLB07 treatment of mergers, which estimated a
dynamical friction time until merging only once the satellite's subhalo is
fully disrupted. Here we estimate this decay time as soon as the mass of a
subhalo drops below that of the galaxy it contains, and we immediately set the
countdown clock for merging. The position and velocity of the satellite galaxy
are thereafter traced by the most bound particle of the subhalo at the time
the merger clock was switched on, modified by a time-dependent
orbit-shrinking factor which models the orbital decay caused by the dynamical
friction (see below). As in DLB07, we adopt the dynamical friction formula
of \cite{Binney1987} to estimate the merging time for a satellite galaxy,
\begin{equation}
t_{\rm friction}=\alpha_{\rm fric}\frac{V_{\rm vir}r_{\rm sat}^2}{Gm_{\rm sat}\ln\Lambda}.
\end{equation}
where $\alpha_{\rm fric}=2.34$. DLB07 found this coefficient to be needed to
reproduce observed luminosity functions at the luminous end. It was also found
to be appropriate in the N-body studies of \cite{Boylan2008}
and \cite{Jiang2008}. Unlike DLB07, we here take $m_{\rm sat}$ to be the sum
of the baryonic mass of the satellite galaxy and the dark matter mass of its
subhalo. The dark matter mass here refers to the mass of the subhalo just
before the merger clock is set. $r_{\rm sat}$ is the distance between the
central and satellite galaxies at the time when we start the merger clock, and
$\ln\Lambda=\ln (1+M_{\rm vir}/m_{\rm sat})$ is the Coulomb logarithm. After a
time $t_{\rm friction}$ the satellite galaxy is assumed to merge with the
central galaxy of the main halo. If a main halo is accreted onto a larger
system and becomes a subhalo, any of its satellites for which the merger clock
is already set are assumed to keep orbiting within this subhalo and to merge
into its central galaxy when the time runs out. In this way, our model allows
satellites to merge into the central galaxies of both main and subdominant
subhalos (although the latter is quite rare).

We have also attempted to model approximately the dynamical friction
driven orbital decay of type 2 galaxies which leads to their eventual
merging with the central galaxy, even though the low-mass subhalo or
the simulation particle with which the galaxy is associated clearly
experiences no such decay.  A simple model in which an ``isothermal''
satellite spirals to the centre of a larger ``isothermal'' host on a
near circular orbit predicts that the radius of the orbit should decay
linearly in time. To mimic this, we multiply the positional offset of
the tracer particle from the central galaxy with which its galaxy is
destined to merge by a factor $(1 - \Delta t/t_{\rm friction})$ to
define the position of the galaxy, where $\Delta t$ is the time since
the merger clock was initialised. The velocity of the galaxy is kept
equal to that of the tracer particle.

Our modelling differentiates between major and minor mergers. Major mergers
are those between galaxies with baryonic masses differing by less than a
factor of three. More extreme mass ratios are treated as minor mergers. During
a major merger, the disks of the progenitors are destroyed completely, leading
to the formation of a spheroidal remnant. In a minor merger, the disk of the
larger progenitor survives and accretes the cold gas and stellar components of
the small galaxy. In both cases, the merger triggers a starburst which we
represent using the ``collisional starburst'' model
of \cite{Somerville2001}. During the merger, a fraction, $e_{\rm burst}$, of the
cold gas of the merging galaxies is converted into stars, where
\begin{equation}
e_{\rm burst}=0.56\left(\frac{M_{\rm minor}}{M_{\rm major}}\right)^{0.7},
\end{equation}
and $M_{\rm minor}$ and $M_{\rm major}$ are the total baryon masses of the
minor and major progenitors, respectively. The coefficient and index here
are consistent with those given by \cite{Cox2008} and \cite{Somerville2008}.
The stars formed during major mergers contribute to the elliptical remenants,
while those formed during minor mergers are added to the disks. Feedback and
chemical enrichment from the starburst are modeled in the same way as for
quiescent star formation, and the strong SN feedback produced by a major
merger can expel almost all the remaining cold gas from the system,
suppressing further star formation until a new gas disk grows.

To summarize, our treatment of mergers differs from that of DLB07 only
in that we switch on the merger clock as soon as the dark matter mass
of a subhalos drops below the baryonic mass of its central galaxy,
that we take into account the baryonic mass of the galaxy when
calculating the dynamical friction time, and that we include an
approximate representation of the shrinkage of orbits caused by
dynamical friction. Many aspects of these recipes are quite crude but
they nevertheless represent reasonably well the results of recent
simulations of both gas-poor and gas-rich galaxy
mergers \citep[e.g.][]{Naab2003,Cox2008}.

 \subsection{Bulge Formation}
\label{sec:bulge}

\begin{figure}
\bc
\hspace{-0.6cm}
\resizebox{8.5cm}{!}{\includegraphics{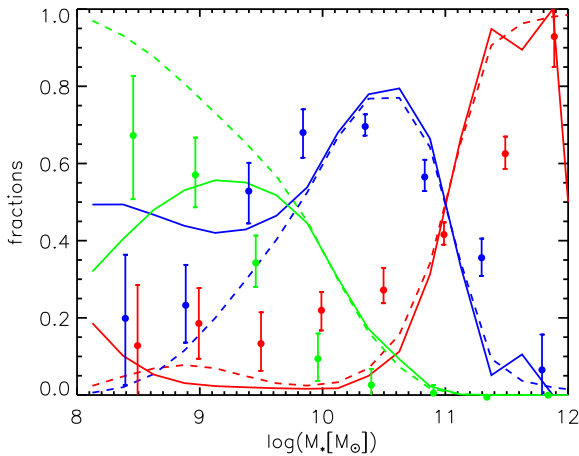}}\\%
\caption{The distribution of galaxies across morphological type as 
a function of stellar mass. Red lines show the fraction of galaxies
with $\frac{M_{\rm bulge}}{M_{\rm total}}>0.7$, which we consider to represent
elliptical galaxies. Blue lines indicate galaxies with
$0.03<\frac{M_{\rm bulge}}{M_{\rm total}}<0.7$ (normal spirals) and green 
indicate $\frac{M_{\rm bulge}}{M_{\rm total}}<0.03$, essentially pure-disk
or extreme late-type galaxies. Model results for the MS are shown with
dashed lines and for the MS-II with solid lines. The symbols give 
observational results for real galaxies from \cite{Conselice2006}.}
\label{fig:morphology}
\ec
\end{figure}

\begin{figure}
\bc
\hspace{-0.6cm}
\resizebox{8.5cm}{!}{\includegraphics{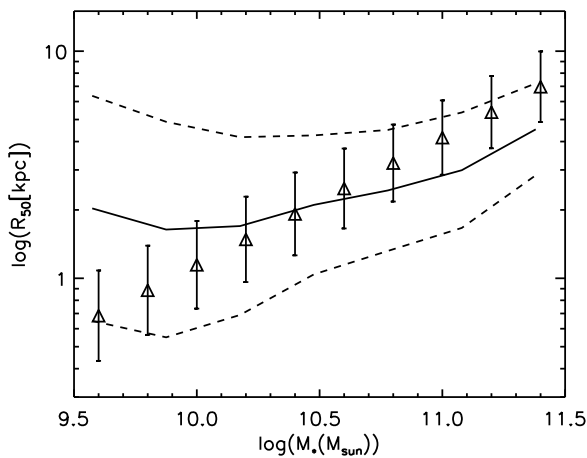}}\\%
\caption{Half-mass radius as a function of stellar mass for early-type 
galaxies, which we define as galaxies with SDSS concentration parameter
$c>2.86$. The solid curve gives the median half-mass radius predicted by our
model at each stellar mass, while dashed curves show the {\it rms} scatter in
$\log R$. Symbols with error bars indicate the median and {\it rms} scatter of
observational estimates taken from the SDSS study of \cite{Shen2003}. }
\label{fig:bulgesize}
\ec
\end{figure}

Three modes of bulge growth are included in our model: major mergers, minor
mergers and disk buckling.

After a major merger, all stars from the progenitors and all the newly
formed stars are assumed to end up in a spheroidal component.  After a
minor merger, the disk of the larger progenitor remains intact but its
bulge acquires all the pre-existing stars from the minor progenitor,
while the newly formed stars are added to the disk. In both cases, the
spheroidal component grows in mass and changes in size.  We use energy
conservation and the virial theorem to estimate the change in size:
\begin{equation}
C\frac{GM_{\rm new,bulge}^2}{R_{\rm new, bulge}}=C\frac{GM_1^2}{R_1}+C\frac{GM_2^2}{R_2}+\alpha_{\rm inter}\frac{GM_1 M_2}{R_1+R_2},
\label{eq:bulgesize}
\end{equation}
where C is a structure parameter relating the binding energy of a galaxy to
its mass and radius, and $\alpha_{\rm inter}$ is a parameter quantifying the
effective interaction energy deposited in the stellar components. $C = 0.49$
for an exponential disk whereas $C = 0.45$ for a bulge with an $r^{1/4}$
profile; to simplify, we adopt $C=0.5$. We also set $\alpha_{\rm inter}=0.5$,
so that $\alpha_{\rm inter}/C = 1$. This is roughly consistent with the
numerical simulation results of \cite{Boylan2005} which give $1.3<\alpha_{\rm
inter}/C<1.7$ for the most probable orbits of dissipationless major mergers of
elliptical galaxies. We prefer a slightly smaller value of $\alpha_{\rm
inter}$ because it gives bulge sizes in better agreement with observation (see
below). A similar formula was used by \cite{Cole2000}, but rather than
following subhalo mass directly, as in our approach, they used analytic
formulae to estimate the dark matter mass of satellites just prior to
merging. This dark matter mass was included in their analogue of
Equ.~\ref{eq:bulgesize}. In our model, the dark matter mass of satellites is
almost always very small (or zero) at the time they merge. Further, we assume
the final merger to be from a tightly bound orbit.  Thus we neglect the
effects of dark matter and use the stellar masses of the two objects in
Equ.~\ref{eq:bulgesize}.

The term on the left-hand side of Equ.~(\ref{eq:bulgesize}) is the
binding energy of the final bulge: $M_{\rm new,bulge}$ is its stellar
mass and $R_{\rm new,bulge}$ is its half-stellar-mass radius. The
first and second terms on the right-hand side are the self-binding
energies of the two individual progenitors, while the third term is
the binding energy invested in their orbit at merger.  For major
mergers, $M_1$ and $M_2$ are the sum of the mass of stars and of the
cold gas converted into stars for the two progenitors, and $R_1$ and
$R_2$ are the corresponding half mass radii. For minor mergers, $M_1$
and $R_1$ are the mass and half-mass radius of the bulge of the major
progenitor, and $M_2$ and $R_2$ are the stellar mass and the
half-stellar-mass radius of the minor progenitor.

Secular evolution is thought to be another important channel for the
formation of galaxy bulges, in particular in systems where the self-gravity of
the disk is dominant. Here we adopt the same simple, schematic criterion
as DLB07 to delineate disk instability:
\begin{equation}
V_{\rm max} < \sqrt{\frac{GM_{*,\rm d}}{3R_{*,\rm d}}}
\label{eq:stability}
\end{equation}
where $M_{*,\rm d}$ and $R_{*,\rm d}$ are the stellar mass and
exponential scale length of the stellar disk, and $V_{\rm max}$, as
usual, is the maximum circular velocity of the DM (sub)halo hosting
the disk. In the original presentation of this
criterion by \cite{Efstathiou1982} the factor of 3 was missing and
$V_{max}$ represented the maximum circular velocity of the combined
disk-halo system. The smaller coefficient used here reflects the facts
that this latter $V_{max}$ is expected to be significantly larger than
the maximum circular velocity of the unperturbed dark halo for
realistic systems near the instability boundary, and that more recent
simulations have shown exponential disks in NFW halos to be somewhat
more stable than would be inferred from the early experiments of
\cite{Efstathiou1982} \citep[see, for example,][]{Sellwood1999,Sellwood2001}.

When the criterion of Equ.~(\ref{eq:stability}) is met, we transfer mass,
$\delta M_*$, from the disk to the bulge to keep the disk marginally
stable. Recall that we assume an exponential density profile for the stellar
disk. Here we further assume that the mass is transferred from the inner part
of the disk and that the bulge formed in this way occupies the corresponding
region (i.e. the bulge half-mass radius equals to the radius of this region):
\begin{equation}
\delta M_*=2\pi\Sigma_{*0}R_{*,\rm d}[R_{*,\rm d}-(R_{\rm b}+R_{*,\rm
  d})\exp(-R_{\rm b}/R_{*,\rm d})],
\label{eq:bulgesize1}
\end{equation}
where $R_{\rm b}$ is the half-mass radius of the newly formed bulge, and
covers the region from which the stellar mass is transferred into the
bulge. We assume that negligible angular momentum is transferred to
the bulge from the disk with these stars so that the angular momentum
of the disk is unchanged.  Since we also assume an unchanged rotation
velocity and an exponential profile, the disk exponential scale length
increases and its central surface density decreases when stars are
transferred to the bulge.

If there is already a bulge present when a disk goes unstable, we assume the
instability to produce a new bulge with half mass radius $R_{\rm b}$ given by
Equ.~(\ref{eq:bulgesize1}), which ``merges'' into the existing bulge in the same
way as in galaxy mergers, simply replacing $M_{\rm 1}$ and $R_{\rm 1}$ with the mass and
half-mass radius of the existing bulge, and replacing $M_2$ and
$R_{\rm 2}$ with
$\delta M_*$ and $R_{\rm b}$. The only difference is that we set
$\alpha_{\rm inter}=2$
in this case, since the interaction energy between the ``old'' and ``new''
bulges is stronger than in the case of galaxy mergers since the two are
concentric.

To illustrate how well these recipes work, Fig.~\ref{fig:morphology}
compares observational data to model predictions for the distribution
of galaxies across morphological type as a function of stellar
mass. Red curves are for galaxies with $M_{\rm Bulge}/M_{\rm
total}\geq 0.7$ (``elliptical galaxies''), blue for galaxies with
$0.7>M_{\rm Bulge}/M_{\rm total}\geq 0.03$ (``normal spirals'') and
green for galaxies with $M_{\rm Bulge}/M_{\rm total}<0.03$ (``pure
disks''). Solid and dashed curves are results based on the MS-II and
the MS, respectively. The two simulations produce convergent results
above $\log M_* = 10$, but they differ progressively at lower stellar
masses because the resolution of the MS is no longer good enough to
follow accurately the detailed formation histories of the galaxies.
The symbols in Fig.~\ref{fig:morphology} are observational results
from \cite{Conselice2006}. These agree well with the models, provided
the MS-II results are taken at low stellar masses. To study the
relative roles in of disk instability and mergers in building bulges,
we calculated a model without the disk instability mode.  This showed
that in our default model, disk instability is a major contributor to
bulge formation in intermediate mass galaxies like the Milky Way. At
both higher and lower masses, mergers are the dominant mechanism; in
particular, massive elliptical galaxies are built by mergers. This is
consistent with the results found by \cite{Delucia2006} using our
previous galaxy formation model.

To illustrate how well our simple recipe reproduces the sizes of the
spheroidal components of galaxies, Fig.~\ref{fig:bulgesize} plots half-mass
radius against stellar mass for early-type galaxies defined to be those with
concentration parameter $c> 2.86$ (see \ref{sec:modeldisk} for how we estimate
$c$; in practice, this limit corresponds approximately to $M_{\rm
Bulge}/M_{\rm total}> 0.20$ ). A solid curve gives our model prediction for
the median half-mass radius at each stellar mass, while dashed curves indicate
the predicted scatter. The symbols are SDSS results for the median and scatter
from \cite{Shen2003}.  Overall, agreement is fair, at best. At lower masses,
our default model predicts a larger median value than is observed, perhaps
reflecting gas dissipation during gas-rich
mergers \citep[e.g.][]{Hopkins2009}. Indeed, recent work suggests that
including gas dissipation may exlain both the steep slope of the size
vs. stellar mass relation \citep{Dekel2006b}, and its small
scatter \citep{Covington2008}. It is also noticeable that the scatter in size
is larger in our simple model than in the SDSS data, particularly at low
mass. These deficiencies actually become worse if we restrict the sample to
more strongly bulge-dominated galaxies, since a significant part of the trend
in this figure is due to the size-stellar mass relation for disks highlighted
in Fig. 2. The small observed scatter has recently been confirmed for a large
sample of visually classified SDSS galaxies by \cite{Nair2010} who emphasise
that a tight relation may be difficult to understand if spheroids are built
stochastically through mergers. Our model confirms that this may be a problem
and that a more detailed theoretical treatment is warranted.

\subsection{Black Hole Growth and AGN feedback}
\label{sec:bh}
There is growing evidence that galactic nuclear activity is closely related to
galaxy formation. Here we follow \cite{Croton2006}, separating black hole
growth into two modes: ``quasar'' mode and ``radio'' mode.

The quasar mode applies to black hole growth during gas-rich mergers. During a
merger, the central black hole of the major progenitor grows both by
absorbing the central black hole of the minor progenitor, and by accreting
cold gas. The total growth in mass is calculated as
\begin{equation}
\delta M_{\rm BH} =
M_{\rm BH,min}+f\left(\frac{M_{\rm min}}{M_{\rm maj}}\right)\left(\frac{M_{\rm cold}} {1+(280 {\rm km/s}/V_{\rm vir})^2}\right),
\end{equation}
where $M_{\rm BH,min}$ is the mass of the black hole in the minor progenitor,
$M_{\rm cold}$ is the total cold gas in the two progenitors, and $M_{\rm min}$ and
$M_{\rm maj}$ are the total baryon masses of the minor and major progenitors,
respectively. Here $f$ is a free parameter, which, following \cite{Croton2006}
we set to 0.03 in order to reproduce the observed local $M_{\rm BH}-M_{\rm bulge}$
relation. Both major mergers and gas-rich minor mergers contribute
significantly to the growth of the black hole. We do not explicitly model
feedback due to this mode, which always coincides with a starburst in the
merging galaxies. Any feedback from accretion onto the black holes can be
thought of as being part of the substantial energy input which we assume this
starburst to produce. As noted above, this is often sufficient to eject all
the gas from the merger remnant.
 
Radio mode growth occurs through hot gas accretion onto central black holes.
The growth rate in this mode is calculated, following \cite{Croton2006}, as
\begin{equation}
\dot{M}_{\rm BH} = \kappa \left(\frac{f_{\rm hot}}{0.1}\right)\left(\frac{V_{\rm vir}}
{200kms^{-1}}\right)^{3}\left(\frac{M_{\rm
BH}}{10^8/{\rm h}M_{\odot}}\right)M_{\odot}/{\rm yr},
\end{equation}
where, for a main subhalo, the hot gas fraction, $f_{\rm hot}$, is the
ratio of hot gas mass, $M_{\rm hot}$, to subhalo dark matter mass
$M_{\rm DM}$, while for a type 1 galaxy in a satellite subhalo,
$f_{\rm hot}$ is the ratio of hot gas mass to dark matter mass within
$R_{\rm strip}$, $\frac{R_{\rm strip}}{R_{\rm DM,infall}}M_{\rm
DM,infall}$. The parameter $\kappa$ sets the efficiency of hot gas
accretion.
\begin{figure}
\bc
\hspace{-0.6cm}
\resizebox{8.5cm}{!}{\includegraphics{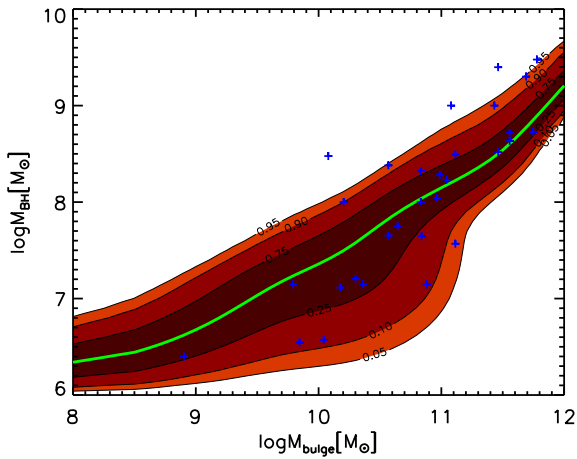}}\\%
\caption{The relation between black hole mass and bulge stellar mass at
$z=0$. Red contours give predictions from our model applied to the MS-II.  The
distributions in black hole mass are normalised to unity at each stellar mass
and the contours indicate their 5, 10, 25, 75, 90 and 95 percentage
points. The green curve represents the median values. Blue crosses are observational data taken from \cite{Haring2004}.}
\label{fig:BH}
\ec
\end{figure}
Again following \cite{Croton2006}, we assume that this hot mode accretion
deposits energy in relativistic jets with 10\% efficiency and that this energy
is then deposited as heat in the hot gas atmosphere, as is observed
directly through radio bubbles in galaxy clusters \citep[e.g.][]{McNamara2007, Birzan2004}. 
Specifically, we assume an energy input rate:
\begin{equation}
\dot{E}_{\rm radio}=0.1\dot{M}_{\rm BH}c^2,
\end{equation}
where $c$ is the speed of light.
The effective (net) mass cooling rate is thus
\begin{equation}
\dot M_{\rm cool,eff}=\dot M_{\rm cool}-2\dot{E}_{\rm radio}/V^2_{\rm vir}.
\end{equation}
Note this specific form of ``radio mode feedback'' is only one possible
description of AGN feedback, and other forms can give quite similar
results \citep[e.g.][]{Bower2006}. Moreover, there is no direct observational
evidence that radio mode feedback can offset cooling in halos with mass as low
as $\sim 10^{13}M_\odot$, where it plays an important role in our models.

In our preferred model the accretion efficiency $\kappa$ is set to be $1.5\times
10^{-5}$ in order to match the high-mass end of the stellar mass
function. This is twice the value adopted in DLB07 ($\kappa =7.5 \times
10^{-6}$).  There are three reasons for this change, in
addition to the fact that the new SDSS stellar mass functions cut off at high mass 
more strongly than the data used in DLB07. The first is that DLB07
assumed the hot gas mass of a halo to be $f_b^{\rm cos}M_{\rm vir}$ minus the baryonic
masses of all the galaxies associated with the FOF group, even those which lie
outside $R_{\rm vir}$. Here we substract only the baryonic masses of the galaxies
that lie inside $R_{\rm vir}$, resulting in higher estimates of $M_{\rm hot}$ and so
larger cooling rates which the radio mode feedback must offset.  The second
reason is that we have introduced a ``disruption'' mechanism which destroys
some type 2 galaxies which previously survived in galaxy clusters. The ISM of
these disrupted galaxies adds additional metal-rich material to the hot gas
atmosphere, again enhancing its predicted cooling rate relative to the
previous model. The final reason is that the enhanced feedback at low mass,
which we have introduced in order to match the observed $z=0$ stellar mass
function, results in more gas remaining available to cool at later times.
Note that our model assumes the hot gas in all systems to be distributed with
$\rho\propto r^{-2}$ at the virial temperature $T_{\rm vir}$. In reality, feedback
both from star formation and from an AGN may well change the profile of the
surrounding hot gas, making it less centrally concentrated and less able to
cool. This would result in less need for feedback at later times.
(See \cite{Bower2008} for a simple model based on this idea.) As may be seen
in Fig.~\ref{fig:BH}, the increased feedback efficiency in our new model does
not significantly affect its fit to the observed relation between the black
hole mass and bulge stellar mass. This is because black hole growth is in any
case dominated by the quasar mode.

Radio mode feedback works in essentially the same way in our model as
in \cite{Croton2006} and DLB07. It is more effective at low redshift and in
massive objects, both because the black hole is more massive, and because the
hot gas fraction is higher there. The effect has a very weak, if any,
dependence on large-scale environment \citep{Croton2008}. Note that our
model differs from DLB07 in that radio mode can also operate in satellite
galaxies at the centres of their own subhalos. In DLB07, such satellite
subhalos no longer retained any hot gas so that radio mode activity was
completely quenched there.

\subsection{Metal Enrichment}
Our treatment of metal enrichment follows that of \cite{DeLucia2004} quite
closely. Here we briefly summarise the various processes we include. As stars
evolve, both heavy elements and a fraction of the initial
stellar mass are returned instantaneously to the cold gas component of the
ISM. The new material is assumed to be fully mixed with the pre-existing cold
gas. A more realistic treatment should take into account the time delay
between star formation and the return of both mass and metals to the
interstellar medium. While the return of mass and metals from SN type II is
indeed effectively instantaneous for the purposes of galaxy evolution, the
same is not true for SN type Ia. In addition, mass loss and metal enrichment
from intermediate mass stars takes place over Gyr timescales and is also
important for a detailed understanding of metallicity patterns in galaxies. We
intend to implement these processes in future work. In our current model,
metals are carried into hot gas atmospheres and ejecta reservoirs when SN
feedback reheats cold disk gas and ejects it. Metals from both these
components can then be stripped from satellite galaxies and added to the
corresponding components of the host system. Reincorporation and cooling can
then take the metals into another (or the same) galaxy again.  A more detailed
description of metal enrichment and the exchange between different components
can be found in \cite{DeLucia2004}.

\subsection{Stellar Population Synthesis and Dust Extinction}

To compare model prediction with observations, we need to calculate the
photometric properties of our model galaxies.  Here we again follow DLB07,
using stellar population synthesis models from \cite{Bruzual2003}. We adopt a
Chabrier initial function which has fewer low-mass stars than a Salpeter IMF
and is a better fit to observational data both in our own Galaxy and in those
nearby early-type galaxies for which detailed dynamical data are available
\citep[e.g.][]{Cappellari2006}. A detailed description can be found
in \cite{DeLucia2004}. We also follow DLB07 and adopt a slab dust model to
account for the extinction of the star light. At higher redshift, we extend
this model as in \cite{Guo2009}. Extinction is modeled as a function of gas
column density, metallicity and redshift. The main difference from DLB07, is
that a redshift dependence is introduced so that for galaxies of given gas
metallicity, the gas-to-dust ratio is higher at high redshift than in the
local universe. This is motivated by observational data on high-redshift
galaxies \citep[e.g.][]{Steidel2004,Quadri2008}. \cite{Kitzbichler2007}
found that such a redshift dependence was needed for the DLB07 galaxy
formation model to reproduce faint galaxy counts and redshift distributions,
while \cite{Guo2009} needed it to reproduce the abundance and clustering pf
colour-selected galaxy populations at redshifts of 2 and 3 (see these papers
for details).

\begin{table*}
\caption{Summary of those parameters of our preferred model which were adjusted
primarily to fit the observed $z=0$ stellar mass function.}

\begin{tabular}{||l||l||c||} 

\hline
Parameter & Description  & Preferred value  \\
\hline
 $\alpha$       & Star formation efficiency (Sec.\ref{sec:sfr}) & 0.02 \\
 $\epsilon$     & Amplitude of SN reheating efficiency (Sec.~\ref{sec:SNback})& 6.5 \\
 $\beta_1$      & Slope of SN reheating efficiency (Sec.~\ref{sec:SNback})& 3.5 \\
  $\eta$     & Amplitude of SN ejection efficiency (Sec.~\ref{sec:SNback})& 0.32 \\
 $\beta_2$      & Slope of SN ejection efficiency (Sec.~\ref{sec:SNback})& 3.5 \\
 $\gamma$ & Ejecta reincorporation efficiency (Sec.~\ref{sec:SNback}) & 0.3 \\
 {$\it \kappa$} & Hot gas accretion efficiency onto black holes (Sec.~\ref{sec:bh})& 1.5 $\times$ 10$^{-5}$\\
\hline
\end{tabular} 
\label{table:para}
\end{table*}

\section{Systematic properties of the galaxy population}
\label{sec:results}
In the last section we set out our new galaxy formation model and clarified
the areas where it significantly alters or extends the earlier model of
DLB07. Several of these extensions involve processes which were not
previously included, notably the separate evolution of the sizes and
orientations of gaseous and stellar disks, the size evolution of spheroids,
tidal and ram-pressure stripping of satellite galaxies, and the disruption of
galaxies to produce intracluster light. To illustrate the effects of these new
ingredients, we have already presented a number of results from a simultaneous
application of the new model to the MS and MS-II. In the current section we
present a wide range of further results, primarily for the low-redshift
universe where recent data now constrain the galaxy population over a range
exceeding six orders of magnitude in stellar mass. By combining the MS and
MS-II we are able to test our model against observation over this full range -
the first time this has been possible using a direct simulation
technique. By combining the two simulations we are also able to check
explicitly how our results are effected by their limited mass resolution  
(as already done, for example, in Fig.~\ref{fig:morphology}).

We begin with a comparison of our model with the observed stellar mass
function of galaxies, because we use this as the primary constraint on the
various parameters in our star-formation and feedback models. We summarize
these parameters and the values we assign to them in our preferred model in
Table~\ref{table:para}. Note that other model parameters (for example, those
in our treatments of cooling, of disk and spheroid sizes, and of stripping,
merging and disruption) also affect the stellar mass function, but we have set
these to agree with other simulation or observational data, whereas the
parameters in Table~\ref{table:para} were chosen primarily to fit the stellar
mass function, and secondarily to ensure that gas-to-star ratios are in
reasonable accord with observation. Because of the coupling between different
parts of the model, an iterative method has to be used to find acceptable
parameter sets.  Those of our preferred model are almost certainly not unique,
but they all lie within the physically plausible range discussed, for example,
by \cite{Croton2006}. Indeed, where the meanings correspond, our parameters
are close to those presented in that paper and DLB07, except in a few cases
which we highlight individually.

\subsection{Stellar Mass  and Luminosity Functions}
\label{sec:GMF}
\begin{figure}
\bc
\hspace{-0.6cm}
\resizebox{8.5cm}{!}{\includegraphics{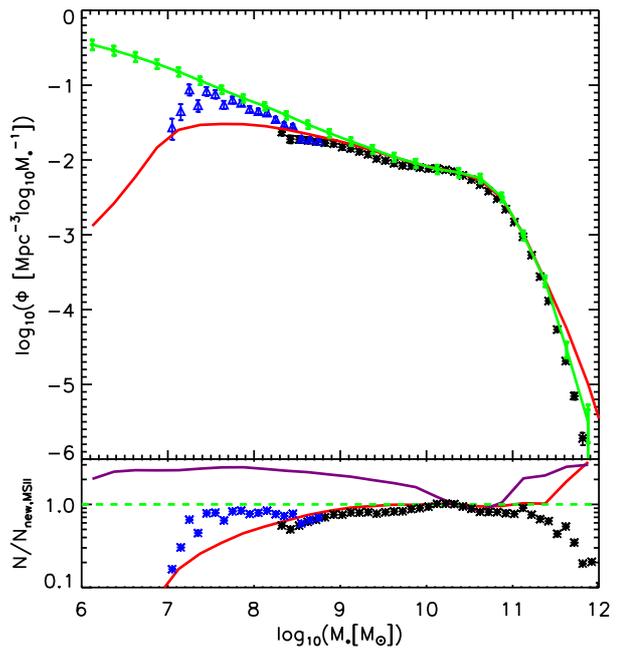}}\\%
\caption{The abundance of galaxies as a function of their stellar mass. 
In the upper panel, green and red curves give the predictions of our
preferred model when applied to the MS-II and the MS, respectively. The error
bars on the MS-II curve show a ``cosmic variance'' uncertainty estimated from
the {\it rms} scatter in the mass functions among 125 disjoint subvolumes of
the MS, each with volume equal to that of the MS-II. Stars with error bars are
the observational result, including cosmic variance uncertainties, for
SDSS/DR7 as given by \cite{Li2009} after a correction to total stellar masses
following \cite{Guo2010}. Blue triangles with error bars are the SDSS/DR4
results of \cite{Baldry2008}. These are corrected for surface brightness
incompleteness, but the error bars do not include cosmic variance
uncertainties which are quite large for these low-mass objects. In the
lower panel, black \citep{Li2009} and blue \citep{Baldry2008} symbols show the
abundance ratio of the SDSS data to our model prediction based on the
MS-II. The red curve is the ratio of our MS and MS-II predictions, while the
purple curve is the ratio of the DLB07 prediction to our MS-II prediction. A
dashed green line indicates a ratio of unity.}
\label{fig:MF}
\ec
\end{figure}

In Fig.~\ref{fig:MF} we compare the predictions of our preferred model to the
observed abundance of galaxies as a function of stellar mass. The solid green
curve is the prediction based on the MS-II, while the solid red curve is based
on the MS. The two converge well above a stellar mass of about
$3\times10^{9}M_{\odot}$, but at lower masses the MS underpredicts abundances
because it does not resolve halos less massive than $2.3\times
M^{10}M_{\odot}$, as compared to $1.9\times 10^{8}$ in the MS-II. At the
highest masses, the two simulations also diverge, but this is mainly due to
cosmic variance and the relatively small volume of the MS-II.  We estimate
this uncertainty by dividing the MS into 125 sub-cubes, each of the same
volume as the MS-II. The {\it rms} scatter among their individual stellar mass
functions is given by the error bars overplotted on the green MS-II
curve. \cite{Boylan2009} show that such differences become more prominent at
high redshift. Black stars are the observed stellar mass function estimated
from SDSS/DR7 by \cite{Li2009}, except that the masses are converted to total
stellar masses as described in Appendix A of \cite{Guo2010}. Note that these
observed stellar masses are also based on the same Chabrier IMF used in our
models, so that IMF uncertainties should not affect the comparison of the
two. The error bars here include cosmic variance uncertainties and are very
small, reflecting the large volume of the survey. Blue triangles are estimates
based on SDSS/DR4 from \cite{Baldry2008}. These include a correction for
surface brightness incompleteness, which becomes significant at these low
masses, but their error bars do not include cosmic variance which is quite
large because of the small volume effectively surveyed for such faint
galaxies. To make the comparison clearer, the lower panel
of Fig.~\ref{fig:MF} shows the SDSS data (the symbols), our MS prediction
(red curve) and the MS predictions of DLB07 (purple curve) all ratioed
to our predictions based on the MS-II.

It is clear from Fig.~\ref{fig:MF} that adopting the MS-II stellar mass
function below about $3\times10^{9}M_{\odot}$ and the MS function at higher
masses results in a very good match to the observational results for our
preferred parameters. The fit extends over the full range of the observations
from about $10^{12}M_{\odot}$ all the way down to about
$2\times10^7M_{\odot}$.  The slope at the low-mass end is around -1.46 in the
model, significantly steeper than the value of -1.155 quoted
by \cite{Li2009}. The \cite{Baldry2008} results suggest that this may
reflect the onset of incompleteness effects at the lowest masses considered
by \cite{Li2009}.  The high resolution of the MS-II allows us to predict
galaxy abundances to substantially lower stellar masses. Here we show our
predictions down to $10^{6}M_{\odot}$ although there are currently no
reliable observations with which to compare them. At this mass, the predicted
number density is around 0.3 Mpc$^{-3}(\log M_*)^{-1}$. Galaxies even less
massive than this {\it can} be observed in the Local Group and we show below
that our model does, in fact, agree quite well with the abundance of Milky Way
satellites as a function of luminosity (see
Sec.~\ref{sec:resultreionization}). 

At high stellar masses, where growth is limited by AGN feedback as
in \cite{Croton2006}, our model overpredicts the abundance found
by \cite{Li2009}. This likely reflects the observational difficulty in
estimating stellar masses for the most luminous cD galaxies in clusters.  As a
result of the problems with dealing with extended envelopes and crowded
fields, SDSS photometry gives luminosities for such galaxies which are
significantly lower (by up to one magnitude) than found in other
investigations \citep[e.g.][]{vonderLinden2007}. As the lower panel of
Fig.\ref{fig:MF} shows, our model agrees with these SDSS data significantly
better than the older model of DLB07.

\begin{figure*}
\bc
\hspace{-0.6cm}
\resizebox{12.5cm}{!}{\includegraphics{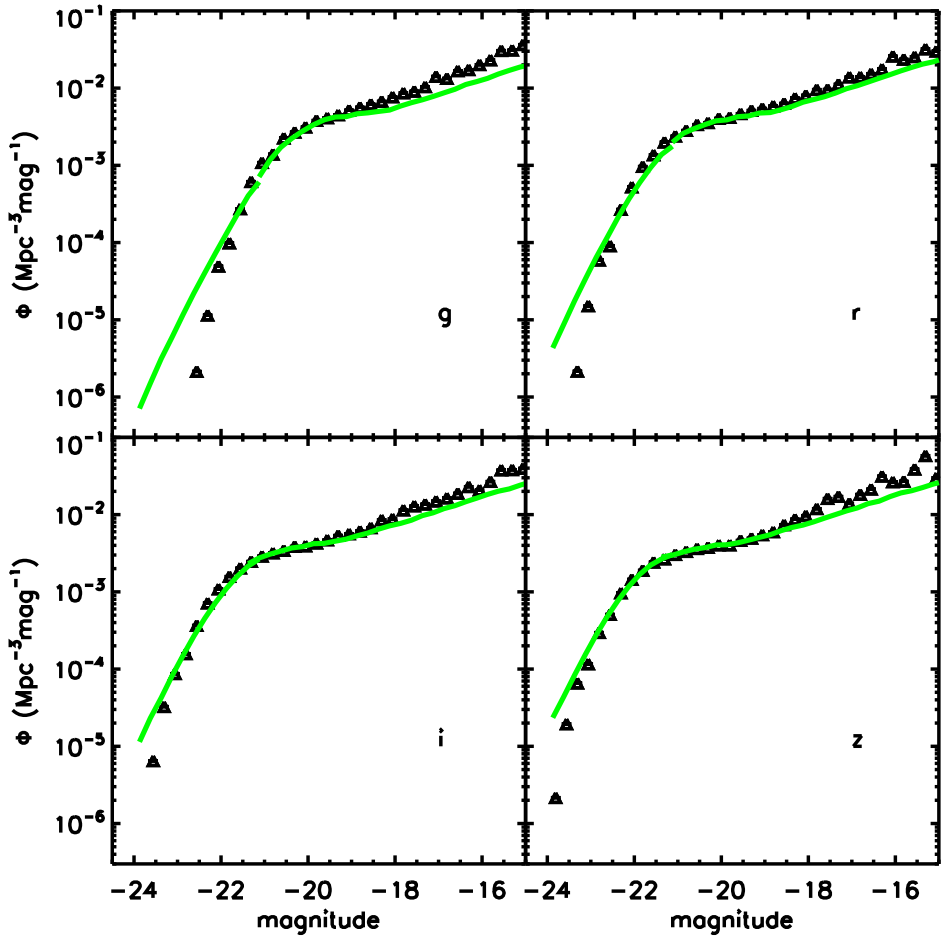}}\\%
\caption{Galaxy luminosity functions in the SDSS $g$, $r$, $i$ and $z$
photometric bands. The smooth green curves are predictions from our preferred
model taken from the MS+MS-II at high luminosities and from the MS-II alone at
absolute magnitudes fainter than about -20. The symbols are observational data
for a low-redshift SDSS sample taken from \cite{Blanton2005}.}
\label{fig:LF}
\ec
\end{figure*}

In Fig.~\ref{fig:LF} we show predictions of this same preferred model
for galaxy luminosity functions in the SDSS $g$, $r$, $i$ and $z$
bands, comparing them with observational data from a low-redshift SDSS
sample taken from \cite{Blanton2005}. In all these plots we have used
results from MS+MS-II at absolute magnitudes brighter than -20, where
results from the two simulations converge, and results from the MS-II
alone at fainter magnitudes. Given that Fig.~\ref{fig:MF} shows our
model to overpredict slightly the abundance of low-mass dwarf
galaxies, it is somewhat surprising that it turns out to underpredict
their abundance as a function of luminosity in all four bands.
Several effects may contribute to this discrepancy. One is that, as we
will see later, the fraction of non-star-forming dwarf galaxies
appears to be significantly larger in the model than in the SDSS data,
so we are probably assigning stellar mass-to-light ratios which are
too large to many dwarfs.  A second is that \cite{Blanton2005}
corrected their luminosity functions for incompleteness at low
surface-brightness, and their corrections may be larger than those
applied by \cite{Baldry2008} in the SDSS mass function estimate
plotted in Fig.~\ref{fig:MF}. Finally, quite small volumes are
surveyed when compiling luminosity functions for dwarfs, even with the
SDSS, so cosmic variance effects may be significant. The discrepancy
could then in part reflect differences in large-scale structure
between the small low-redshift volume surveyed by \cite{Blanton2005} and
DR7. The model also noticeably overpredicts the abundance of
very luminous galaxies in the $g$-band. These are massive galaxies
undergoing merger-induced starbursts, and it is likely that our simple
dust model is failing to predict enough extinction for these systems.

\subsection{The stellar mass -- halo mass relation}
\label{sec:starhalo}

\begin{figure}
\bc
\hspace{-0.6cm}
\resizebox{8.5cm}{!}{\includegraphics{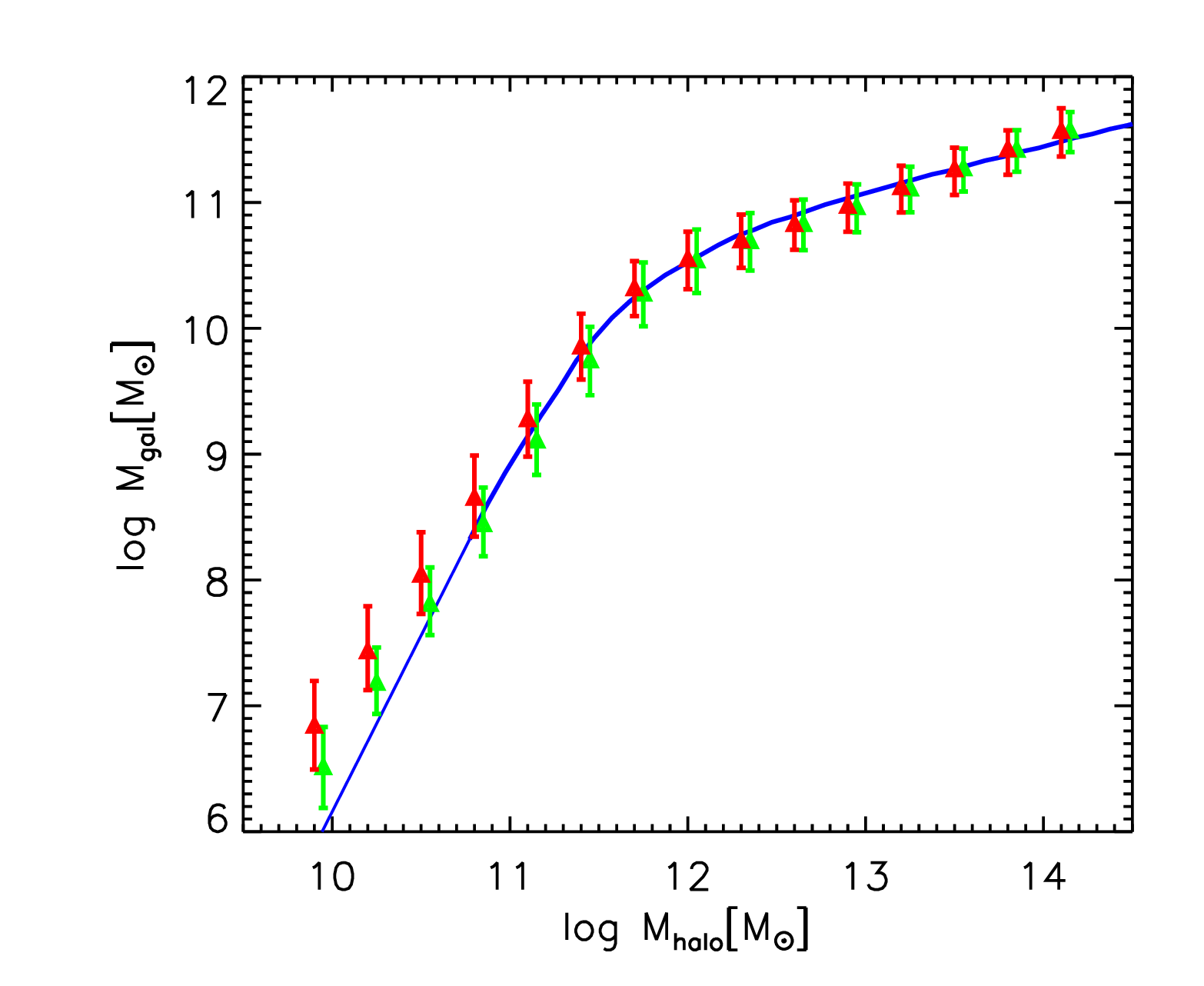}}\\%
\caption{Galaxy stellar mass as a function of maximum past halo mass. The
latter is the largest mass ever attained by the dark matter subhalo centred on
the galaxy over its full history. This is almost always the mass of the
subhalo at the last time its central galaxy was type 0, i.e. the present
subhalo mass for current type 0 galaxies and the subhalo mass just before
infall for current type 1 and 2 galaxies. Symbols with error bars show
predictions from our preferred model applied to the MS-II for $\log M_* < 10.$
and applied to the MS at higher masses. Green symbols are for central galaxies
(type 0) while red symbols are for satellites (types 1 and 2). The blue curve
is the relation derived directly from the SDSS stellar mass function and from
subhalo abundances in the MS and MS-II under the assumption that the two
quantities are monotonically related without scatter \citep{Guo2010}.}
\label{fig:gf}
\ec
\end{figure}

Simplified models for populating dark matter only simulations with galaxies
often assume a simple relation between the stellar mass of a galaxy and the
mass of the halo surrounding it -- more massive halos should contain more
massive galaxies at their centres. For such a model to represent galaxy
clustering even approximately, it must also place galaxies at the centres of
satellite subhalos, and the resolution of the simulation must therefore be
good enough that a subhalo corresponding to every galaxy can be
identified. Since tidal stripping often substantially reduces the masses of
satellite subhalos, but plausibly has little effect on the galaxies at their
centres, the stellar masses of such galaxies should be much more closely
related to the maximum masses ever attained by their halos than to their
current masses. This argument has led many authors to consider models which
populate simulations with galaxies assuming a simple monotonic relation
between the stellar mass of a galaxy and this maximum past halo mass 
\citep{Vale2004,Kravtsov2004,Conroy2006,Wetzel2009,Moster2010,Guo2010}. 
For cosmological simulations of high resolution, matching the
(sub)halo abundance as a function of maximum past mass to the observed
galaxy abundance as a function of stellar mass allows one to derive an
(assumed) monotonic relation beween the two masses. By using this
relation to populate the simulation, one can then predict the spatial
distribution of galaxies for detailed comparison with observation.

The MS-II simulation provides an unparalleled opportunity to carry
through this programme because, in combination with the MS, it gives a
much more precise estimate of the abundance of (sub)halos as a
function of maximum past mass than has previously been
available. \cite{Guo2010} matched an estimate based on both the MS and
the MS-II to the SDSS stellar mass function of \cite{Li2009},
producing the relation between stellar mass and maximum past halo mass
which we show as a blue curve in Fig.~\ref{fig:gf}. For comparison,
green and red symbols show the median value and the $\pm 1\sigma$
scatter of stellar mass predicted by our preferred model at given past
maximum halo mass for $z=0$ central and satellite galaxies,
respectively. Variations in assembly history and environmental
influence ensure that there is significant scatter in the relation for
our model; the {\it rms} scatter in $\log M_*$, is 0.17, 0.20, 0.24
and 0.31 for $\log M_{\rm halo} = 14,$ 13, 12 and 11 respectively. At
low mass, there is a noticeable offset between the predictions for
satellite and central galaxies, with satellites having systematically
larger stellar masses for given maximum past halo mass. This behaviour
was also present in the DLB07 model \citep[see,][]{Wang2006} and can be
traced to the fact that low-mass satellite galaxies typically
achieved their maximum halo mass at $z\sim 1$, whereas for the
corresponding centrals this is typically around $z\sim 0$. Since halos
are 8 times denser at $z=1$ than at $z=0$, their escape velocities at
given mass are roughly 40\% larger at the higher redshift and this
reduces the efficiency with which SN feedback can expel gas,
increasing the retention of baryons for star formation.

The median stellar mass predicted by our model at each maximum past
halo mass is very close to the \cite{Guo2010} relation at halo masses
above about $10^{11}M_\odot$, but lies noticeably above it at lower
masses. This is because \cite{Guo2010} extrapolated the \cite{Li2009}
stellar mass function to masses below $10^{8.3}M_\odot$ using their
quoted slope of $-1.15$, which predicts significantly fewer low-mass
dwarfs than the \cite{Baldry2008} function which we plot in
Fig.~\ref{fig:MF} and use to set the parameters of our preferred
model. In their own comparison of a similar relation to observational
data on satellite galaxy dynamics,
\cite{More2009} estimated the scatter in $\log L$ for relatively
massive halos ($\log M_{\rm vir} \sim 13$) to be $0.16\pm 0.04$.  This is in
good agreement with the scatter actually produced by our galaxy
formation model, but is considerably smaller than that predicted, for example,
by the models of \cite{Bower2006} or \cite{Font2008}.

\subsection{Gas-phase metal abundances}
\label{sec:metals}
 
\begin{figure}
\bc
\hspace{-0.6cm}
\resizebox{8.5cm}{!}{\includegraphics{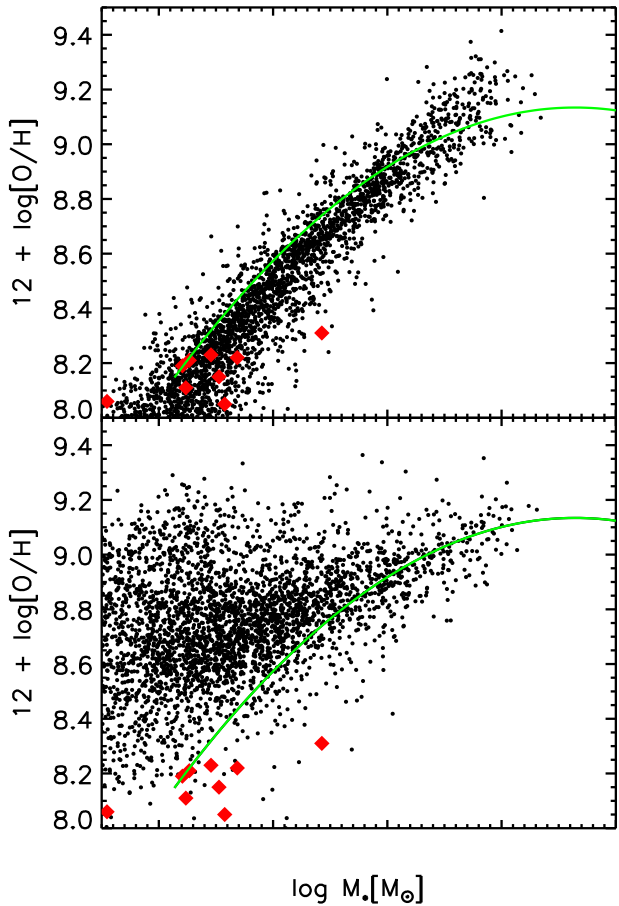}}\\%
\caption{Cold gas metallicity as a function of stellar mass. The top panel 
shows results for star-forming galaxies when our preferred model is applied to
the MS-II.  The bottom panel shows similar results but based instead on the
DLB07 model. In both panels, the solid curves represent observational results
for the SDSS from \cite{Tremonti2004}, while red diamonds are taken
from \cite{Lee2006}}
\label{fig:metals}
\ec
\end{figure}

In Fig.~\ref{fig:metals} we show the metallicity of the cold ISM gas
as a function of stellar mass for star-forming galaxies in our
preferred model. Here, we define as star-forming those galaxies with a
specific star formation rate $\dot{M}_*/M_* > 10^{-11}{\rm yr}^{-1}$.
Observational data from \cite{Tremonti2004} and \cite{Lee2006} are
represented by the solid green curve and the red diamonds,
respectively. When estimating the oxygen abundance of model galaxies
for comparison with these observations, we, for consistency, use the
same nucleosynthetic yields and solar abundances as DLB07.  Recent
work has suggested that these may need to be
revised \citep{Asplund2006,Delahaye2006} but the strong-line
metallicity measurements underlying the observational results in
Fig.~\ref{fig:metals} have substantial and controversial
uncertainties, so we prefer to keep our previous assumptions so that
the models can be easily compared.  Results for star-forming galaxies
in the MS-II are shown as small black dots in the upper panel of
Fig.~\ref{fig:metals}, which shows that our model appears to reproduce
the tight observed relation between gas metallicity and stellar mass
quite well. This is mainly due to our introducing a velocity
dependence in our SN feedback prescription, which leads to less star
formation and to more effective ejection of metals from low-mass
galaxies, thus to lower metallicities.
   
For comparison, in the bottom panel of Fig.~\ref{fig:metals} we show
the predictions obtained when the DLB07 model is applied to the MS-II.
In this model the SN feedback efficiency is assumed to be independent
of circular velocity ($\beta_1=\beta_2=0$) leading to a weaker
dependence of metallicity on stellar mass at low masses than our
current preferred model, as well as to an overabundance of dwarf
galaxies (see Fig.~\ref{fig:delucia}).  The better apparent agreement
of the DLB07 model with dwarf galaxy properties found in earlier
papers turns out to have been due largely to the limited resolution of
the MS.

\subsection{Galaxy colours}

The colours of galaxies are influenced strongly by dust, by their
star-formation histories, particularly by current and recent star
formation, and by the metallicities of their stars.
This makes colours especially difficult to predict with models of the kind we are
discussing, because they are sensitive not only to the details of stellar
population synthesis, but also to assumptions about the production and
quantity of dust, and about its distribution relative to the different
stellar populations. While population synthesis models have solid theoretical
foundations, are well developed and tested, and are probably reliable in most
situations, the opposite is true for dust modelling. For this reason, rather
than predicting the luminosities and colours of galaxies directly, it is often
safer to make model predictions for physical properties like stellar mass and
star formation rate, and to compare these with distributions inferred from
observation using methods designed to be as insensitive as possible to dust.
 
\begin{figure}
\bc
\hspace{-0.6cm}
\resizebox{8.5cm}{!}{\includegraphics{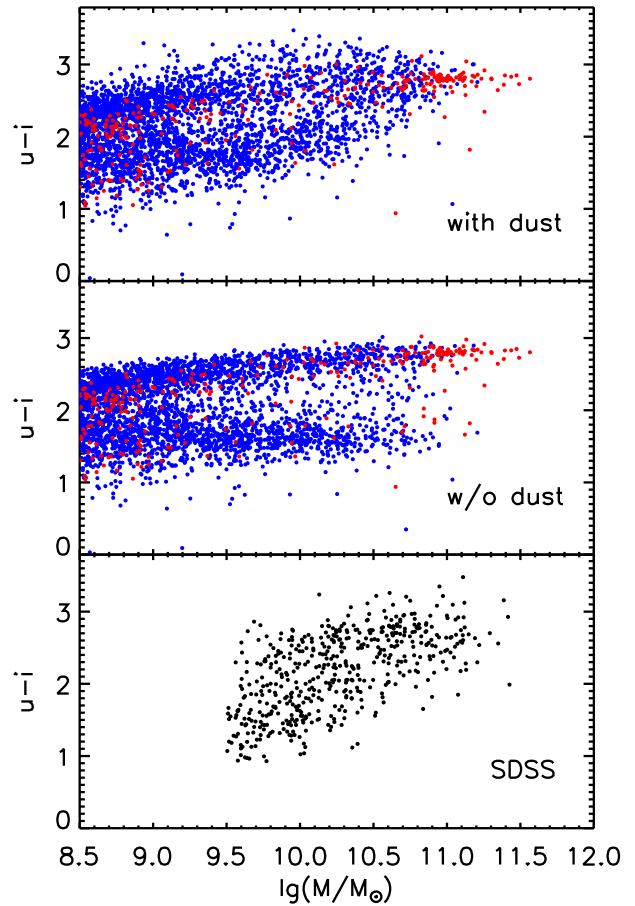}}\\%
\caption{$u-i$ colour as a function of stellar mass for galaxies in our
preferred model applied to the MS-II. The upper and central panels are for model
colours including and excluding dust extinction effects, respectively. In each
panel, red and blue dots refer to bulge-dominated and disk-dominated galaxies,
respectively, with the split set at equal stellar masses for the two
components. The bottom panel is for a volume-limited subset of
SDSS/DR4 with no distinction by morphology.}
\label{fig:color}
\ec
\end{figure}

In Fig.~\ref{fig:color}, we show a scatter plot of SDSS $u-i$ colour against
stellar mass for model galaxies at $z=0$. The upper panel includes dust
extinction effects, while the middle one does not. Blue dots represent
galaxies with dominant disks ($M_{\rm bulge}< M_{\rm disk}$), and red dots
galaxies with dominant bulges ($M_{\rm bulge} > M_{\rm disk}$). A clear split
of the population into a red sequence and a blue cloud is visible in both
plots. When dust effects are included, our model predicts the reddest galaxies
to be passive disk systems scattered up from the red sequence. It is notable
that we predict substantial numbers of disk galaxies on the red sequence,
particularly at intermediate stellar masses.  This appears consistent with the
fact that S0 galaxies substantially outnumber ellipticals in this stellar mass
range in the local universe \citep[e.g.][]{Dressler1980}, although real S0's
rarely have as much dust and gas as our model is assigning them. As in DLB07,
the most massive galaxies are bulge-dominated and lie on the red
sequence. There are also a few massive bulge-dominated galaxies with bluer
colours, corresponding to elllipticals that have undergone a recent
star-formation event, the equivalent of the E+A galaxies seen
locally \citep[e.g.][]{Zabludoff1996}. Finally at low stellar masses we
predict both sequences to be well populated.  As we will see, the fraction of
passive dwarf galaxies in our model appears larger than observed. To compare with observation, we show results from SDSS/DR4 in the bottom
   panel. These have been down-sampled to correspond to a volume-limited
   subset with stellar masses above $10^{9.5}M_{\odot}$ as in
   \cite{Weinmann2009}.  The numbers are quite small because the reddest
   galaxies at this lower mass limit fall within the spectroscopic sample only
   at very low redshift. Clear differences with the models appear in the slope
   of the red sequence and in the number of fainter red-sequence galaxies.

\begin{figure}
\bc
\hspace{-0.6cm}
\resizebox{8.5cm}{!}{\includegraphics{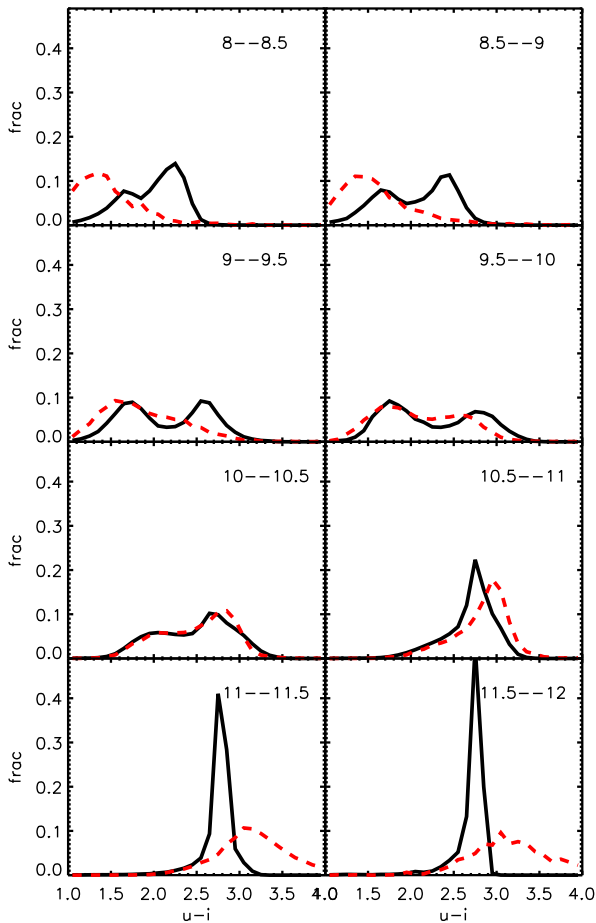}}\\%
\caption{$u-i$ colour distributions as a function of stellar mass. Solid black
curves show the distributions predicted by our preferred model (including
extinction effects) applied to the MS (above $\log M_* = 10.0$) and the MS-II
(at lower masses), while dashed red curves are distributions compiled from
SDSS/DR7. The range in $\log M_*/M_{\odot}$ corresponding to each panel is
indicated at top right.}

\label{fig:colorui}
\ec
\end{figure}

To explore this discrepancy with observation in more detail,
Fig.~\ref{fig:colorui} shows the distributions of $u-i$ (including dust extinction) for galaxies in 8 stellar mass ranges spanning four orders of
magnitude in stellar mass. The solid histograms are constructed from our
preferred model applied to the MS (for $\log M_* > 10.0$) and to the MS-II (at
lower masses) while the dashed histograms are compiled from SDSS/DR7 including
$1/V_{\rm max}$ corrections so that they correspond to volume-limited
statistics. All histograms are normalised to have unit integral. For galaxies
in the stellar mass range $9.5<\log M_*<11.0$ which contains the bulk of all
stars, our predictions for the $u-i$ distribution are in reasonable agreement
with observation, despite our over-simplified dust model. At lower masses, the
fraction of red galaxies is clearly larger in our model than observed. A
substantial fraction of dwarfs (roughly half) are predicted to finish their
star formation early and to become passive. The observed fraction of such
passive dwarfs is substantially smaller.  At the highest masses, the SDSS
galaxies are redder than our model predicts. In the model most of these galaxies 
have mean stellar ages greater than 10 Gyr and stellar metallicities of order
0.5 $Z_\odot$. The real galaxies are more metal-rich, but for the population 
synthesis model we are using, a 12 Gyr old population with
twice solar metallicity has $u-i = 3.07$, thus metallicity and age
effects are insufficient to explain the discrepancy and no significant
dust effects are expected. Photometric or K-correction problems may be
affecting these galaxies which are typically at $z\sim 0.2$. Note that at lower mass,
the red tails of the distributions correspond to the (unrealistically)
reddened passive disk galaxies seen in the upper panel of
Fig.~\ref{fig:color}.  This tail is absent at the highest masses where the
galaxies no longer have gas disks.

\subsection{Tully-Fisher Relation}

\begin{figure}
\bc
\hspace{-0.6cm}
\resizebox{8.5cm}{!}{\includegraphics{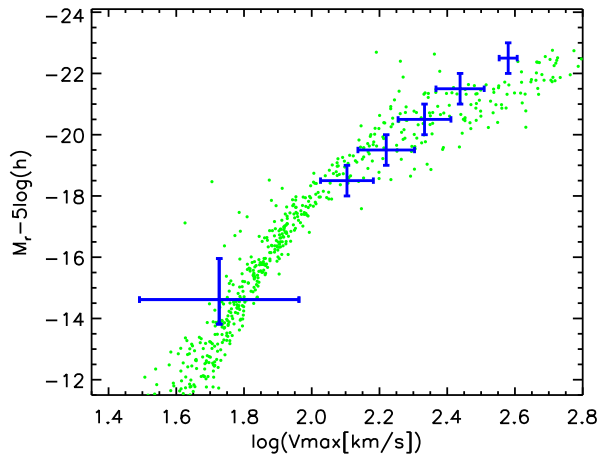}}\\%
\caption{$r$-band Tully-Fisher relation. Blue symbols with error bars are
observational results for isolated disk galaxies taken from \cite{Blanton2008}
and from \cite{Springob2007}. The vertical bar on each symbol shows the bin in
absolute magnitude considered, while the horizontal bar is centred on the
median and shows the {\it rms} scatter of $\log V_{\rm max}$ for the galaxies
within that bin. Green dots are results for central (type 0) late-type galaxies from our
preferred model applied to the MS (brighter than -21) and to the MS-II (for
fainter galaxies). For the model galaxies $V_{\rm max}$ is the maximum circular
velocity of the hosting dark halo.}
\label{fig:TF}
\ec
\end{figure}

There has been a long-standing debate about the ability of galaxy formation
models in a CDM context to reproduce simultaneously the observed abundance and
Tully-Fisher (TF) relation of disk
galaxies \citep{Kauffmann1993,Cole1994,Navarro2000,Cole2000,Blanton2008}. We
have shown above that our preferred model reproduces the observed galaxy
luminosity functions in four SDSS bands at $z=0$. In this section, we study
whether it simultaneously produces a relation between $r$-band luminosity and
maximum circular velocity which is consistent with that observed for isolated
spiral and irregular systems. Early semi-analytic work on the TF
relation \citep[e.g.][]{Kauffmann1993,Cole1994,Somerville1999} took the disk
rotation velocity to be $V_{\rm vir}$, the circular velocity of its halo at
the virial radius. In the DLB07 model a reasonable match to the observed
relation was instead found by identifying the disk rotation velocity with the
{\it maximum} circular velocity of its halo \citep[see][and its
erratum]{Croton2006}. On the other hand, \cite{Cole2000} found that if baryon
condensation is assumed to cause halo contraction according to the standard
simple formula \citep{Barnes1984,Blumenthal1986} the models are no longer able
to reproduce the Tully-Fisher relation and the luminosity function
simultaneously. As discussed in Sec.~\ref{sec:modeldisk}, the simplified model
of adiabatic contraction adopted by \cite{Cole2000} appears to overestimate
the effect of baryons, and at least some recent simulations suggest that the
maximum halo circular velocity found in an equivalent dark matter only
simulation may be a good approximation to the disk rotation
velocity \citep[e.g.][]{Tissera2010}.  Here we use this maximum circular
velocity as our disk rotation velocity surrogate for the TF relation.

We concentrate on central galaxies in the model and compare to observations of
isolated systems, because, as noted by \cite{Blanton2008} and others
\cite{Einasto1974}, dwarf satellite galaxies appear systematically
gas-poor and to have systematically lower rotation velocities relative to
isolated dwarfs of similar stellar mass. This is presumably related to the
various stripping mechanisms discussed above. In order to keep the test
simple, it seems wise to concentrate on galaxies where such effects are
absent.
 
We select central galaxies in our model for which the $r$-band absolute
magnitude of the bulge is at least 1.5 magnitudes fainter than that of the
galaxy as a whole, and, as before, we assume the rotation velocity of the disk
to be $V_{\rm max}$, the maximum circular velocity of its host halo.  In massive
spirals, where baryons dominate in the visible regions, this may
underestimate the rotation velocity because we do not take the mass of the
baryons into account. In dwarf galaxies it may, in contrast, overestimate the
rotation velocity because baryonic effects are weaker and the observable HI may
not extend out to the maximum of the halo circular velocity curve. For
simplicity, we neglect such effects here. The Tully-Fisher relation predicted
in our preferred model by these assumptions is shown using green dots in
Fig.~\ref{fig:TF}. At absolute magnitudes above -21 the data are taken from
the MS, while for fainter galaxies they are taken from the
MS-II. Observational data for relatively bright galaxies from
\cite{Springob2007} and for isolated dwarfs from \cite{Blanton2008} are shown
by blue symbols. The vertical bar on each symbol represents the absolute
magnitude bin considered and is positioned at the median $\log V_{\rm max}$ of the
observed galaxies in that bin. The horizontal bar shows the $\pm 1\sigma$
scatter in  $\log V_{\rm max}$ within the bin.

It is striking that our model, although clearly not a power law,
nevertheless agrees reasonably well with the data over
an absolute magnitude range of about eight magnitudes. There is no evidence
for any major problem, even for dwarf galaxies with $M_r \sim -15$. This
is somewhat unexpected, and is due in part to the fact that 
\cite{Blanton2008} excluded dwarf satellite (as opposed to central) galaxies for which the measured
rotation velocities {\it are} significantly lower at the faintest magnitudes.
A more careful comparison does show some discrepancies, however. At high
circular velocities ($V_{\rm max}\sim 250$km/s or more) model galaxies have a
larger scatter in luminosity than the observations. The brightest real
galaxies have smaller rotation velocities than we predict, perhaps because we
are stopping star formation too efficiently in at least some massive
systems. At the lowest luminosities the simulation predicts slightly higher
rotation velocities and considerably less scatter than is observed. This may
reflect the fact that HI data often do not reach the peak of the rotation
curve in these systems, although the current sample of isolated dwarfs is
probably too sparse to draw reliable conclusions.

\subsection{Profiles and mass functions in rich clusters}

\begin{figure}                                                                                 
\bc                                                                                            
\hspace{-0.6cm}                                                                                
\resizebox{8.5cm}{!}{\includegraphics{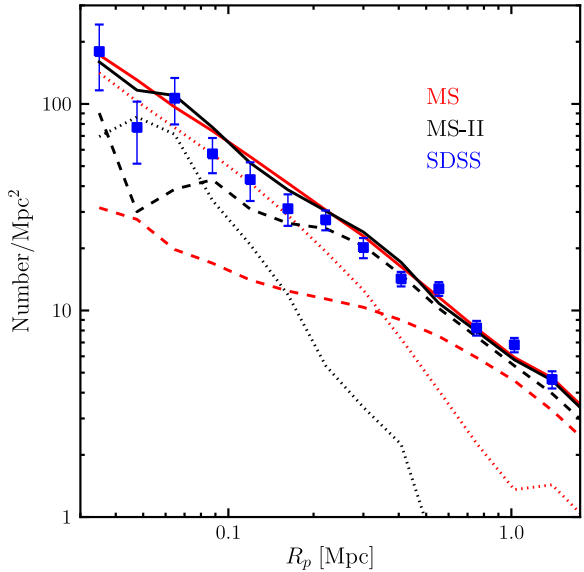}}\\%
\caption{Projected galaxy number density profiles for samples of massive
clusters from the MS (red lines) the MS-II (black lines) and from the
SDSS (blue symbols with error bars).  Observational and model clusters
are selected in the same way and are not scaled before stacking (see text for
details). Solid lines are for all model galaxies with $M_*>
1.2 \times 10^{10}M_\odot$, while dashed and dotted lines split them into
galaxies with surviving dark matter subhalos and orphans,
respectively. Note the excellent agreement in mean profile between the MS
and MS-II despite the very different number of orphans in the two
simulations.  The SDSS profiles here have been corrected for the
spectroscopic incompleteness of the survey, which varies as a function
of projected radius and reaches 60\% near cluster centre. The error bars reflect the uncertainty in
the mean estimated from the scatter among the 31 SDSS cluster
profiles.}

\label{fig:clust_prof}                                                                          
\ec                                                                                            
\end{figure}

An important aspect of our galaxy formation models is associated with
the disruption and merging of substructures. When tidal effects
destroy a dark matter subhalo, we continue to follow the properties of
its central galaxy, tracking its position and velocity using those of
the particle which was most bound to the subhalo when it was last
seen. Such ``orphan'' galaxies may merge with another galaxy (usually
the central galaxy of the main system) or may themselves be tidally
destroyed, when specific conditions are satisfied (see
sections \ref{sec:disrup} and
\ref{sec:merge}).  These procedures account for the fact that dark
matter subhalos are often prematurely disrupted in our simulations
both for numerical reasons (resolution may be insufficient to follow
tidal stripping down to the scale of the central galaxy) and for
astrophysical reasons (dissipation associated with galaxy formation
may make the stellar components more resistant to disruption).  Thus
at any given time our galaxy catalogues contain a population of orphan
(or type 2) galaxies which are concentrated in the inner regions of
massive halos. 

The large size of our two simulations and the factor of 125 difference
in their mass resolution makes it possible to carry out convincing
tests of these procedures for the first time. Appendix A presents the
fraction of galaxies of different types in our preferred model.  For
stellar masses in the range $9.5< \log M_*/M_\odot < 11$ where both
simulations have good statistics, they show similar fractions of all
galaxies to be satellites, but the fraction of these satellites which
are orphans changes from 52\% in the MS to 25\% in the MS-II (at $\log
M_*/M_\odot = 9.5$) or from 27\% to 17\% (at $\log M_*/M_\odot = 11$).
Here we test for numerical convergence in a considerably more extreme
situation by comparing the number density profiles predicted for rich
clusters in the MS and the MS-II. Another sensitive test, based on
counts of close pairs, is presented below in section \ref{sec:correl}.

In order to facilitate comparison with real clusters from the SDSS, we have
implemented a simple ``observational'' cluster finder on our simulations,
designed to find objects with virial masses in the range $14 < \log M_{\rm
  vir}/M_\odot < 14.5$. We take all galaxies with stellar masses above $1.2
\times 10^{10}M_\odot$ and we view their distribution in ``redshift space''
where the $x$ and $y$ coordinate directions are considered transverse to the
``line-of-sight'' and the $z$ peculiar velocity is added to the Hubble constant
times the $z$-coordinate to produce a pseudo-recession velocity. We then
consider all galaxies as potential cluster centres, and we count neighbours within
a surrounding cylinder of radius $r_p = 1.5$~Mpc and line-of-sight velocity
difference $\pm 1200$~km/s, weighting by an ``optimal'' filter $F(r_p)$ which
we take to be an NFW approximation to the projected mass distribution of the
target clusters. Potential centres are ranked by this weighted neighbour count
and those lying within the cylinder of a higher ranked neighbour are
eliminated. The MS is then used to relate the corresponding unweighted counts
$N_c$ to halo mass in order to identify the count range $45 \leq N_c \leq 105$
corresponding to $14 < \log M_{\rm vir}/M_\odot < 14.5$. This algorithm can be
used almost unmodified on a stellar-mass-limited sample of 39 600 SDSS galaxies
from DR7 with $0.01<z<0.06$ and $M_*>1.2 \times 10^{10}M_\odot$. The only
complication is that the SDSS spectroscopy becomes significantly incomplete in
the inner regions of clusters so that a completeness correction must be
applied.  This can be estimated from the overall spectroscopic completeness as
a function of $r_p$ within the stacked regions.  These procedures select 2251,
61 and 31 clusters in the MS, the MS-II and the SDSS, respectively\footnote{In
  order to improve the statistics, we include three orthogonal projections of
  the MS-II data, so the mean number of clusters per MS-II volume is 20.3.}. The effective SDSS volume searched is $6\times10^6$
Mpc$^3$; given the expected cosmic variance expected for the cluster count in a volume of this size ($\sim$ 25\% 
{\it rms}), and the rather large amplitude $\sigma_8 = 0.9$ adopted in
the simulations, the observed and simulated cluster abundances appear
quite consistent.

In Fig.~\ref{fig:clust_prof} we show mean projected number density profiles for
stacks of the clusters in these different sets. Solid lines show the mean
profiles for the two simulations, while dashed and dotted profiles split these
profiles into galaxies with and without associated dark matter subhalos. Red
curves refer to the MS and black curves to the MS-II.  The agreement in the
total profiles is remarkable -- certainly better than one might have expected
since the dashed and dotted profiles show that orphans make a much larger
contribution to the MS profiles (where they dominate for $r_p<350$ kpc) than to
the MS-II profiles (where they dominate only for $r_p< 80$ kpc). Within a
projected radius of 1.5~Mpc, 37\% of all cluster galaxies more massive than
$10^{10}M_\odot$ are orphans in the MS but only 14\% in the MS-II. The fact
that the total profiles agree so well thus demonstrates that the survival times
and positions that we assign to our orphans are appropriate.

The SDSS clusters in Fig.~\ref{fig:clust_prof} are shown by the blue
symbols with error bars indicating the uncertainty in the mean profile
due to cluster to cluster variations. The agreement with the
simulations is quite good, although there may be an indication that
the SDSS clusters are somewhat less concentrated than our models. This
may be an indication that the $\sigma_8$ value adopted in the
simulations is somewhat too high (see also Section 4.9 below).

\begin{figure}
\bc
\hspace{-0.6cm}
\resizebox{8.5cm}{!}{\includegraphics{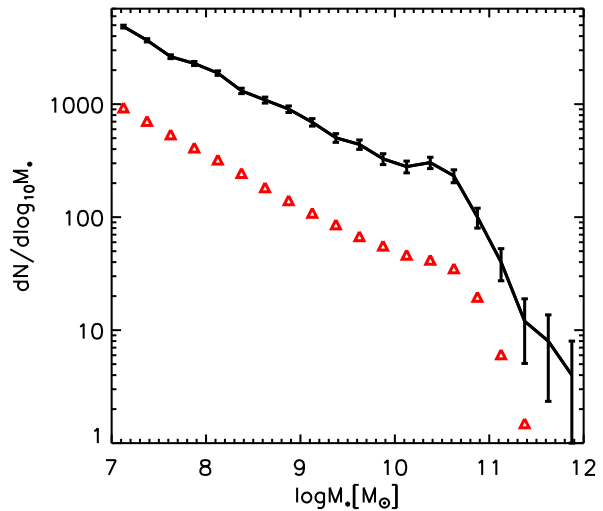}}\\%
\caption{The stellar mass function of galaxies in a rich cluster.
The solid curve links counts in 0.25dex bins for galaxies within
$R_{\rm vir} = 2$Mpc of the the centre of the most massive cluster in the
MS-II according to our preferred galaxy formation model. Error bars
indicate Poisson uncertainties in these counts. Red open triangles
represent the general stellar mass function of galaxies constructed
from the MS-II as a whole.  This has been renormalized arbitrarily to
allow its shape to be compared to that of the cluster stellar mass
function.}
\label{fig:MFcluster}
\ec
\end{figure}
The biggest halo in the MS-II has a mass of $\sim 10^{14.8}$M$_{\rm \odot}$,
similar to that of the Coma cluster, and contains over 119 million particles.
Its substructures are thus very well resolved. Here we use this
biggest halo to investigate whether the galaxy stellar mass function inside 
clusters is expected to differ significantly from that of the galaxy
population as a whole. It is well known that the most massive galaxies
occur exclusively in rich clusters, and that cluster populations have 
systematically different star formation histories and morphologies 
to field galaxies. Nearby clusters also appear to contain a population of
small dwarf ellipticals which are not found in less dense environments
\citep[e.g.][]{Binggeli1990}. Thus it is interesting to see whether our
galaxy formation model predicts differences which might correspond to these
observations, and, in particular, to see if the relative number of dwarf
galaxies in a rich cluster is predicted to differ from that in the ``field''.

We study this in Fig.~\ref{fig:MFcluster}.  The solid curve is the stellar
mass function for galaxies within $R_{\rm vir}=2$Mpc of the centre of this
massive cluster, with error bars indicating the Poisson uncertainty in the
count in each bin.  The slope at the low mass end is around -1.4, which is
higher than the observed $r$- or $R$-band slope for galaxies in the Coma
cluster: $\sim$ 1.16 \citep{Beijersbergen2002,Mobasher2003}, but perhaps consistent
with recent observational estimates based on the SDSS data for nearby
X-ray-selected clusters \citep{Popesso2006}. At very faint magnitudes the
slope in the Coma cluster may be steeper \citep{Adami2007,Jenkins2007,
Milne2007}. Given the large dispersion in observational results, our model seems quite compatible with the data.  The triangles show the overall stellar
mass function of the MS-II, renormalized for ease of comparison with the
cluster result. The shapes of the two stellar mass functions are very similar,
both the faint-end slope and the break at high mass. This echoes the results found
for the infrared luminosity function of the Coma cluster by \cite{Bai2006}.
This is interesting, since both observations and the simulations of this paper show substantial
differences in colour and morphology between clusters and the field.  In the
simulations over 95\% of cluster galaxies within $R_{\rm vir}$ are passive. This
fraction seems overly large in comparison to observation 
\citep[e.g.][]{Hansen2009}, again reflecting the fact that the passive galaxy 
fraction in general is somewhat too high in our model.

\subsection{Intracluster Light}

\begin{figure}
\bc
\hspace{-0.6cm}
\resizebox{8.5cm}{!}{\includegraphics{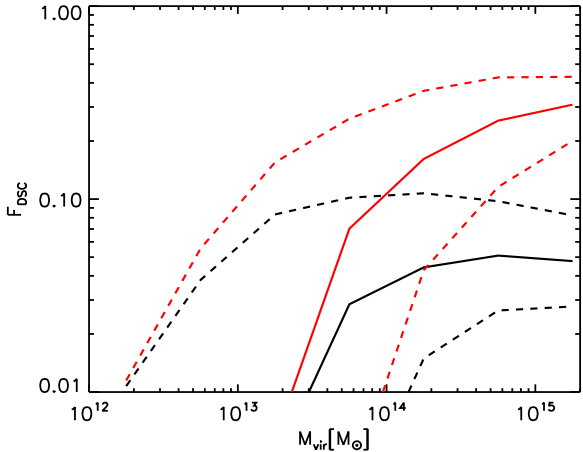}}\\%
\caption{The stellar mass fraction in intergalactic stars as a function 
of virial mass for clusters. The solid black line shows the fraction of all
stars within $R_{\rm vir}$ which are assigned to the intracluster component when
our preferred model is applied to the MS. Dashed black lines show the 16 and
84\% points of the distribution of this fraction. Solid and dashed red lines
show the same statistics but for the fraction of stars in the main subhalo of
each cluster which are associated with its diffuse component, rather than with
its central galaxy.}
\label{fig:ICM}
\ec
\end{figure}
Recent observations of diffuse intracluster light and of intracluster
stars \citep{Zibetti2005,Gerhard2005,Mihos2005,
Gonzalez2005,Aguerri2006,Gonzalez2007,McGee2010} indicate that a
significant fraction of all cluster stars lie between the galaxies,
but they disagree about the exact amount. It seems likely that such
stars must be the remains of disrupted galaxies, and our model now
includes a treatment of the tidal disruption process.  In
Fig.~\ref{fig:ICM}, we show the fraction of cluster stars in the
intergalactic component as a function of cluster virial mass.  We
consider two different fractions here. The black lines refer to the
fraction by mass of all stars within $R_{\rm vir}$ which are assigned
to the intergalactic component. The solid curve is the median value at
each $M_{\rm vir}$, while the dashed lines indicate the 16 and 84\%
points of the distribution. This intracluster fraction increases with
cluster mass and has a large scatter in low-mass clusters. In our
preferred model (here applied to the MS) around 5-10\% of all stars in
clusters with $M_{\rm vir}>5\times 10^{14}M_{\odot}$) are in the
intracluster component and the dependence on cluster mass is quite
weak. In less massive systems this fraction drops very rapidly,
reaching 1\% in groups of mass 3$\times10^{13}M_{\odot}$.  Both the
trends and the value are within the scatter of the observational
results cited above.

Fig.~\ref{fig:ICM} also shows another fraction of interest. The red curves
show the median and the 20 and 80\% points of the distribution of the fraction
of all the stars in the main subhalo which are associated with the diffuse
component, rather than with the central galaxy.  This can be considered as a
proxy for the fraction of the stellar mass of the cD galaxy which is
associated with its extended envelope.  This fraction also increases with
cluster mass, ranging from $\sim 10\%$ in clusters with
$M_{\rm vir} \sim10^{14}M_{\odot}$ to 30\% in clusters with $M_{\rm vir}\sim
1.4 \times 10^{15}M_{\odot}$. Thus, in the richest clusters, the mass in
intergalactic stars is comparable to the stellar mass of the main body of the
central galaxy, or, alternatively, the extended envelope of the cD galaxy
contains about half of its stars.  In galaxy groups, this fraction decreases
rapidly with decreasing virial mass, reaching 1\% in groups of mass 2$\times
10^{13}M_{\odot}$.
\subsection{Luminosity function of Milky Way satellites}
\label{sec:resultreionization}

\begin{figure}
\bc
\hspace{-0.6cm}
\resizebox{6.5cm}{!}{\includegraphics{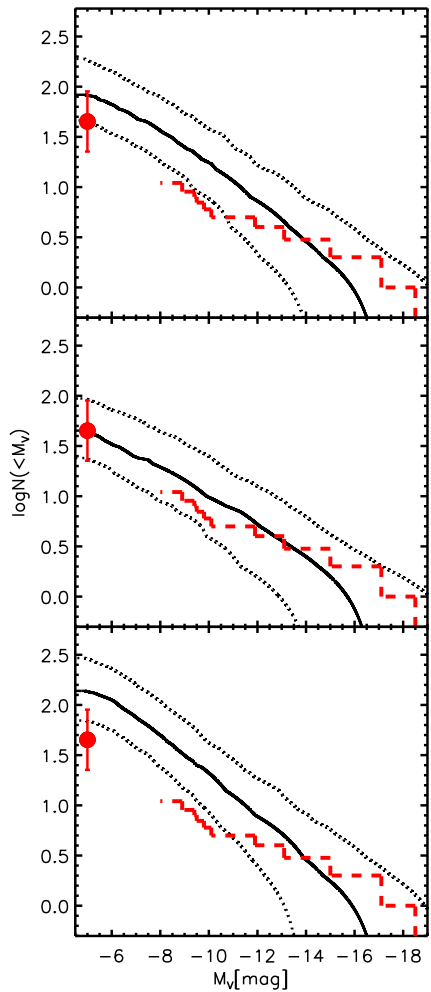}}\\%
\caption{Cumulative luminosity functions for the Milky Way satellite system, 
defined to consist of all galaxies within 280~kpc of the Galactic Centre.
Simulated ``Milky Ways'' are taken to be disk-dominated central galaxies with
stellar masses between 4 and 8$\times 10^{10}M_\odot$. Solid curves give the
median satellite count predicted above each absolute magnitude, while dotted
curves delineate the 10\% and 90\% tails of the count distribution. The upper
panel gives results for our preferred model applied to the MS-II. This assumes
the effects of reionization to be as advocated by \cite{Okamoto2008}. In the
central panel we show what happens if we instead use the reionization
prescription of \cite{Gnedin2000}, keeping all other model parameters fixed.
Reionization effects are weaker in the \cite{Okamoto2008} model than in
that of \cite{Gnedin2000}. More detailed discussion of these two recipes can
be found in Sec.~\ref{sec:reionize}. For the lower panel, reionization is
assumed to have no effect on galaxy formation.  In each panel the cumulative
luminosity function for the 11 ``classical'' satellites of the Milky Way is
shown as a stepped red curve ending at $M_V\sim -8$. The abundance of
satellites with $M_V<-5$ estimated by \cite{Koposov2008} is indicated by a
large filled red circle.  Because of the substantial and uncertain
completeness correction needed to make this estimate, we have arbitrarily
assigned it an error bar of a factor of two.}
\label{fig:LFMW}
\ec
\end{figure}

\begin{figure}
\bc
\hspace{-0.6cm}. 
\resizebox{8.5cm}{!}{\includegraphics{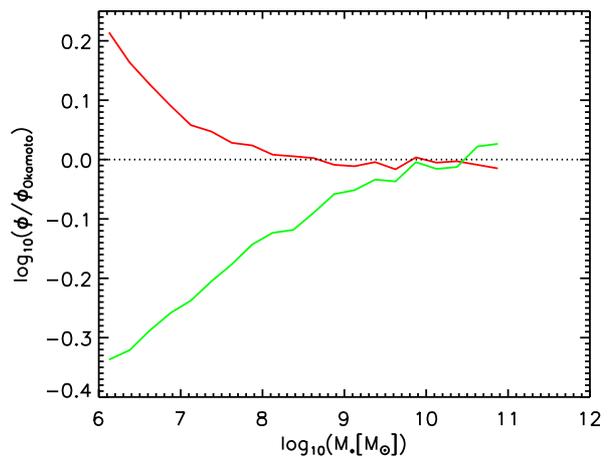}}\\%
\caption{The effects of reionization on the low-mass end of the 
stellar mass function of galaxies. The red curve is the ratio of the stellar
mass function predicted for the MS-II by a model excluding the effects of
reionization to that predicted by our preferred model which is identical
except that reionization is included following the prescription
of \cite{Okamoto2008}. Reionization changes the abundance of galaxies only at
stellar masses below $10^8M_\odot$. Effects are stronger if the prescriptions
of \cite{Gnedin2000} are used instead, as in DLB07. This is shown by the green
curve which gives the ratio of the abundances predicted for this model to those
predicted by our preferred model. Above $10^8M_\odot$ the effects remain below
20\%.} 
\label{fig:LFreio}
\ec
\end{figure}

The abundance of the very lowest mass galaxies can be measured observationally
only in the Local Group, in particular, in the halo of the Milky Way. The
apparent discrepancy between the relatively small number of observed
satellites and the many dark matter subhalos predicted in a $\Lambda$CDM
cosmogony has been promoted as ``the missing satellite problem'', a possible
flaw in the concordance structure formation model \citep{Moore1999,
Klypin1999}, despite earlier suggestions that it might rather reflect the
astrophysics of galaxy formation in weak potential
wells \citep{Kauffmann1993}. Over the last decade new observational results,
primarily from the SDSS, have increased the directly observed number of
satellites by almost a factor of two and the estimated total number of
satellites by about a factor of four \citep[e.g.][]{Koposov2008}.  At the same
time, improved simulations have increased the predicted number of subhalos by
a factor of 1000 \citep[e.g.][]{Springel2008}. Thus the discrepancy has
grown. Our galaxy formation models make it possible to address this issue in
the context of the more general problem of matching the low-mass end of the
stellar mass function of galaxies. This is because the MS-II contains several
thousand isolated galaxies similar in mass to the Milky Way, and its
resolution turns out to be (just) sufficient to get predictions for objects
with stellar masses comparable to those of the observed Milky Way stellites.

In the MS-II, there are around 7000 halos with virial mass within a factor of
three of that estimated for the halo of the Milky Way (see \cite{Boylan2009b} for an analysis of the properties of these
halos and their substructure). In order to make a
detailed comparison, we select all disk-dominated ($M_{*,\rm
  disk}>M_{*,\rm bulge}$)
central galaxies with total stellar mass between 4 and 8 $\times
10^{10}M_{\odot}$. (The stellar mass of the Milky Way is estimated to be
$5\times 10^{10}M_{\odot}$ \citep{Flynn2006}.)  This provides us with a sample
of 1603 ``Milky Ways'' which have median halo mass $M_{\rm vir} = 1.30\times$
10$^{12}$M$_{\odot}$ with lower and upper quartiles at 0.90 and 2.18 
$\times$10$^{12}$M$_{\odot}$ . For the purposes of this
section, all galaxies within 280~kpc of each ``Milky Way''are defined to be
its satellites. Fig.~\ref{fig:LFMW} shows the cumulative $V$-band luminosity
function of these satellite systems in our preferred model and in two
variations with different assumptions about reionization. Specifically, we
plot the median and the 10 and 90\% points of the distribution of satellite
counts as a function of limiting absolute magnitude, $M_V$. A dashed red curve
plotted for $M_V<-8$ represents the cumulative luminosity function of the 11
``classical'' Milky Way satellites. To this limit, the observed sample is
thought to be (almost) complete. We also use a large filled red circle to
indicate the estimate of 45 Milky Way satellites with $M_V<-5$ and $r<280$~kpc
from \cite{Koposov2008}. This estimate required a large and uncertain
incompleteness correction, so we have arbitrarily assigned it an error bar of
a factor of two.

The top panel of Fig.~\ref{fig:LFMW} shows results for our preferred model
which assumes the \cite{Okamoto2008} prescriptions when estimating the effects
of reionization. The predicted satellite abundance is consistent with
observation all the way from bright LMC/M33-like systems down to $M_V \sim
-11$, even though model parameters were set to match the general galaxy
stellar mass function rather than Local Group data. For fainter systems, the
observational estimates are close to the lower 10\% point of the predicted
counts, but, as just noted, the Koposov estimate has a substantial intrinsic
uncertainty. In addition the classical satellite count may well have missed a
one or two systems behind the Galactic Plane. As the middle panel shows, if we
substitute the \cite{Gnedin2000} parameters used by DLB07 for those
of \cite{Okamoto2008}, the predicted number of faint galaxies is reduced, and
the match to the observational estimates is almost perfect. However, 
\cite{Okamoto2008} and \cite{Hoeft2006} argue that the simulations of
\cite{Gnedin2000} substantially overestimated the extent to which an 
ionizing background suppresses the accretion of gas onto small haloes. If, on
the other hand, we neglect the effects of reionization altogether, the bottom
panel shows the disagreement with the observational data to worsen only at the
faintest magnitudes.  The median count of satellites with $M_V<-5$ is
predicted to be about four times the Koposov estimate, but brighter than
$M_V\sim -10$, the abundances are almost unchanged from our preferred model.
Thus, if \cite{Okamoto2008} are right, reionization has a significant effect
only on the very faintest galaxies. This is consistent with results of
previous work \citep{Bullock2000,Somerville2002, Benson2003b, Gnedin2006,
Okamoto2010}

We explore this point further in Fig.~\ref{fig:LFreio}, which shows how
reionization modelling affects the low-mass end of the overall stellar mass
function. We plot the factor by which the galaxy abundance in the MS-II is
changed as a function of stellar mass if our preferred model, which uses
the \cite{Okamoto2008} reionization parameters, is altered to use those
of \cite{Gnedin2000}, as in DLB07 (green line), or to neglect the effects of
reionization altogether (red line). In our preferred model, reionization
affects the abundance of galaxies noticeably only below about
$10^7M_\odot$. The stronger effects implied by the \cite{Gnedin2000} recipe,
reduce the abundance by about 20\% already at $10^8M_\odot$, but remain small
for more massive systems.Thus we conclude that reionization has very little
effect on galaxies similar to the brighter Local Group dwarfs, but may
significantly affect the abundance of the fainter dwarf spheroidals.

\subsection{Correlation Functions}
\label{sec:correl}
\begin{figure}
\bc
\hspace{-0.6cm}
\resizebox{8.5cm}{!}{\includegraphics{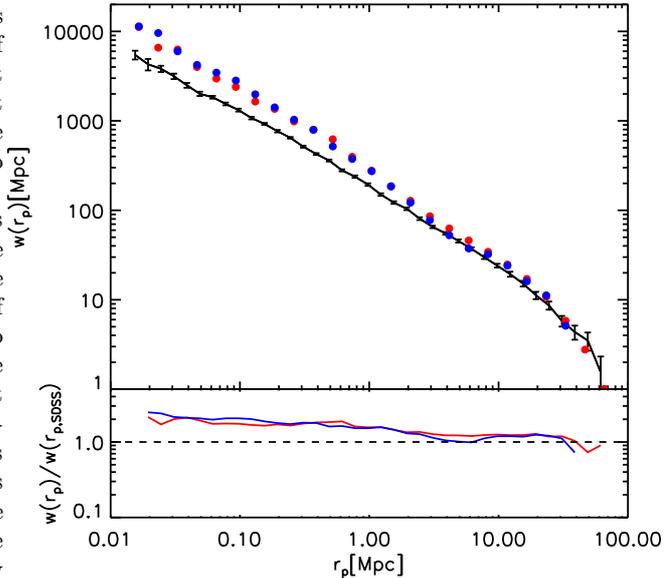}}\\%
\caption{The projected autocorrelation function of stellar mass (upper
panel). Blue and red circles show results from our preferred model applied to
the MS-II and to the MS respectively. Numerical convergence is excellent, even
on scales below 100~kpc. An estimate from the final release of the SDSS is
shown by a black solid line joining points with error bars which include both
counting noise and cosmic variance \citep{Li2009}. On large scales our model
overstimates the observed amplitude of clustering by 10 to 20\%. On small
scales the discrepancy rises to a factor of two. In the lower panel we show
the ratio of the two model autocorrelation functions to the SDSS estimate. }

\label{fig:masscor}
\ec
\end{figure}

\begin{figure*}
\bc
\hspace{-0.6cm}
\resizebox{12cm}{!}{\includegraphics{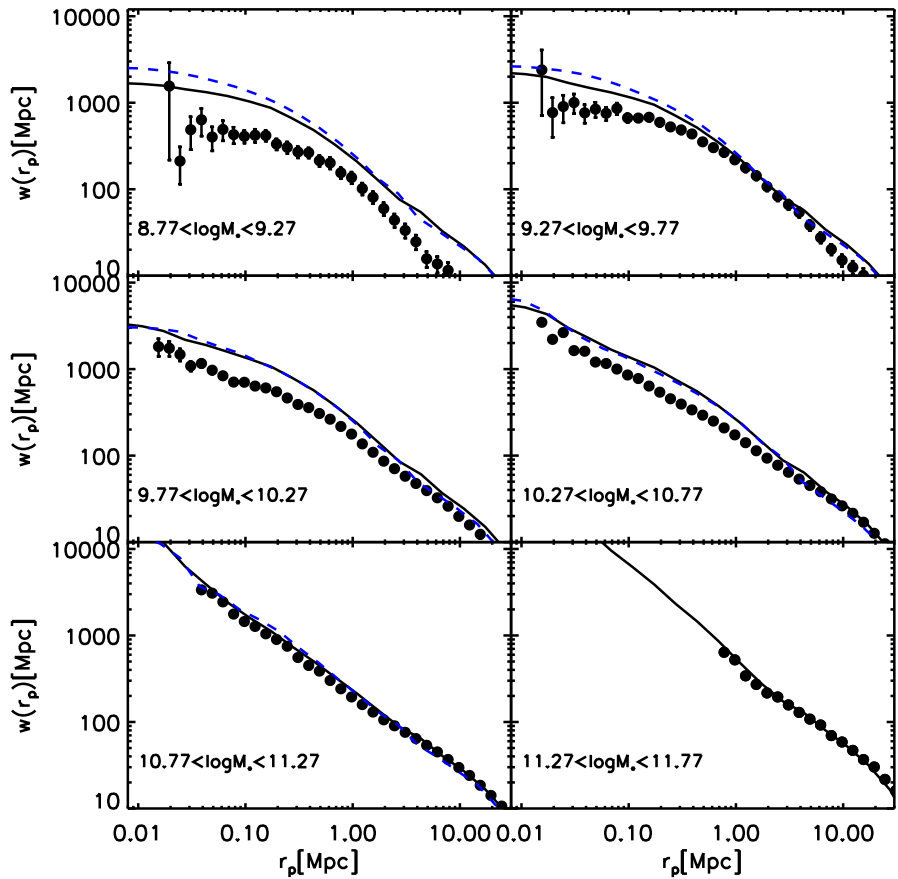}}\\%
\caption{Projected autocorrelation functions for galaxies in different 
stellar mass ranges. Black solid and blue dashed curves give results for our
preferred model applied to the MS and the MS-II, respectively. Symbols with
error bars are results for SDSS/DR7 calculated using the same techniques as
in \cite{Li2006}. The two simulations give convergent results for $M_* >
6\times 10^9M_\odot$. At lower mass the MS underestimates the correlations on
small scales but still matches the MS-II for $r_p> 1$~Mpc. The model agrees
quite well with the SDSS at all separations for $M_* > 6\times
10^{10}M_\odot$, overestimating the correlations slightly on small
scales, but at smaller masses the correlations are overestimated
substantially, particularly at small separations.}
\label{fig:cor}
\ec
\end{figure*}

\begin{figure*}
\bc
\hspace{-0.6cm}
\resizebox{12cm}{!}{\includegraphics{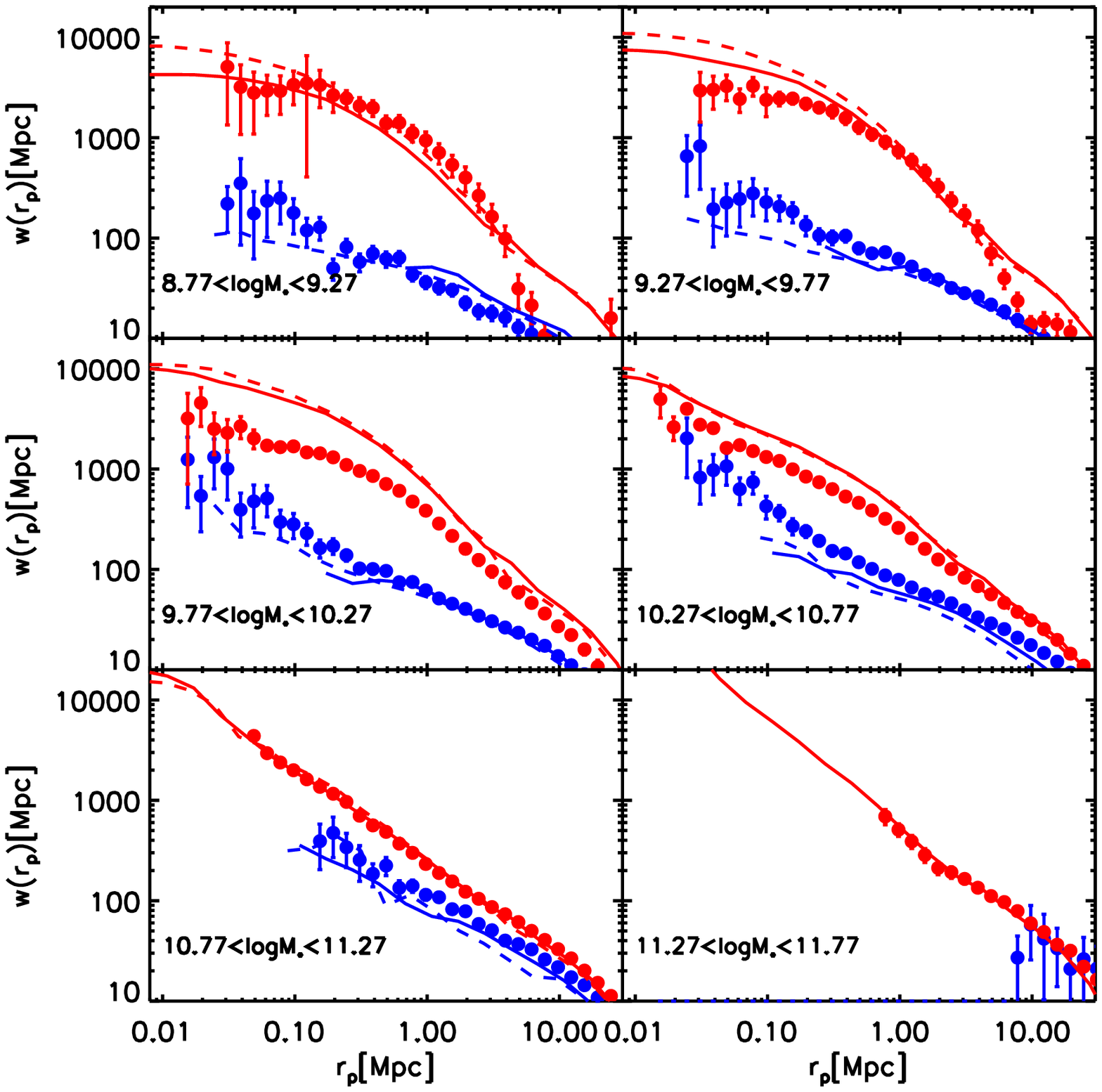}}\\%
\caption{Projected autocorrelation functions for galaxies as a function of 
colour and stellar mass. As in Fig.~\ref{fig:cor}, solid and dashed curves are
for our preferred model applied to the MS and to the MS-II, respectively.
Symbols with error bars are again derived from SDSS/DR7 using the techniques
of \cite{Li2006}. In each mass range, the galaxies are split into passive
(red) and active (blue) subsamples according to their $g-r$ colour. The colours
of the symbols and curves correspond to those of the populations.
Qualitatively, the agreement between models and observations is good, with
quantitative agreement at both high ($M_* > 6\times 10^{10}M_\odot$) and low
($M_* < 6\times 10^9M_\odot$) stellar mass and a somewhat stronger dependence
of clustering on colour than is observed at intermediate stellar masses.}
\label{fig:corcolor}
\ec
\end{figure*}

The SDSS has revolutionised our knowledge of the nearby galaxy population not
only by providing quantitatively reliable data for galaxy abundances as a
function of luminosity, stellar mass and colour over the full range from
dwarfs to cD galaxies, but also by providing precise measurements of the
spatial clustering of galaxies as a function of their luminosity and colour on
scales from 20~kpc to 30~Mpc and beyond. With simulations the size of the MS
and the MS-II, our galaxy formation models make equally precise predictions
for the clustering of simulated galaxies as a function of their physical
properties. Comparing observation and simulation provides powerful constraints
on the galaxy formation modelling. No modern semi-analytic or hydrodynamic
simulation of the formation of the galaxy population should be considered
viable unless it demonstrates at least adequate agreement, not only with
stellar mass, luminosity and color distributions, but also with clustering as a function
of galaxy properties.

In Fig.~\ref{fig:masscor} we compare the projected autocorrelation of stellar
mass in the final release of the SDSS to the results we obtain for our
preferred galaxy formation model.  In the upper panel, red and blue symbols
are results from the MS and MS-II, respectively; the black solid line shows
the SDSS/DR7 measurement from \cite{Li2009}. The error bars on the latter
include the effects of counting noise and cosmic variance and are impressively
small.  This is remarkable because small-scale correlations are dominated by
the distribution of satellite galaxies near halo centre, where one might have
expected resolution effects to cause substantial differences. For example, the
number of type 2 (orphan) galaxies differs substantially between the two
simulations (see Appendix A). In part, the agreement reflects the fact that,
as \cite{Li2009} show, the main contribution to the autocorrelation of stellar
mass comes from galaxies with individual stellar masses similar to the Milky
Way, and thus well above the resolution limit of the MS (see the stellar mass
functions in Sec.~\ref{sec:GMF} and the mass-dependent correlation functions
presented below). For $r_p> 2$~Mpc, where the correlations are produced by
galaxies inhabiting different halos (thus typically both type 0 galaxies), the
model autocorrelation function is 10 to 20\% higher than that observed. The small difference between the MS and the MS-II on these scales may reflect
the effect of cosmic variance due to the relative small box size of the
MS-II. On smaller scales where the correlations are dominated by galaxy pairs
inhabiting the same halo (thus typically type 0 -- type 1, or type 0 -- type 2
pairs) the discrepancy grows, reaching a factor of 2 at $r_p< 100$~kpc.  This
suggests an overdominance of 1-halo relative to 2-halo pairs in comparison to
the observations, arguing, perhaps, for a lower value of $\sigma_8$ than used
in the MS cosmology \citep[see ][]{Li2009}. 

We investigate the source of this discrepancy further in Fig.~\ref{fig:cor},
which shows projected autocorrelation functions for galaxies in a set of
disjoint stellar mass ranges, as indicated by the labels in each panel. Black
solid and blue dashed curves give the predictions obtained by applying our
preferred galaxy formation model to the MS and to the MS-II,
respectively. Corresponding $1/V_{\rm max}$-weighted estimates from the full SDSS/DR7, obtained using
the techniques of \cite{Li2006}, are shown by symbols with error bars. Here
the errors are estimated from a set of 80 mock SDSS surveys and so should, in principle, include cosmic
variance effects. This becomes a significant issue at the smallest masses. No result
is shown for the MS-II in the most massive bin, because it contains too few
galaxies to give a meaningful estimate. Results from the two simulations
converge for galaxies more massive than $6\times 10^9M_{\odot}$. For smaller
masses the MS underpredicts the correlations on small scales but still agrees
with the MS-II for $r_p>1$~Mpc. This indicates that resolution limitations
begin to affect satellite galaxies in the MS at higher stellar mass than
central galaxies.  

For $M_* \geq 6\times 10^{10}M_{\odot}$ the model autocorrelations agree with
the SDSS at all separations to better than about 20\%. For $M_* > 6
\times 10^9M_\odot$, simulation and observation continue to agree at about
the 20\% level for $r_p > 2$~Mpc.  This shows that the relation between halo
mass and central galaxy mass shown in Fig.~\ref{fig:gf} leads to
autocorrelations for central galaxies as a function of their stellar mass
which are in good agreement with observation. The small remaining off-set
may indicate a fluctuation amplitude somewhat smaller than the $\sigma_8 =
0.9$ adopted in the simulations. At yet smaller masses the large-scale
correlation amplitude estimated from the SDSS disagrees with the model. Plots
of the distribution of these galaxies on the sky show that their correlations
are dominated by a very small number of structures (just the Coma and Virgo
clusters in the lowest mass bin) which are particularly pronounced in the
minority red population. In these very shallow samples, correlation estimates
are also significantly distorted by peculiar velocity effects (e.g. the
finger-of-god of the Coma cluster and Virgocentric infall). Proper accounting
for these effects is beyond the scope of this paper.

At smaller separations ($r_p \leq 1$~Mpc) Fig.~\ref{fig:cor} shows substantial
discrepancies between model and observation for stellar masses below $6\times
10^{10}M_{\odot}$, indicating that there are more satellite--central pairs in
the model than in the real data. Since the overall abundance of
galaxies as a function of stellar mass matches observation very well (see
Fig.~\ref{fig:MF}), this discrepancy indicates that too large a fraction of
the model galaxies are satellites. Again this is a clear indication favoring a
lower value of $\sigma_8$ which would result in a lower abundance of the
high-mass halos which host two or more galaxies in these stellar mass
ranges \citep[cf][]{vandenBosch2007}.

Additional insight into possible errors in our treatment of the astrophysics
of galaxy evolution can be obtained by studying clustering as a function of
star formation activity. To this end, Fig.~\ref{fig:corcolor} repeats
Fig.~\ref{fig:cor} but with the galaxies in each mass range divided into
``passive'' (red) and ``actively star-forming'' (blue) subsamples according to
their $g-r$ colours, as in \cite{Li2006}.\footnote{For the simulations we take
the division at the minimum of the ``green valley'' in a plot similar to
Fig.~\ref{fig:color}.}  Lines and symbols are as in Fig.~\ref{fig:cor}, except
that they are coloured according to the colour of the corresponding galaxy
population. As expected, red galaxies are more clustered than blue galaxies on
all scales and at all stellar masses. It is encouraging that the effects are
qualitatively similar in the models and in the SDSS data.  Indeed, at large
separation ($r_p>2$~Mpc) there is reasonable quantitative agreement for both
populations at all but the smallest stellar masses, while at large stellar
mass ($M_*\geq 6\times 10^{10}M_{\odot}$) there is good agreement at all
separations.  For active galaxies, this simply indicates once more that our
halo mass -- central galaxy mass relation leads to the right large-scale
correlations as a function of $M_*$ for type 0 galaxies. For passive galaxies
the situation is more complex, since most of the lower mass objects are
satellites rather than centrals.  Apparently, at given stellar mass, their
distribution across halos of different mass is similar in the simulation and
in the real world.  For the two lowest mass bins the large-scale correlations
are again distorted by the small volume and peculiar velocity distortion
effects discussed above.

At small separations the simulations overpredict the autocorrelations of
passive galaxies for stellar masses in the range $6\times 10^{9}M_{\odot}< M_*
< 6\times 10^{10}M_{\odot}$, but, curiously, the MS-II again matches the real
data at lower mass. Small-scale correlations of active galaxies are
underpredicted in all our lower stellar mass bins, showing that our model
still has somewhat too few blue satellite galaxies. An interesting example is
provided by our lowest stellar mass bin. For $2~{\rm Mpc} > r_p > 200~{\rm
kpc}$ the MS-II model fits the SDSS data quite well in
Fig.~\ref{fig:corcolor}, yet lies substantially above them in
Fig.~\ref{fig:cor}. This is because our model overpredicts the fraction of
passive galaxies in this mass range (see Fig.~\ref{fig:color}). Similar
apparent discrepancies between the model/observation comparison in
Fig.~\ref{fig:corcolor} and that in Fig.~\ref{fig:cor} are visble at a lower
level in other mass ranges, and again reflect the slightly different
weightings applied in the two cases when going from colour-differentiated to
``total'' results.

\subsection{Some properties at higher redshift}
\label{sec:hz}
So far we have only discussed properties of our models at $z\sim 0$.  This is
because our observational knowledge of the galaxy population is still much
more complete, more precise and less subject to systematic error in the nearby
universe than at high redshift, despite the enormous recent progress in
amassing data for relatively large, objectively selected samples of distant
galaxies.  Nevertheless, a viable galaxy formation model must be consistent
not only with the present-day galaxy population, but also with that at all
earlier times, so a comparison of our models with high-redshift populations is
a critical part of assessing how realistically they treat the astrophysics of
galaxy formation. Such work is complicated by the strong selection effects and
the substantial observational uncertainties which affect the measurement of
physical properties for faint and distant galaxies. As a result, detailed
comparison is beyond the scope of the present paper.  Earlier versions of our
models have been compared to the evolution of the cosmic star-formation rate
density by \cite{Croton2006}, to the evolution of brightest cluster galaxies
out to $z=1$ by \cite{DeLucia2007}, to the galaxy counts, luminosity functions
and redshift distributions inferred from deep magnitude-limited redshift
surveys by \cite{Kitzbichler2007} and to the abundances, redshift
distributions, stellar mass distributions and clustering of colour-selected
samples of $z\sim 2$ and $z\sim 3$ galaxies by \cite{Guo2009}. The current models can
be expected to give similar results to this previous work and to be sensitive
to many of the same uncertainties, notably to the treatment of dust
obscuration. In this section we will limit ourselves to presenting two of the
least uncertain model predictions at high redshift.

\begin{figure}
\bc
\hspace{-0.6cm}
\resizebox{8.5cm}{!}{\includegraphics{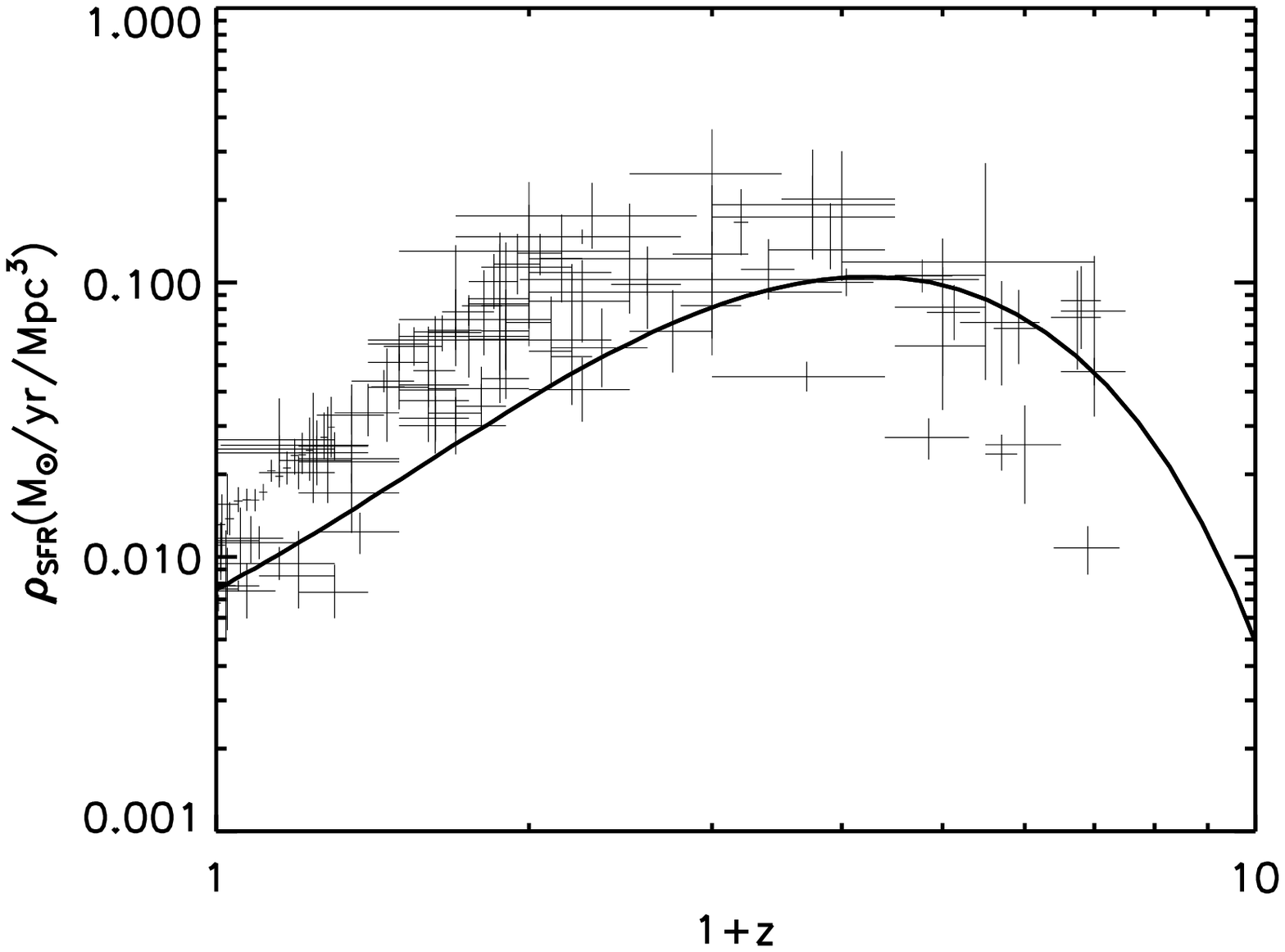}}\\%
\caption{Cosmic star formation rate density as a function of redshift. The
crosses are individual observational estimates compiled by \cite{Hopkins2007}
while the solid curve is obtained from our preferred model applied to the MS.}
\label{fig:madau}
\ec
\end{figure}

In Fig.~\ref{fig:madau} we compare the evolution of the cosmic star formation
rate density predicted by our preferred model to a compilation of
observational estimates taken from \cite{Hopkins2007}. The most obvious
feature of this plot is a clear off-set between the model and the
observations. At all redshifts the model lies a factor of two or more below
the centre of the cloud of observational points. This is a reflection of the
well known fact that if one integrates observational estimates of the star
formation rate density with respect to time, one substantially overpredicts
the observed stellar mass density, not only at $z=0$ but also at all higher
redshifts \citep[e.g.][]{Wilkins2008}. We have chosen to adjust our model to
fit the SDSS stellar mass function, so we necessarily fail to fit observational
estimates of the evolution of the star formation rate density. In our model
the rate of star formation peaks at $z\sim 3$ and has already declined again
by a factor of 3 at $z\sim 1$, whereas the observations suggest a more
constant star formation rate density over this time interval. Given the large
scatter in the observational estimates and the discrepancy just discussed, it
is difficult to know how seriously to take this difference. As we shall see in
the next paragraph, however, there are other indications that galaxy formation
occurs too early in our model, particularly for low-mass galaxies.

\begin{figure*}
\bc
\hspace{-0.6cm}
\resizebox{12.5cm}{!}{\includegraphics{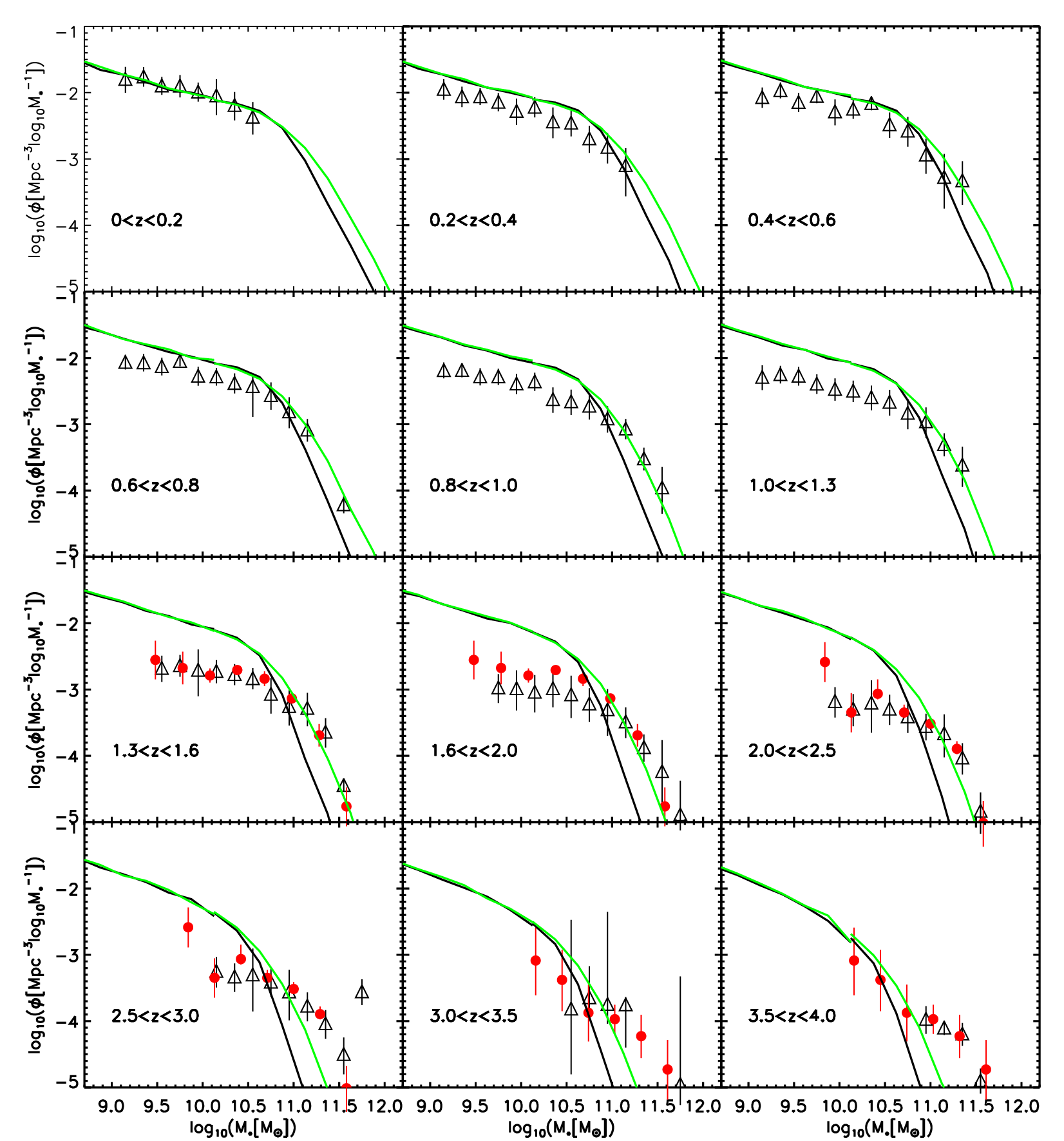}}\\%
\caption{Stellar mass functions for a series of redshift intervals indicated
by the labels in each panel. Observational data are taken
from \cite{Perez2008} and from \cite{Marchesini2009}. \cite{Marchesini2009}
compiled their mass functions for wider bins than \cite{Perez2008} so in each
panel we plot the \cite{Marchesini2009} results for the wider bin that
includes the indicated redshift range. For the triangles representing the
\cite{Perez2008} data we use the error bars quoted in their paper.
For the filled circles representing the \cite{Marchesini2009} results we use
the error estimates which include counting statistics, cosmic variance,
photometric redshift uncertainties and photometric errors, but exclude
systematic uncertainties due to the IMF and other stellar population modelling
issues. The mass scales of these observational results have been shifted to
correct approximately to the Chabrier IMF assumed in our modelling. Black
curves are the functions measured directly from the MS and the MS-II for our
preferred galaxy formation model, while green curves show the result of
convolving with a gaussian of dispersion 0.25~dex in $\log M_*$ in order to
represent uncertainties in the individual observational stellar mass
determinations.}
\label{fig:hzmf}
\ec
\end{figure*}

Stellar masses for high-redshift galaxies are notoriously difficult to
estimate because of the faintness of the images, the strong effects of dust,
and the fact that the observed optical and near-IR bands correspond to the
rest-frame blue and ultraviolet. The situation has improved considerably with
the availability of deep data at 3.6 to 8$\mu$ from Spitzer, and according to
the careful error analysis of \cite{Marchesini2009}, masses with realistic
error bars can now be estimated out to at least $z\sim 4$. In
Fig.~\ref{fig:hzmf} we compare the stellar mass functions predicted by our
preferred model to recent observational estimates based on combined very deep
optical, near-IR and Spitzer photometry from \cite{Perez2008}
and \cite{Marchesini2009}. We have shifted all these observational estimates
so that they correspond to the same Chabrier Initial Mass Function used in our
models. As \cite{Marchesini2009} describe, even with this excellent data
coverage substantial random errors remain in the stellar masses estimated for
individual galaxies \citep[see also ][]{Fontanot2009}. To account roughly for
this, we convolve the stellar mass functions predicted by our preferred model
with a gaussian of dispersion 0.25~dex in $\log M_*$ before comparing them
with the observations.

Our model parameters are adjusted so that they fit the observed stellar mass
function at $z\sim 0$.  This good agreement is maintained out to redshifts
somewhat less than unity.  At higher redshift, the massive end of our
predicted mass functions remains consistent with observation, once it has been
convolved with the  observational mass estimation
uncertainties, but the abundance of lower mass galaxies ( $M_* \sim
10^{10}M_{\odot}$) is substantially overpredicted.\footnote{If the systematic
error ranges discussed by \cite{Marchesini2009} are considered appropriate,
this overprediction appears only marginally significant, but a large part of
these systematic errors are due to possible IMF variations which we exclude
for the present discussion.} At face value, the discepancy suggests that
low-mass galaxies form considerably earlier in our model than in the real
universe. This is consistent both with the overly high redshift of the peak of
the star formation rate density (see Fig.~\ref{fig:madau}) and the overly
large fraction of passive galaxies in the $z\sim 0$ low-mass population (see
Fig.\ref{fig:colorui}). The problem is not specific to the details of our
model. It has been seen in the comparison of earlier models (both our own and
those of others) to this and other similar datasets \citep[e.g.][]{Fontana2006,
Marchesini2009,LoFaro2009,Fontanot2009,Cirasuolo2010}.  As several of these
authors emphasise in their own discussion, the problem seems most likely to
lie in the way star formation is treated in the models, particularly at high
redshift.

\section{Discussion}
\label{sec:conclusion}
New observational data at low redshift give precise measures of the abundance
and clustering of galaxies as a function of their physical properties (stellar
mass, luminosity, size, star formation rate, nuclear activity...)  over a
range of almost five orders of magnitude in stellar mass ($7< \log M_*/M_\odot
< 12$). Abundances of even lower mass galaxies are measured reasonably
reliably in the Local Group. In addition, the explosion of data from ultra-deep
surveys is beginning to provide convincing results for the general galaxy
population at much earlier cosmic epochs.  Matching such a wealth of data over
such a large dynamic range is an extraordinary challenge for any {\it a priori}
galaxy formation model. By combining results from the MS and the MS-II, and by
updating and readjusting our treatments of the many relevant astrophysical
processes, we have made a model which has the necessary dynamic range and
statistical power to confront the full range of abundance and clustering data
available at low redshift. The MS-I gives good statistics for rare, high-mass
galaxies, while the MS-II provides well-resolved assembly histories for
low-mass systems.

In this paper we have extended and modified our earlier treatments of the
transition between the rapid infall and cooling flow regimes of gas accretion,
of the sizes of bulges and of gaseous and stellar disks, of supernova feedback
in low-mass galaxies, of the transition between central and satellite status
as galaxies fall into larger systems, and of the stripping of gas and stars
once they have become satellites.  For physically plausible values of its
parameters, the new model fits both the abundance and the large-scale
clustering of low-$z$ galaxies as a function of stellar mass, luminosity and
(to a lesser extent) colour. At high mass the efficiency of star formation is
limited by AGN feedback, as proposed by \cite{Croton2006}. At low mass,
consistency with the observed SDSS luminosity and stellar mass functions
requires supernova feedback to be significantly more efficient and the
reincorporation of ejected gas to be considerably less efficient than in
DLB07.  This enhanced SN feedback also leads to reasonable agreement with the
abundance of faint satellites around the Milky Way, suggesting that
reionisation influences the formation of, at most, the very smallest
galaxies \citep[see also ][]{Li2010,Maccio2010}.

For galaxies of high stellar mass, our preferred model also fits both the
colour distribution and the small-scale clustering of SDSS galaxies ($\log
M_*/M_\odot > 9.5$ for the colours and $>10.5$ for the clustering). At lower
stellar mass, the model predicts a substantial fraction of red, passive
galaxies which are not present in the SDSS data, and a clustering strength
which rises progressively above that observed for $r_p < 1$~Mpc. (Note,
however, that the {\it difference} in clustering between active and passive
galaxies is still modelled quite accurately.)  Given that our model matches
both the stellar mass function and the mass-dependent large-scale clustering
data from SDSS, this excessive small-scale clustering implies that too large a
fraction of our galaxies are satellites at each stellar mass. Since individual
groups and clusters in our model have galaxy occupation numbers and radial
distributions in quite good agreement with observation, the discrepant
small-scale correlations suggest that massive halos are overabundant in our
simulations, i.e. that $\sigma_8=0.9$ is too
large \citep[c.f.][]{vandenBosch2007}. We intend to test this explicitly
in future work by using the rescaling techniques of \cite{Angulo2010} on the
MS and MS-II so that they can be used to construct galaxy formation models
very similar to those of this paper, but for cosmologies other than that
originally assumed for the simulations, for example, cosmologies with lower
values of $\sigma_8$, as suggested by more recent WMAP results.

The excessive passive fraction at low stellar mass implies that our preferred
model is quenching star formation in small halos in order to limit the
total production of stars, whereas real objects form stars at a steady but low
rate until the present day. This is also the principal reason why the
model continues to have too few blue satellites, despite our improved
treatment of stripping effects -- at low stellar masses ($\log M_*/M_\odot <
10$) there are too few star-forming galaxies everywhere. Low-mass star-forming
galaxies in the model fit on the observed Tully-Fisher relation for isolated
galaxies just as well as their giant cousins, and their large-scale clustering
is also correct. Thus dwarfs appear to be forming in the proper dark halos.
The overly early truncation of their star formation is very likely related to
the fact that while the model correctly fits the observed abundance of massive
galaxies ($M_*\sim 10^{11}M_\odot$) out to $z\sim 4$, it overpredicts the
observed abundance of lower mass systems ($M_*\sim 10^{10}M_\odot$) by
progressively larger amounts beyond $z\sim 0.6$. Lower mass galaxies clearly
complete their formation too early in the model.

With the increased resolution provided by the MS-II we are able to show that
the stellar mass function of galaxies in rich clusters is predicted to be very
similar in shape to that in the general field, even down to $M_*\sim
10^7M_\odot$. Almost all galaxies within the virial radius of a relaxed
cluster are predicted to be passive, but this may be an overestimate for the
reasons discussed in the last paragraph. Our new treatment of galaxy
disruption suggests that 5\% to 10\% of all cluster stars should be be part of
the intracluster light, and that this fraction should increase with cluster
mass and show substantial cluster to cluster variation. 

The predictions for the luminosity functions and radial number count
profiles of clusters are very similar in the MS and the MS-II and
agree quite well with observation {\it provided} the substantial
population of galaxies without surviving dark matter subhalos is
included. Such orphan galaxies account for almost half of all cluster
members with $M_*>10^{10}M_*$ in the MS, and
for about 13\% in the MS-II.  Without them the abundance of galaxies
in the inner cluster would be substantially underpredicted. This
demonstrates that, even at MS-II resolution, schemes that place
galaxies in subhalos in a high-resolution simulation without
accounting for subhalos which have been tidally disrupted but whose
galaxies have survived \citep[e.g.][]{Vale2004,Conroy2006,
Wetzel2009,Moster2010,Guo2010} will not correctly reproduce the observed
structure of galaxy clusters. This argument was already presented by
\cite{Gao2004}.

The degree to which our physically based model reproduces the observed
abundance and clustering properties of the $z\sim 0$ galaxy population
is impressive, but there are clear and significant discrepancies, and
a comparison with high-redshift populations, although barely started
here, also shows substantial discrepancies. Further work is needed to
understand the source of these problems. Our simple recipes for
complex astrophysical processes may turn out to be inappropriate when
more accurate treatments become feasible. In addition, processes other
than those we discuss may produce similar behaviour, making them
operationally indistinguishable at the present level of description.
Finally, there are undoubtedly degeneracies among the model parameters
we have adjusted, making our specific model non-unique \citep[see, for
example, ][]{Henriques2009,Bower2010, Neistein2009}.  Such
degeneracies can only be lifted, and the recipes improved, by
increasing the range, variety and precision of the data used to
constrain the model.

The clearest physical indication from the results presented in this
paper is that our current treatment of star formation, although
similar to that used both in other phenomenological models and in
direct simulations of galaxy formation, is significantly in error,
producing overly efficient star formation at early times and in small
galaxies. We have tried simple modifications of these recipes but have
not so far identified one which leads to substantially improved
results. A better astrophysical understanding of large-scale star
formation is probably required.

\section*{Acknowledgements}                                                                                         
The halo/subhalo merger trees for the Millennium and Millennium-II
Simulations are publicly available at
http://www.mpa-garching.mpg.de/millennium, as are galaxy catalogues,
galaxy merger trees and light-cone mock catalogues for previous
galaxy formation models implemented on the MS. Galaxy catalogues for
our new models implemented on both the MS and the MS-II will be made
available on this same site as soon as this paper is accepted for
publication.  This Millennium site was created as part of the
activities of the German Astrophysical Virtual Observatory. SW and GL
acknowledge support from the European Research Council as part of the
Galformod Project. GDL acknowledges
financial support from the European Research Council under the European Community's Seventh Framework Programme (FP7/2007-2013)/ERC grant agreement n. 202781

\bibliographystyle{mn2e}

\bibliography{SAM}

\begin{thebibliography}{}

\bibitem[\protect\citeauthoryear{{Abadi}, {Navarro}, {Fardal}, {Babul} \&
  {Steinmetz}}{{Abadi} et~al.}{2010}]{Abadi2010}
{Abadi} M.~G.,  {Navarro} J.~F.,  {Fardal} M.,  {Babul} A.,    {Steinmetz} M.,
  2010, \mnras, 407, 435

\bibitem[\protect\citeauthoryear{{Adami}, {Picat}, {Durret}, {Mazure},
  {Pell{\'o}} \& {West}}{{Adami} et~al.}{2007}]{Adami2007}
{Adami} C.,  {Picat} J.~P.,  {Durret} F.,  {Mazure} A.,  {Pell{\'o}} R.,
  {West} M.,  2007, \aap, 472, 749

\bibitem[\protect\citeauthoryear{{Aguerri}, {Castro-Rodr{\'{\i}}guez},
  {Napolitano}, {Arnaboldi} \& {Gerhard}}{{Aguerri} et~al.}{2006}]{Aguerri2006}
{Aguerri} J.~A.~L.,  {Castro-Rodr{\'{\i}}guez} N.,  {Napolitano} N.,
  {Arnaboldi} M.,    {Gerhard} O.,  2006, \aap, 457, 771

\bibitem[\protect\citeauthoryear{{Allen}, {Rapetti}, {Schmidt}, {Ebeling},
  {Morris} \& {Fabian}}{{Allen} et~al.}{2008}]{Allen2008}
{Allen} S.~W.,  {Rapetti} D.~A.,  {Schmidt} R.~W.,  {Ebeling} H.,  {Morris}
  R.~G.,    {Fabian} A.~C.,  2008, \mnras, 383, 879

\bibitem[\protect\citeauthoryear{{Angulo} \& {White}}{{Angulo} \&
  {White}}{2010}]{Angulo2010}
{Angulo} R.~E.,  {White} S.~D.~M.,  2010, \mnras, 405, 143

\bibitem[\protect\citeauthoryear{{Asplund}, {Grevesse} \& {Sauval}}{{Asplund}
  et~al.}{2006}]{Asplund2006}
{Asplund} M.,  {Grevesse} N.,    {Sauval} A.~J.,  2006, Communications in
  Asteroseismology, 147, 76

\bibitem[\protect\citeauthoryear{{Bai}, {Rieke}, {Rieke}, {Hinz}, {Kelly} \&
  {Blaylock}}{{Bai} et~al.}{2006}]{Bai2006}
{Bai} L.,  {Rieke} G.~H.,  {Rieke} M.~J.,  {Hinz} J.~L.,  {Kelly} D.~M.,
  {Blaylock} M.,  2006, \apj, 639, 827

\bibitem[\protect\citeauthoryear{{Baldry}, {Balogh}, {Bower}, {Glazebrook},
  {Nichol}, {Bamford} \& {Budavari}}{{Baldry} et~al.}{2006}]{Baldry2006}
{Baldry} I.~K.,  {Balogh} M.~L.,  {Bower} R.~G.,  {Glazebrook} K.,  {Nichol}
  R.~C.,  {Bamford} S.~P.,    {Budavari} T.,  2006, \mnras, 373, 469

\bibitem[\protect\citeauthoryear{{Baldry}, {Glazebrook} \& {Driver}}{{Baldry}
  et~al.}{2008}]{Baldry2008}
{Baldry} I.~K.,  {Glazebrook} K.,    {Driver} S.~P.,  2008, \mnras, 388, 945

\bibitem[\protect\citeauthoryear{{Barnes}}{{Barnes}}{1984}]{Barnes1984}
{Barnes} J.,  1984, \mnras, 208, 873

\bibitem[\protect\citeauthoryear{{Baugh}}{{Baugh}}{2006}]{Baugh2006}
{Baugh} C.~M.,  2006, Reports on Progress in Physics, 69, 3101

\bibitem[\protect\citeauthoryear{{Baugh}, {Lacey}, {Frenk}, {Granato}, {Silva},
  {Bressan}, {Benson} \& {Cole}}{{Baugh} et~al.}{2005}]{Baugh2005}
{Baugh} C.~M.,  {Lacey} C.~G.,  {Frenk} C.~S.,  {Granato} G.~L.,  {Silva} L.,
  {Bressan} A.,  {Benson} A.~J.,    {Cole} S.,  2005, \mnras, 356, 1191

\bibitem[\protect\citeauthoryear{{Beijersbergen}, {Hoekstra}, {van Dokkum} \&
  {van der Hulst}}{{Beijersbergen} et~al.}{2002}]{Beijersbergen2002}
{Beijersbergen} M.,  {Hoekstra} H.,  {van Dokkum} P.~G.,    {van der Hulst} T.,
   2002, \mnras, 329, 385

\bibitem[\protect\citeauthoryear{{Belokurov} \& et al.}{{Belokurov} \&
  et~al.}{2007}]{Belokurov2007}
{Belokurov} V.,  et al. 2007, \apj, 654, 897

\bibitem[\protect\citeauthoryear{{Benson}}{{Benson}}{2010}]{Bensonarxiv2010}
{Benson} A.~J.,  2010, ArXiv e-prints

\bibitem[\protect\citeauthoryear{{Benson} \& {Bower}}{{Benson} \&
  {Bower}}{2010a}]{Benson2010}
{Benson} A.~J.,  {Bower} R.,  2010a, ArXiv e-prints

\bibitem[\protect\citeauthoryear{{Benson} \& {Bower}}{{Benson} \&
  {Bower}}{2010b}]{BensonBower2010}
{Benson} A.~J.,  {Bower} R.,  2010b, \mnras, 405, 1573

\bibitem[\protect\citeauthoryear{{Benson}, {Bower}, {Frenk}, {Lacey}, {Baugh}
  \& {Cole}}{{Benson} et~al.}{2003}]{Benson2003a}
{Benson} A.~J.,  {Bower} R.~G.,  {Frenk} C.~S.,  {Lacey} C.~G.,  {Baugh} C.~M.,
     {Cole} S.,  2003, \apj, 599, 38

\bibitem[\protect\citeauthoryear{{Benson}, {Frenk}, {Baugh}, {Cole} \&
  {Lacey}}{{Benson} et~al.}{2003}]{Benson2003b}
{Benson} A.~J.,  {Frenk} C.~S.,  {Baugh} C.~M.,  {Cole} S.,    {Lacey} C.~G.,
  2003, \mnras, 343, 679

\bibitem[\protect\citeauthoryear{{Benson}, {Lacey}, {Baugh}, {Cole} \&
  {Frenk}}{{Benson} et~al.}{2002}]{Benson2002}
{Benson} A.~J.,  {Lacey} C.~G.,  {Baugh} C.~M.,  {Cole} S.,    {Frenk} C.~S.,
  2002, \mnras, 333, 156

\bibitem[\protect\citeauthoryear{{Benson}, {Pearce}, {Frenk}, {Baugh} \&
  {Jenkins}}{{Benson} et~al.}{2001}]{Benson2001}
{Benson} A.~J.,  {Pearce} F.~R.,  {Frenk} C.~S.,  {Baugh} C.~M.,    {Jenkins}
  A.,  2001, \mnras, 320, 261

\bibitem[\protect\citeauthoryear{{Binggeli}, {Tarenghi} \&
  {Sandage}}{{Binggeli} et~al.}{1990}]{Binggeli1990}
{Binggeli} B.,  {Tarenghi} M.,    {Sandage} A.,  1990, \aap, 228, 42

\bibitem[\protect\citeauthoryear{{Binney} \& {Tremaine}}{{Binney} \&
  {Tremaine}}{1987}]{Binney1987}
{Binney} J.,  {Tremaine} S.,  1987, {Galactic dynamics}

\bibitem[\protect\citeauthoryear{{Binney} \& {Tremaine}}{{Binney} \&
  {Tremaine}}{2008}]{Binney2008}
{Binney} J.,  {Tremaine} S.,  2008, {Galactic Dynamics: Second Edition}.
Princeton University Press

\bibitem[\protect\citeauthoryear{{Birnboim} \& {Dekel}}{{Birnboim} \&
  {Dekel}}{2003}]{Birnboim2003}
{Birnboim} Y.,  {Dekel} A.,  2003, \mnras, 345, 349

\bibitem[\protect\citeauthoryear{{B{\^i}rzan}, {Rafferty}, {McNamara}, {Wise}
  \& {Nulsen}}{{B{\^i}rzan} et~al.}{2004}]{Birzan2004}
{B{\^i}rzan} L.,  {Rafferty} D.~A.,  {McNamara} B.~R.,  {Wise} M.~W.,
  {Nulsen} P.~E.~J.,  2004, \apj, 607, 800

\bibitem[\protect\citeauthoryear{{Blanton}, {Geha} \& {West}}{{Blanton}
  et~al.}{2008}]{Blanton2008}
{Blanton} M.~R.,  {Geha} M.,    {West} A.~A.,  2008, \apj, 682, 861

\bibitem[\protect\citeauthoryear{{Blanton}, {Lupton}, {Schlegel}, {Strauss},
  {Brinkmann}, {Fukugita} \& {Loveday}}{{Blanton} et~al.}{2005}]{Blanton2005}
{Blanton} M.~R.,  {Lupton} R.~H.,  {Schlegel} D.~J.,  {Strauss} M.~A.,
  {Brinkmann} J.,  {Fukugita} M.,    {Loveday} J.,  2005, \apj, 631, 208

\bibitem[\protect\citeauthoryear{{Blumenthal}, {Faber}, {Flores} \&
  {Primack}}{{Blumenthal} et~al.}{1986}]{Blumenthal1986}
{Blumenthal} G.~R.,  {Faber} S.~M.,  {Flores} R.,    {Primack} J.~R.,  1986,
  \apj, 301, 27

\bibitem[\protect\citeauthoryear{{Blumenthal}, {Faber}, {Primack} \&
  {Rees}}{{Blumenthal} et~al.}{1984}]{Blumenthal1984}
{Blumenthal} G.~R.,  {Faber} S.~M.,  {Primack} J.~R.,    {Rees} M.~J.,  1984,
  \nat, 311, 517

\bibitem[\protect\citeauthoryear{{Bode}, {Ostriker} \& {Turok}}{{Bode}
  et~al.}{2001}]{Bode2001}
{Bode} P.,  {Ostriker} J.~P.,    {Turok} N.,  2001, \apj, 556, 93

\bibitem[\protect\citeauthoryear{{Bower}, {Benson}, {Malbon}, {Helly}, {Frenk},
  {Baugh}, {Cole} \& {Lacey}}{{Bower} et~al.}{2006}]{Bower2006}
{Bower} R.~G.,  {Benson} A.~J.,  {Malbon} R.,  {Helly} J.~C.,  {Frenk} C.~S.,
  {Baugh} C.~M.,  {Cole} S.,    {Lacey} C.~G.,  2006, \mnras, 370, 645

\bibitem[\protect\citeauthoryear{{Bower}, {McCarthy} \& {Benson}}{{Bower}
  et~al.}{2008}]{Bower2008}
{Bower} R.~G.,  {McCarthy} I.~G.,    {Benson} A.~J.,  2008, \mnras, 390, 1399

\bibitem[\protect\citeauthoryear{{Bower}, {Vernon}, {Goldstein}, {Benson},
  {Lacey}, {Baugh}, {Cole} \& {Frenk}}{{Bower} et~al.}{2010}]{Bower2010}
{Bower} R.~G.,  {Vernon} I.,  {Goldstein} M.,  {Benson} A.~J.,  {Lacey} C.~G.,
  {Baugh} C.~M.,  {Cole} S.,    {Frenk} C.~S.,  2010, arXiv:1004.0711
  [astro-ph]

\bibitem[\protect\citeauthoryear{{Boyarsky}, {Lesgourgues}, {Ruchayskiy} \&
  {Viel}}{{Boyarsky} et~al.}{2009a}]{Boyarsky2009a}
{Boyarsky} A.,  {Lesgourgues} J.,  {Ruchayskiy} O.,    {Viel} M.,  2009a,
  Journal of Cosmology and Astro-Particle Physics, 5, 12

\bibitem[\protect\citeauthoryear{{Boyarsky}, {Lesgourgues}, {Ruchayskiy} \&
  {Viel}}{{Boyarsky} et~al.}{2009b}]{Boyarsky2009b}
{Boyarsky} A.,  {Lesgourgues} J.,  {Ruchayskiy} O.,    {Viel} M.,  2009b,
  Physical Review Letters, 102, 201304

\bibitem[\protect\citeauthoryear{{Boylan-Kolchin}, {Ma} \&
  {Quataert}}{{Boylan-Kolchin} et~al.}{2005}]{Boylan2005}
{Boylan-Kolchin} M.,  {Ma} C.,    {Quataert} E.,  2005, \mnras, 362, 184

\bibitem[\protect\citeauthoryear{{Boylan-Kolchin}, {Ma} \&
  {Quataert}}{{Boylan-Kolchin} et~al.}{2008}]{Boylan2008}
{Boylan-Kolchin} M.,  {Ma} C.-P.,    {Quataert} E.,  2008, \mnras, 383, 93

\bibitem[\protect\citeauthoryear{{Boylan-Kolchin}, {Springel}, {White} \&
  {Jenkins}}{{Boylan-Kolchin} et~al.}{2010}]{Boylan2009b}
{Boylan-Kolchin} M.,  {Springel} V.,  {White} S.~D.~M.,    {Jenkins} A.,  2010,
  MNRAS, in press; arXiv:0911.4484 [astro-ph]

\bibitem[\protect\citeauthoryear{{Boylan-Kolchin}, {Springel}, {White},
  {Jenkins} \& {Lemson}}{{Boylan-Kolchin} et~al.}{2009}]{Boylan2009}
{Boylan-Kolchin} M.,  {Springel} V.,  {White} S.~D.~M.,  {Jenkins} A.,
  {Lemson} G.,  2009, \mnras, 398, 1150

\bibitem[\protect\citeauthoryear{{Bruzual} \& {Charlot}}{{Bruzual} \&
  {Charlot}}{2003}]{Bruzual2003}
{Bruzual} G.,  {Charlot} S.,  2003, \mnras, 344, 1000

\bibitem[\protect\citeauthoryear{{Bullock}, {Kravtsov} \& {Weinberg}}{{Bullock}
  et~al.}{2000}]{Bullock2000}
{Bullock} J.~S.,  {Kravtsov} A.~V.,    {Weinberg} D.~H.,  2000, \apj, 539, 517

\bibitem[\protect\citeauthoryear{{Cappellari}, {Bacon}, {Bureau}, {Damen},
  {Davies}, {de Zeeuw}, {Emsellem}, {Falc{\'o}n-Barroso}, {Krajnovi{\'c}},
  {Kuntschner}, {McDermid}, {Peletier}, {Sarzi}, {van den Bosch} \& {van de
  Ven}}{{Cappellari} et~al.}{2006}]{Cappellari2006}
{Cappellari} M.,  {Bacon} R.,  {Bureau} M.,  {Damen} M.~C.,  {Davies} R.~L.,
  {de Zeeuw} P.~T.,  {Emsellem} E.,  {Falc{\'o}n-Barroso} J.,  {Krajnovi{\'c}}
  D.,  {Kuntschner} H.,  {McDermid} R.~M.,  {Peletier} R.~F.,  {Sarzi} M.,
  {van den Bosch} R.~C.~E.,    {van de Ven} G.,  2006, \mnras, 366, 1126

\bibitem[\protect\citeauthoryear{{Cattaneo}, {Blaizot}, {Weinberg}, {Kere{\v
  s}}, {Colombi}, {Dav{\'e}}, {Devriendt}, {Guiderdoni} \& {Katz}}{{Cattaneo}
  et~al.}{2007}]{Cattaneo2007}
{Cattaneo} A.,  {Blaizot} J.,  {Weinberg} D.~H.,  {Kere{\v s}} D.,  {Colombi}
  S.,  {Dav{\'e}} R.,  {Devriendt} J.,  {Guiderdoni} B.,    {Katz} N.,  2007,
  \mnras, 377, 63

\bibitem[\protect\citeauthoryear{{Cirasuolo}, {McLure}, {Dunlop}, {Almaini},
  {Foucaud} \& {Simpson}}{{Cirasuolo} et~al.}{2010}]{Cirasuolo2010}
{Cirasuolo} M.,  {McLure} R.~J.,  {Dunlop} J.~S.,  {Almaini} O.,  {Foucaud} S.,
     {Simpson} C.,  2010, \mnras, 401, 1166

\bibitem[\protect\citeauthoryear{{Cole}, {Aragon-Salamanca}, {Frenk}, {Navarro}
  \& {Zepf}}{{Cole} et~al.}{1994}]{Cole1994}
{Cole} S.,  {Aragon-Salamanca} A.,  {Frenk} C.~S.,  {Navarro} J.~F.,    {Zepf}
  S.~E.,  1994, \mnras, 271, 781

\bibitem[\protect\citeauthoryear{{Cole}, {Lacey}, {Baugh} \& {Frenk}}{{Cole}
  et~al.}{2000}]{Cole2000}
{Cole} S.,  {Lacey} C.~G.,  {Baugh} C.~M.,    {Frenk} C.~S.,  2000, \mnras,
  319, 168

\bibitem[\protect\citeauthoryear{{Colless} \& et al.}{{Colless} \&
  et~al.}{2001}]{Colless2001}
{Colless} M.,  et al. 2001, \mnras, 328, 1039

\bibitem[\protect\citeauthoryear{{Conroy}, {Wechsler} \& {Kravtsov}}{{Conroy}
  et~al.}{2006}]{Conroy2006}
{Conroy} C.,  {Wechsler} R.~H.,    {Kravtsov} A.~V.,  2006, \apj, 647, 201

\bibitem[\protect\citeauthoryear{{Conselice}}{{Conselice}}{2006}]{Conselice200%
6}
{Conselice} C.~J.,  2006, \mnras, 373, 1389

\bibitem[\protect\citeauthoryear{{Couchman} \& {Rees}}{{Couchman} \&
  {Rees}}{1986}]{Couchman1986}
{Couchman} H.~M.~P.,  {Rees} M.~J.,  1986, \mnras, 221, 53

\bibitem[\protect\citeauthoryear{{Covington}, {Dekel}, {Cox}, {Jonsson} \&
  {Primack}}{{Covington} et~al.}{2008}]{Covington2008}
{Covington} M.,  {Dekel} A.,  {Cox} T.~J.,  {Jonsson} P.,    {Primack} J.~R.,
  2008, \mnras, 384, 94

\bibitem[\protect\citeauthoryear{{Cox}, {Jonsson}, {Somerville}, {Primack} \&
  {Dekel}}{{Cox} et~al.}{2008}]{Cox2008}
{Cox} T.~J.,  {Jonsson} P.,  {Somerville} R.~S.,  {Primack} J.~R.,    {Dekel}
  A.,  2008, \mnras, 384, 386

\bibitem[\protect\citeauthoryear{{Croton} \& {Farrar}}{{Croton} \&
  {Farrar}}{2008}]{Croton2008}
{Croton} D.~J.,  {Farrar} G.~R.,  2008, \mnras, 386, 2285

\bibitem[\protect\citeauthoryear{{Croton}, {Springel}, {White}, {De Lucia},
  {Frenk}, {Gao}, {Jenkins}, {Kauffmann}, {Navarro} \& {Yoshida}}{{Croton}
  et~al.}{2006}]{Croton2006}
{Croton} D.~J.,  {Springel} V.,  {White} S.~D.~M.,  {De Lucia} G.,  {Frenk}
  C.~S.,  {Gao} L.,  {Jenkins} A.,  {Kauffmann} G.,  {Navarro} J.~F.,
  {Yoshida} N.,  2006, \mnras, 365, 11

\bibitem[\protect\citeauthoryear{{Davis}, {Efstathiou}, {Frenk} \&
  {White}}{{Davis} et~al.}{1985}]{Davis1985}
{Davis} M.,  {Efstathiou} G.,  {Frenk} C.~S.,    {White} S.~D.~M.,  1985, \apj,
  292, 371

\bibitem[\protect\citeauthoryear{{De Lucia} \& {Blaizot}}{{De Lucia} \&
  {Blaizot}}{2007}]{DeLucia2007}
{De Lucia} G.,  {Blaizot} J.,  2007, \mnras, 375, 2

\bibitem[\protect\citeauthoryear{{De Lucia}, {Kauffmann} \& {White}}{{De Lucia}
  et~al.}{2004}]{DeLucia2004}
{De Lucia} G.,  {Kauffmann} G.,    {White} S.~D.~M.,  2004, \mnras, 349, 1101

\bibitem[\protect\citeauthoryear{{De Lucia}, {Springel}, {White}, {Croton} \&
  {Kauffmann}}{{De Lucia} et~al.}{2006}]{Delucia2006}
{De Lucia} G.,  {Springel} V.,  {White} S.~D.~M.,  {Croton} D.,    {Kauffmann}
  G.,  2006, \mnras, 366, 499

\bibitem[\protect\citeauthoryear{{Dekel}, {Birnboim}, {Engel}, {Freundlich},
  {Goerdt}, {Mumcuoglu}, {Neistein}, {Pichon}, {Teyssier} \& {Zinger}}{{Dekel}
  et~al.}{2009}]{Dekel2009}
{Dekel} A.,  {Birnboim} Y.,  {Engel} G.,  {Freundlich} J.,  {Goerdt} T.,
  {Mumcuoglu} M.,  {Neistein} E.,  {Pichon} C.,  {Teyssier} R.,    {Zinger} E.,
   2009, \nat, 457, 451

\bibitem[\protect\citeauthoryear{{Dekel} \& {Cox}}{{Dekel} \&
  {Cox}}{2006}]{Dekel2006b}
{Dekel} A.,  {Cox} T.~J.,  2006, \mnras, 370, 1445

\bibitem[\protect\citeauthoryear{{Delahaye} \& {Pinsonneault}}{{Delahaye} \&
  {Pinsonneault}}{2006}]{Delahaye2006}
{Delahaye} F.,  {Pinsonneault} M.~H.,  2006, \apj, 649, 529

\bibitem[\protect\citeauthoryear{{Diemand}, {Kuhlen} \& {Madau}}{{Diemand}
  et~al.}{2007}]{Diemand2007}
{Diemand} J.,  {Kuhlen} M.,    {Madau} P.,  2007, \apj, 657, 262

\bibitem[\protect\citeauthoryear{{Doroshkevich}, {Zel'Dovich} \&
  {Novikov}}{{Doroshkevich} et~al.}{1967}]{Doroshkevich1967}
{Doroshkevich} A.~G.,  {Zel'Dovich} Y.~B.,    {Novikov} I.~D.,  1967, Soviet
  Astronomy, 11, 233

\bibitem[\protect\citeauthoryear{{Dressler}}{{Dressler}}{1980}]{Dressler1980}
{Dressler} A.,  1980, \apj, 236, 351

\bibitem[\protect\citeauthoryear{{Dunkley}, {Komatsu}, {Nolta}, {Spergel},
  {Larson}, {Hinshaw}, {Page}, {Bennett}, {Gold}, {Jarosik}, {Weiland},
  {Halpern}, {Hill}, {Kogut}, {Limon}, {Meyer}, {Tucker}, {Wollack} \&
  {Wright}}{{Dunkley} et~al.}{2009}]{Dunkley2009}
{Dunkley} J.,  {Komatsu} E.,  {Nolta} M.~R.,  {Spergel} D.~N.,  {Larson} D.,
  {Hinshaw} G.,  {Page} L.,  {Bennett} C.~L.,  {Gold} B.,  {Jarosik} N.,
  {Weiland} J.~L.,  {Halpern} M.,  {Hill} R.~S.,  {Kogut} A.,  {Limon} M.,
  {Meyer} S.~S.,  {Tucker} G.~S.,  {Wollack} E.,    {Wright} E.~L.,  2009,
  \apjs, 180, 306

\bibitem[\protect\citeauthoryear{{Efstathiou}}{{Efstathiou}}{1992}]{Efstathiou%
1992}
{Efstathiou} G.,  1992, \mnras, 256, 43P

\bibitem[\protect\citeauthoryear{{Efstathiou}, {Lake} \&
  {Negroponte}}{{Efstathiou} et~al.}{1982}]{Efstathiou1982}
{Efstathiou} G.,  {Lake} G.,    {Negroponte} J.,  1982, \mnras, 199, 1069

\bibitem[\protect\citeauthoryear{{Einasto}, {Saar}, {Kaasik} \&
  {Chernin}}{{Einasto} et~al.}{1974}]{Einasto1974}
{Einasto} J.,  {Saar} E.,  {Kaasik} A.,    {Chernin} A.~D.,  1974, \nat, 252,
  111

\bibitem[\protect\citeauthoryear{{Flynn}, {Holmberg}, {Portinari}, {Fuchs} \&
  {Jahrei{\ss}}}{{Flynn} et~al.}{2006}]{Flynn2006}
{Flynn} C.,  {Holmberg} J.,  {Portinari} L.,  {Fuchs} B.,    {Jahrei{\ss}} H.,
  2006, \mnras, 372, 1149

\bibitem[\protect\citeauthoryear{{Font}, {Bower}, {McCarthy}, {Benson},
  {Frenk}, {Helly}, {Lacey}, {Baugh} \& {Cole}}{{Font} et~al.}{2008}]{Font2008}
{Font} A.~S.,  {Bower} R.~G.,  {McCarthy} I.~G.,  {Benson} A.~J.,  {Frenk}
  C.~S.,  {Helly} J.~C.,  {Lacey} C.~G.,  {Baugh} C.~M.,    {Cole} S.,  2008,
  \mnras, 389, 1619

\bibitem[\protect\citeauthoryear{{Fontana}, {Salimbeni}, {Grazian},
  {Giallongo}, {Pentericci}, {Nonino}, {Fontanot}, {Menci}, {Monaco},
  {Cristiani}, {Vanzella}, {de Santis} \& {Gallozzi}}{{Fontana}
  et~al.}{2006}]{Fontana2006}
{Fontana} A.,  {Salimbeni} S.,  {Grazian} A.,  {Giallongo} E.,  {Pentericci}
  L.,  {Nonino} M.,  {Fontanot} F.,  {Menci} N.,  {Monaco} P.,  {Cristiani} S.,
   {Vanzella} E.,  {de Santis} C.,    {Gallozzi} S.,  2006, \aap, 459, 745

\bibitem[\protect\citeauthoryear{{Fontanot}, {De Lucia}, {Monaco}, {Somerville}
  \& {Santini}}{{Fontanot} et~al.}{2009}]{Fontanot2009}
{Fontanot} F.,  {De Lucia} G.,  {Monaco} P.,  {Somerville} R.~S.,    {Santini}
  P.,  2009, \mnras, 397, 1776

\bibitem[\protect\citeauthoryear{{Forcada-Miro} \& {White}}{{Forcada-Miro} \&
  {White}}{1997}]{Forcada-Miro1997}
{Forcada-Miro} M.~I.,  {White} S.~D.~M.,  1997, arXiv:astro-ph/9712204

\bibitem[\protect\citeauthoryear{{Fu}}{{Fu}}{2008}]{Fu2008}
{Fu} L. e.~a.,  2008, \aap, 479, 9

\bibitem[\protect\citeauthoryear{{Gao}, {De Lucia}, {White} \& {Jenkins}}{{Gao}
  et~al.}{2004}]{Gao2004}
{Gao} L.,  {De Lucia} G.,  {White} S.~D.~M.,    {Jenkins} A.,  2004, \mnras,
  352, L1

\bibitem[\protect\citeauthoryear{{Gerhard}, {Arnaboldi}, {Freeman},
  {Kashikawa}, {Okamura} \& {Yasuda}}{{Gerhard} et~al.}{2005}]{Gerhard2005}
{Gerhard} O.,  {Arnaboldi} M.,  {Freeman} K.~C.,  {Kashikawa} N.,  {Okamura}
  S.,    {Yasuda} N.,  2005, \apjl, 621, L93

\bibitem[\protect\citeauthoryear{{Gnedin}}{{Gnedin}}{2000}]{Gnedin2000}
{Gnedin} N.~Y.,  2000, \apj, 542, 535

\bibitem[\protect\citeauthoryear{{Gnedin} \& {Kravtsov}}{{Gnedin} \&
  {Kravtsov}}{2006}]{Gnedin2006}
{Gnedin} N.~Y.,  {Kravtsov} A.~V.,  2006, \apj, 645, 1054

\bibitem[\protect\citeauthoryear{{Gnedin}, {Kravtsov}, {Klypin} \&
  {Nagai}}{{Gnedin} et~al.}{2004}]{Gnedin2004}
{Gnedin} O.~Y.,  {Kravtsov} A.~V.,  {Klypin} A.~A.,    {Nagai} D.,  2004, \apj,
  616, 16

\bibitem[\protect\citeauthoryear{{Gonzalez}, {Zabludoff} \&
  {Zaritsky}}{{Gonzalez} et~al.}{2005}]{Gonzalez2005}
{Gonzalez} A.~H.,  {Zabludoff} A.~I.,    {Zaritsky} D.,  2005, \apj, 618, 195

\bibitem[\protect\citeauthoryear{{Gonzalez}, {Zaritsky} \&
  {Zabludoff}}{{Gonzalez} et~al.}{2007}]{Gonzalez2007}
{Gonzalez} A.~H.,  {Zaritsky} D.,    {Zabludoff} A.~I.,  2007, \apj, 666, 147

\bibitem[\protect\citeauthoryear{{Guo}, {White}, {Li} \&
  {Boylan-Kolchin}}{{Guo} et~al.}{2010}]{Guo2010}
{Guo} Q.,  {White} S.,  {Li} C.,    {Boylan-Kolchin} M.,  2010, \mnras, 404,
  1111

\bibitem[\protect\citeauthoryear{{Guo} \& {White}}{{Guo} \&
  {White}}{2009}]{Guo2009}
{Guo} Q.,  {White} S.~D.~M.,  2009, \mnras, 396, 39

\bibitem[\protect\citeauthoryear{{Hambrick}, {Ostriker}, {Naab} \&
  {Johansson}}{{Hambrick} et~al.}{2009}]{Hambrick2009}
{Hambrick} D.~C.,  {Ostriker} J.~P.,  {Naab} T.,    {Johansson} P.~H.,  2009,
  \apj, 705, 1566

\bibitem[\protect\citeauthoryear{{Hansen}, {Sheldon}, {Wechsler} \&
  {Koester}}{{Hansen} et~al.}{2009}]{Hansen2009}
{Hansen} S.~M.,  {Sheldon} E.~S.,  {Wechsler} R.~H.,    {Koester} B.~P.,  2009,
  \apj, 699, 1333

\bibitem[\protect\citeauthoryear{{H{\"a}ring} \& {Rix}}{{H{\"a}ring} \&
  {Rix}}{2004}]{Haring2004}
{H{\"a}ring} N.,  {Rix} H.,  2004, \apjl, 604, L89

\bibitem[\protect\citeauthoryear{{Hatton}, {Devriendt}, {Ninin}, {Bouchet},
  {Guiderdoni} \& {Vibert}}{{Hatton} et~al.}{2003}]{Hatton2003}
{Hatton} S.,  {Devriendt} J.~E.~G.,  {Ninin} S.,  {Bouchet} F.~R.,
  {Guiderdoni} B.,    {Vibert} D.,  2003, \mnras, 343, 75

\bibitem[\protect\citeauthoryear{{Hayashi}, {Navarro}, {Taylor}, {Stadel} \&
  {Quinn}}{{Hayashi} et~al.}{2003}]{Hayashi2003}
{Hayashi} E.,  {Navarro} J.~F.,  {Taylor} J.~E.,  {Stadel} J.,    {Quinn} T.,
  2003, \apj, 584, 541

\bibitem[\protect\citeauthoryear{{Henriques}, {Thomas}, {Oliver} \&
  {Roseboom}}{{Henriques} et~al.}{2009}]{Henriques2009}
{Henriques} B.~M.~B.,  {Thomas} P.~A.,  {Oliver} S.,    {Roseboom} I.,  2009,
  \mnras, 396, 535

\bibitem[\protect\citeauthoryear{{Hoeft}, {Yepes}, {Gottl{\"o}ber} \&
  {Springel}}{{Hoeft} et~al.}{2006}]{Hoeft2006}
{Hoeft} M.,  {Yepes} G.,  {Gottl{\"o}ber} S.,    {Springel} V.,  2006, \mnras,
  371, 401

\bibitem[\protect\citeauthoryear{{Hopkins}}{{Hopkins}}{2007}]{Hopkins2007}
{Hopkins} A.~M.,  2007, in {J.~Afonso, H.~C.~Ferguson, B.~Mobasher, \&
  R.~Norris} ed., Deepest Astronomical Surveys Vol.~380 of Astronomical Society
  of the Pacific Conference Series, {The Star Formation History of the
  Universe}.
pp 423--+

\bibitem[\protect\citeauthoryear{{Hopkins}, {Hernquist}, {Cox}, {Keres} \&
  {Wuyts}}{{Hopkins} et~al.}{2009}]{Hopkins2009}
{Hopkins} P.~F.,  {Hernquist} L.,  {Cox} T.~J.,  {Keres} D.,    {Wuyts} S.,
  2009, \apj, 691, 1424

\bibitem[\protect\citeauthoryear{{Jeltema}, {Binder} \& {Mulchaey}}{{Jeltema}
  et~al.}{2008}]{Jeltema2008}
{Jeltema} T.~E.,  {Binder} B.,    {Mulchaey} J.~S.,  2008, \apj, 679, 1162

\bibitem[\protect\citeauthoryear{{Jenkins}, {Hornschemeier}, {Mobasher},
  {Alexander} \& {Bauer}}{{Jenkins} et~al.}{2007}]{Jenkins2007}
{Jenkins} L.~P.,  {Hornschemeier} A.~E.,  {Mobasher} B.,  {Alexander} D.~M.,
  {Bauer} F.~E.,  2007, \apj, 666, 846

\bibitem[\protect\citeauthoryear{{Jiang}, {Jing}, {Faltenbacher}, {Lin} \&
  {Li}}{{Jiang} et~al.}{2008}]{Jiang2008}
{Jiang} C.~Y.,  {Jing} Y.~P.,  {Faltenbacher} A.,  {Lin} W.~P.,    {Li} C.,
  2008, \apj, 675, 1095

\bibitem[\protect\citeauthoryear{{Kang}, {Jing}, {Mo} \& {B{\"o}rner}}{{Kang}
  et~al.}{2005}]{Kang2005}
{Kang} X.,  {Jing} Y.~P.,  {Mo} H.~J.,    {B{\"o}rner} G.,  2005, \apj, 631, 21

\bibitem[\protect\citeauthoryear{{Kauffmann}}{{Kauffmann}}{1996}]{Kauffmann199%
6}
{Kauffmann} G.,  1996, \mnras, 281, 487

\bibitem[\protect\citeauthoryear{{Kauffmann}, {Colberg}, {Diaferio} \&
  {White}}{{Kauffmann} et~al.}{1999}]{Kauffmann1999}
{Kauffmann} G.,  {Colberg} J.~M.,  {Diaferio} A.,    {White} S.~D.~M.,  1999,
  \mnras, 303, 188

\bibitem[\protect\citeauthoryear{{Kauffmann}, {White} \&
  {Guiderdoni}}{{Kauffmann} et~al.}{1993}]{Kauffmann1993}
{Kauffmann} G.,  {White} S.~D.~M.,    {Guiderdoni} B.,  1993, \mnras, 264, 201

\bibitem[\protect\citeauthoryear{{Kazantzidis}, {Mayer}, {Mastropietro},
  {Diemand}, {Stadel} \& {Moore}}{{Kazantzidis} et~al.}{2004}]{Kazantzidis2004}
{Kazantzidis} S.,  {Mayer} L.,  {Mastropietro} C.,  {Diemand} J.,  {Stadel} J.,
     {Moore} B.,  2004, \apj, 608, 663

\bibitem[\protect\citeauthoryear{{Kennicutt}
  Jr.}{{Kennicutt}}{1998}]{Kennicutt1998}
{Kennicutt} Jr. R.~C.,  1998, \apj, 498, 541

\bibitem[\protect\citeauthoryear{{Kitzbichler} \& {White}}{{Kitzbichler} \&
  {White}}{2007}]{Kitzbichler2007}
{Kitzbichler} M.~G.,  {White} S.~D.~M.,  2007, \mnras, 376, 2

\bibitem[\protect\citeauthoryear{{Klypin}, {Gottl{\"o}ber}, {Kravtsov} \&
  {Khokhlov}}{{Klypin} et~al.}{1999}]{Klypin1999}
{Klypin} A.,  {Gottl{\"o}ber} S.,  {Kravtsov} A.~V.,    {Khokhlov} A.~M.,
  1999, \apj, 516, 530

\bibitem[\protect\citeauthoryear{{Komatsu}, {Smith}, {Dunkley}, {Bennett},
  {Gold}, {Hinshaw}, {Jarosik} \& {Larson et al.}}{{Komatsu}
  et~al.}{2010}]{Komatsu2010}
{Komatsu} E.,  {Smith} K.~M.,  {Dunkley} J.,  {Bennett} C.~L.,  {Gold} B.,
  {Hinshaw} G.,  {Jarosik} N.,    {Larson et al.} 2010, ArXiv e-prints

\bibitem[\protect\citeauthoryear{{Koposov}, {Belokurov}, {Evans}, {Hewett},
  {Irwin}, {Gilmore}, {Zucker}, {Rix}, {Fellhauer}, {Bell} \&
  {Glushkova}}{{Koposov} et~al.}{2008}]{Koposov2008}
{Koposov} S.,  {Belokurov} V.,  {Evans} N.~W.,  {Hewett} P.~C.,  {Irwin} M.~J.,
   {Gilmore} G.,  {Zucker} D.~B.,  {Rix} H.-W.,  {Fellhauer} M.,  {Bell} E.~F.,
     {Glushkova} E.~V.,  2008, \apj, 686, 279

\bibitem[\protect\citeauthoryear{{Kravtsov}, {Gnedin} \& {Klypin}}{{Kravtsov}
  et~al.}{2004}]{Kravtsov2004}
{Kravtsov} A.~V.,  {Gnedin} O.~Y.,    {Klypin} A.~A.,  2004, \apj, 609, 482

\bibitem[\protect\citeauthoryear{{Larson}}{{Larson}}{1974}]{Larson1974}
{Larson} R.~B.,  1974, \mnras, 169, 229

\bibitem[\protect\citeauthoryear{{Lee}, {Skillman}, {Cannon}, {Jackson},
  {Gehrz}, {Polomski} \& {Woodward}}{{Lee} et~al.}{2006}]{Lee2006}
{Lee} H.,  {Skillman} E.~D.,  {Cannon} J.~M.,  {Jackson} D.~C.,  {Gehrz} R.~D.,
   {Polomski} E.~F.,    {Woodward} C.~E.,  2006, \apj, 647, 970

\bibitem[\protect\citeauthoryear{{Li}, {Kauffmann}, {Jing}, {White},
  {B{\"o}rner} \& {Cheng}}{{Li} et~al.}{2006}]{Li2006}
{Li} C.,  {Kauffmann} G.,  {Jing} Y.~P.,  {White} S.~D.~M.,  {B{\"o}rner} G.,
   {Cheng} F.~Z.,  2006, \mnras, 368, 21

\bibitem[\protect\citeauthoryear{{Li} \& {White}}{{Li} \&
  {White}}{2009}]{Li2009}
{Li} C.,  {White} S.~D.~M.,  2009, \mnras, 398, 2177

\bibitem[\protect\citeauthoryear{{Li}, {De Lucia} \& {Helmi}}{{Li}
  et~al.}{2010}]{Li2010}
{Li} Y.,  {De Lucia} G.,    {Helmi} A.,  2010, \mnras, 401, 2036

\bibitem[\protect\citeauthoryear{{Lo Faro}, {Monaco}, {Vanzella}, {Fontanot},
  {Silva} \& {Cristiani}}{{Lo Faro} et~al.}{2009}]{LoFaro2009}
{Lo Faro} B.,  {Monaco} P.,  {Vanzella} E.,  {Fontanot} F.,  {Silva} L.,
  {Cristiani} S.,  2009, \mnras, 399, 827

\bibitem[\protect\citeauthoryear{{Macci{\`o}}, {Kang}, {Fontanot},
  {Somerville}, {Koposov} \& {Monaco}}{{Macci{\`o}} et~al.}{2010}]{Maccio2010}
{Macci{\`o}} A.~V.,  {Kang} X.,  {Fontanot} F.,  {Somerville} R.~S.,  {Koposov}
  S.,    {Monaco} P.,  2010, \mnras, 402, 1995

\bibitem[\protect\citeauthoryear{{Marchesini}, {van Dokkum}, {F{\"o}rster
  Schreiber}, {Franx}, {Labb{\'e}} \& {Wuyts}}{{Marchesini}
  et~al.}{2009}]{Marchesini2009}
{Marchesini} D.,  {van Dokkum} P.~G.,  {F{\"o}rster Schreiber} N.~M.,  {Franx}
  M.,  {Labb{\'e}} I.,    {Wuyts} S.,  2009, \apj, 701, 1765

\bibitem[\protect\citeauthoryear{{McCarthy}, {Frenk}, {Font}, {Lacey}, {Bower},
  {Mitchell}, {Balogh} \& {Theuns}}{{McCarthy} et~al.}{2008}]{McCarthy2008}
{McCarthy} I.~G.,  {Frenk} C.~S.,  {Font} A.~S.,  {Lacey} C.~G.,  {Bower}
  R.~G.,  {Mitchell} N.~L.,  {Balogh} M.~L.,    {Theuns} T.,  2008, \mnras,
  383, 593

\bibitem[\protect\citeauthoryear{{McGee} \& {Balogh}}{{McGee} \&
  {Balogh}}{2010}]{McGee2010}
{McGee} S.~L.,  {Balogh} M.~L.,  2010, \mnras, pp~L21+

\bibitem[\protect\citeauthoryear{{McNamara} \& {Nulsen}}{{McNamara} \&
  {Nulsen}}{2007}]{McNamara2007}
{McNamara} B.~R.,  {Nulsen} P.~E.~J.,  2007, \araa, 45, 117

\bibitem[\protect\citeauthoryear{{Mihos}, {Harding}, {Feldmeier} \&
  {Morrison}}{{Mihos} et~al.}{2005}]{Mihos2005}
{Mihos} J.~C.,  {Harding} P.,  {Feldmeier} J.,    {Morrison} H.,  2005, \apjl,
  631, L41

\bibitem[\protect\citeauthoryear{{Milne}, {Pritchet}, {Poole}, {Gwyn},
  {Kavelaars}, {Harris} \& {Hanes}}{{Milne} et~al.}{2007}]{Milne2007}
{Milne} M.~L.,  {Pritchet} C.~J.,  {Poole} G.~B.,  {Gwyn} S.~D.~J.,
  {Kavelaars} J.~J.,  {Harris} W.~E.,    {Hanes} D.~A.,  2007, \aj, 133, 177

\bibitem[\protect\citeauthoryear{{Mo}, {Mao} \& {White}}{{Mo}
  et~al.}{1998}]{Mo1998}
{Mo} H.~J.,  {Mao} S.,    {White} S.~D.~M.,  1998, \mnras, 295, 319

\bibitem[\protect\citeauthoryear{{Mobasher}, {Colless}, {Carter}, {Poggianti},
  {Bridges}, {Kranz}, {Komiyama}, {Kashikawa}, {Yagi} \& {Okamura}}{{Mobasher}
  et~al.}{2003}]{Mobasher2003}
{Mobasher} B.,  {Colless} M.,  {Carter} D.,  {Poggianti} B.~M.,  {Bridges}
  T.~J.,  {Kranz} K.,  {Komiyama} Y.,  {Kashikawa} N.,  {Yagi} M.,    {Okamura}
  S.,  2003, \apj, 587, 605

\bibitem[\protect\citeauthoryear{{Moore}, {Ghigna}, {Governato}, {Lake},
  {Quinn}, {Stadel} \& {Tozzi}}{{Moore} et~al.}{1999}]{Moore1999}
{Moore} B.,  {Ghigna} S.,  {Governato} F.,  {Lake} G.,  {Quinn} T.,  {Stadel}
  J.,    {Tozzi} P.,  1999, \apjl, 524, L19

\bibitem[\protect\citeauthoryear{{More}, {van den Bosch}, {Cacciato}, {Mo},
  {Yang} \& {Li}}{{More} et~al.}{2009}]{More2009}
{More} S.,  {van den Bosch} F.~C.,  {Cacciato} M.,  {Mo} H.~J.,  {Yang} X.,
  {Li} R.,  2009, \mnras, 392, 801

\bibitem[\protect\citeauthoryear{{Moster}, {Somerville}, {Maulbetsch}, {van den
  Bosch}, {Macci{\`o}}, {Naab} \& {Oser}}{{Moster} et~al.}{2010}]{Moster2010}
{Moster} B.~P.,  {Somerville} R.~S.,  {Maulbetsch} C.,  {van den Bosch} F.~C.,
  {Macci{\`o}} A.~V.,  {Naab} T.,    {Oser} L.,  2010, \apj, 710, 903

\bibitem[\protect\citeauthoryear{{Naab} \& {Burkert}}{{Naab} \&
  {Burkert}}{2003}]{Naab2003}
{Naab} T.,  {Burkert} A.,  2003, \apj, 597, 893

\bibitem[\protect\citeauthoryear{{Nair}, {van den Bergh} \& {Abraham}}{{Nair}
  et~al.}{2010}]{Nair2010}
{Nair} P.~B.,  {van den Bergh} S.,    {Abraham} R.~G.,  2010, \apj, 715, 606

\bibitem[\protect\citeauthoryear{{Navarro} \& {Steinmetz}}{{Navarro} \&
  {Steinmetz}}{1997}]{Navarro1997}
{Navarro} J.~F.,  {Steinmetz} M.,  1997, \apj, 478, 13

\bibitem[\protect\citeauthoryear{{Navarro} \& {Steinmetz}}{{Navarro} \&
  {Steinmetz}}{2000}]{Navarro2000}
{Navarro} J.~F.,  {Steinmetz} M.,  2000, \apj, 528, 607

\bibitem[\protect\citeauthoryear{{Neistein} \& {Weinmann}}{{Neistein} \&
  {Weinmann}}{2009}]{Neistein2009}
{Neistein} E.,  {Weinmann} S.~M.,  2009, ArXiv 0911.3147

\bibitem[\protect\citeauthoryear{{Noordermeer}, {van der Hulst}, {Sancisi},
  {Swaters} \& {van Albada}}{{Noordermeer} et~al.}{2005}]{Noordermeer2005}
{Noordermeer} E.,  {van der Hulst} J.~M.,  {Sancisi} R.,  {Swaters} R.~A.,
  {van Albada} T.~S.,  2005, \aap, 442, 137

\bibitem[\protect\citeauthoryear{{Okamoto}, {Frenk}, {Jenkins} \&
  {Theuns}}{{Okamoto} et~al.}{2010}]{Okamoto2010}
{Okamoto} T.,  {Frenk} C.~S.,  {Jenkins} A.,    {Theuns} T.,  2010, \mnras,
  406, 208

\bibitem[\protect\citeauthoryear{{Okamoto}, {Gao} \& {Theuns}}{{Okamoto}
  et~al.}{2008}]{Okamoto2008}
{Okamoto} T.,  {Gao} L.,    {Theuns} T.,  2008, \mnras, 390, 920

\bibitem[\protect\citeauthoryear{{Paschos}, {Jena}, {Tytler}, {Kirkman} \&
  {Norman}}{{Paschos} et~al.}{2009}]{Paschos2009}
{Paschos} P.,  {Jena} T.,  {Tytler} D.,  {Kirkman} D.,    {Norman} M.~L.,
  2009, \mnras, 399, 1934

\bibitem[\protect\citeauthoryear{{Peng} \& et al.}{{Peng} \&
  et~al.}{2010}]{Peng2010}
{Peng} Y.,  et al. 2010, arXiv:1003.4747 [astro-ph]

\bibitem[\protect\citeauthoryear{{Percival}}{{Percival}}{2010}]{Percival2010}
{Percival} W.~J. e.~a.,  2010, \mnras, 401, 2148

\bibitem[\protect\citeauthoryear{{P{\'e}rez-Gonz{\'a}lez}, {Rieke}, {Villar},
  {Barro}, {Blaylock}, {Egami}, {Gallego}, {Gil de Paz}, {Pascual}, {Zamorano}
  \& {Donley}}{{P{\'e}rez-Gonz{\'a}lez} et~al.}{2008}]{Perez2008}
{P{\'e}rez-Gonz{\'a}lez} P.~G.,  {Rieke} G.~H.,  {Villar} V.,  {Barro} G.,
  {Blaylock} M.,  {Egami} E.,  {Gallego} J.,  {Gil de Paz} A.,  {Pascual} S.,
  {Zamorano} J.,    {Donley} J.~L.,  2008, \apj, 675, 234

\bibitem[\protect\citeauthoryear{{Popesso}, {Biviano}, {B{\"o}hringer} \&
  {Romaniello}}{{Popesso} et~al.}{2006}]{Popesso2006}
{Popesso} P.,  {Biviano} A.,  {B{\"o}hringer} H.,    {Romaniello} M.,  2006,
  \aap, 445, 29

\bibitem[\protect\citeauthoryear{{Quadri}, {Williams}, {Lee}, {Franx}, {van
  Dokkum} \& {Brammer}}{{Quadri} et~al.}{2008}]{Quadri2008}
{Quadri} R.~F.,  {Williams} R.~J.,  {Lee} K.,  {Franx} M.,  {van Dokkum} P.,
  {Brammer} G.~B.,  2008, \apjl, 685, L1

\bibitem[\protect\citeauthoryear{{Rees} \& {Ostriker}}{{Rees} \&
  {Ostriker}}{1977}]{Rees1977}
{Rees} M.~J.,  {Ostriker} J.~P.,  1977, \mnras, 179, 541

\bibitem[\protect\citeauthoryear{{Sawala}, {Scannapieco}, {Maio} \&
  {White}}{{Sawala} et~al.}{2010}]{Sawala2010}
{Sawala} T.,  {Scannapieco} C.,  {Maio} U.,    {White} S.,  2010, \mnras, 402,
  1599

\bibitem[\protect\citeauthoryear{{Sellwood} \& {Evans}}{{Sellwood} \&
  {Evans}}{2001}]{Sellwood2001}
{Sellwood} J.~A.,  {Evans} N.~W.,  2001, \apj, 546, 176

\bibitem[\protect\citeauthoryear{{Sellwood} \& {Moore}}{{Sellwood} \&
  {Moore}}{1999}]{Sellwood1999}
{Sellwood} J.~A.,  {Moore} E.~M.,  1999, \apj, 510, 125

\bibitem[\protect\citeauthoryear{{Shen}, {Mo}, {White}, {Blanton}, {Kauffmann},
  {Voges}, {Brinkmann} \& {Csabai}}{{Shen} et~al.}{2003}]{Shen2003}
{Shen} S.,  {Mo} H.~J.,  {White} S.~D.~M.,  {Blanton} M.~R.,  {Kauffmann} G.,
  {Voges} W.,  {Brinkmann} J.,    {Csabai} I.,  2003, \mnras, 343, 978

\bibitem[\protect\citeauthoryear{{Silk} \& {Rees}}{{Silk} \&
  {Rees}}{1998}]{Silk1998}
{Silk} J.,  {Rees} M.~J.,  1998, \aap, 331, L1

\bibitem[\protect\citeauthoryear{{Somerville}}{{Somerville}}{2002}]{Somerville%
2002}
{Somerville} R.~S.,  2002, \apjl, 572, L23

\bibitem[\protect\citeauthoryear{{Somerville}, {Hopkins}, {Cox}, {Robertson} \&
  {Hernquist}}{{Somerville} et~al.}{2008}]{Somerville2008}
{Somerville} R.~S.,  {Hopkins} P.~F.,  {Cox} T.~J.,  {Robertson} B.~E.,
  {Hernquist} L.,  2008, \mnras, 391, 481

\bibitem[\protect\citeauthoryear{{Somerville} \& {Primack}}{{Somerville} \&
  {Primack}}{1999}]{Somerville1999}
{Somerville} R.~S.,  {Primack} J.~R.,  1999, \mnras, 310, 1087

\bibitem[\protect\citeauthoryear{{Somerville}, {Primack} \&
  {Faber}}{{Somerville} et~al.}{2001}]{Somerville2001}
{Somerville} R.~S.,  {Primack} J.~R.,    {Faber} S.~M.,  2001, \mnras, 320, 504

\bibitem[\protect\citeauthoryear{{Spergel}, {Verde}, {Peiris}, {Komatsu} \& et
  al.}{{Spergel} et~al.}{2003}]{Spergel2003}
{Spergel} D.~N.,  {Verde} L.,  {Peiris} H.~V.,  {Komatsu} E.,    et al. 2003,
  \apjs, 148, 175

\bibitem[\protect\citeauthoryear{{Springel} \& {Hernquist}}{{Springel} \&
  {Hernquist}}{2003}]{Springel2003}
{Springel} V.,  {Hernquist} L.,  2003, \mnras, 339, 312

\bibitem[\protect\citeauthoryear{{Springel}, {Wang}, {Vogelsberger}, {Ludlow},
  {Jenkins}, {Helmi}, {Navarro}, {Frenk} \& {White}}{{Springel}
  et~al.}{2008}]{Springel2008}
{Springel} V.,  {Wang} J.,  {Vogelsberger} M.,  {Ludlow} A.,  {Jenkins} A.,
  {Helmi} A.,  {Navarro} J.~F.,  {Frenk} C.~S.,    {White} S.~D.~M.,  2008,
  \mnras, 391, 1685

\bibitem[\protect\citeauthoryear{{Springel}, {White}, {Jenkins}, {Frenk},
  {Yoshida}, {Gao}, {Navarro}, {Thacker}, {Croton}, {Helly}, {Peacock}, {Cole},
  {Thomas}, {Couchman}, {Evrard}, {Colberg} \& {Pearce}}{{Springel}
  et~al.}{2005}]{Springel2005}
{Springel} V.,  {White} S.~D.~M.,  {Jenkins} A.,  {Frenk} C.~S.,  {Yoshida} N.,
   {Gao} L.,  {Navarro} J.,  {Thacker} R.,  {Croton} D.,  {Helly} J.,
  {Peacock} J.~A.,  {Cole} S.,  {Thomas} P.,  {Couchman} H.,  {Evrard} A.,
  {Colberg} J.,    {Pearce} F.,  2005, \nat, 435, 629

\bibitem[\protect\citeauthoryear{{Springel}, {White}, {Tormen} \&
  {Kauffmann}}{{Springel} et~al.}{2001}]{Springel2001}
{Springel} V.,  {White} S.~D.~M.,  {Tormen} G.,    {Kauffmann} G.,  2001,
  \mnras, 328, 726

\bibitem[\protect\citeauthoryear{{Springob}, {Masters}, {Haynes}, {Giovanelli}
  \& {Marinoni}}{{Springob} et~al.}{2007}]{Springob2007}
{Springob} C.~M.,  {Masters} K.~L.,  {Haynes} M.~P.,  {Giovanelli} R.,
  {Marinoni} C.,  2007, \apjs, 172, 599

\bibitem[\protect\citeauthoryear{{Steidel}, {Shapley}, {Pettini}, {Adelberger},
  {Erb}, {Reddy} \& {Hunt}}{{Steidel} et~al.}{2004}]{Steidel2004}
{Steidel} C.~C.,  {Shapley} A.~E.,  {Pettini} M.,  {Adelberger} K.~L.,  {Erb}
  D.~K.,  {Reddy} N.~A.,    {Hunt} M.~P.,  2004, \apj, 604, 534

\bibitem[\protect\citeauthoryear{{Steinmetz}}{{Steinmetz}}{1999}]{Steinmetz199%
9}
{Steinmetz} M.,  1999, \apss, 269, 513

\bibitem[\protect\citeauthoryear{{Sun}, {Jones}, {Forman}, {Vikhlinin},
  {Donahue} \& {Voit}}{{Sun} et~al.}{2007}]{Sun2007}
{Sun} M.,  {Jones} C.,  {Forman} W.,  {Vikhlinin} A.,  {Donahue} M.,    {Voit}
  M.,  2007, \apj, 657, 197

\bibitem[\protect\citeauthoryear{{Sutherland} \& {Dopita}}{{Sutherland} \&
  {Dopita}}{1993}]{Sutherland1993}
{Sutherland} R.~S.,  {Dopita} M.~A.,  1993, \apjs, 88, 253

\bibitem[\protect\citeauthoryear{{Tissera}, {White}, {Pedrosa} \&
  {Scannapieco}}{{Tissera} et~al.}{2010}]{Tissera2010}
{Tissera} P.~B.,  {White} S.~D.~M.,  {Pedrosa} S.,    {Scannapieco} C.,  2010,
  \mnras, 406, 922

\bibitem[\protect\citeauthoryear{{Toomre}}{{Toomre}}{1964}]{Toomre1964}
{Toomre} A.,  1964, \apj, 139, 1217

\bibitem[\protect\citeauthoryear{{Tremonti}, {Heckman}, {Kauffmann},
  {Brinchmann}, {Charlot}, {White}, {Seibert}, {Peng}, {Schlegel}, {Uomoto},
  {Fukugita} \& {Brinkmann}}{{Tremonti} et~al.}{2004}]{Tremonti2004}
{Tremonti} C.~A.,  {Heckman} T.~M.,  {Kauffmann} G.,  {Brinchmann} J.,
  {Charlot} S.,  {White} S.~D.~M.,  {Seibert} M.,  {Peng} E.~W.,  {Schlegel}
  D.~J.,  {Uomoto} A.,  {Fukugita} M.,    {Brinkmann} J.,  2004, \apj, 613, 898

\bibitem[\protect\citeauthoryear{{Vale} \& {Ostriker}}{{Vale} \&
  {Ostriker}}{2004}]{Vale2004}
{Vale} A.,  {Ostriker} J.~P.,  2004, \mnras, 353, 189

\bibitem[\protect\citeauthoryear{{van den Bosch}, {Yang}, {Mo}, {Weinmann},
  {Macci{\`o}}, {More}, {Cacciato}, {Skibba} \& {Kang}}{{van den Bosch}
  et~al.}{2007}]{vandenBosch2007}
{van den Bosch} F.~C.,  {Yang} X.,  {Mo} H.~J.,  {Weinmann} S.~M.,
  {Macci{\`o}} A.~V.,  {More} S.,  {Cacciato} M.,  {Skibba} R.,    {Kang} X.,
  2007, \mnras, 376, 841

\bibitem[\protect\citeauthoryear{{Viel}, {Becker}, {Bolton}, {Haehnelt},
  {Rauch} \& {Sargent}}{{Viel} et~al.}{2008}]{Viel2008}
{Viel} M.,  {Becker} G.~D.,  {Bolton} J.~S.,  {Haehnelt} M.~G.,  {Rauch} M.,
  {Sargent} W.~L.~W.,  2008, Physical Review Letters, 100, 041304

\bibitem[\protect\citeauthoryear{{Viel}, {Bolton} \& {Haehnelt}}{{Viel}
  et~al.}{2009}]{Viel2009}
{Viel} M.,  {Bolton} J.~S.,    {Haehnelt} M.~G.,  2009, \mnras, 399, L39

\bibitem[\protect\citeauthoryear{{Vikhlinin}, {Kravtsov}, {Burenin}, {Ebeling},
  {Forman}, {Hornstrup}, {Jones}, {Murray}, {Nagai}, {Quintana} \&
  {Voevodkin}}{{Vikhlinin} et~al.}{2009}]{Vikhlinin2009}
{Vikhlinin} A.,  {Kravtsov} A.~V.,  {Burenin} R.~A.,  {Ebeling} H.,  {Forman}
  W.~R.,  {Hornstrup} A.,  {Jones} C.,  {Murray} S.~S.,  {Nagai} D.,
  {Quintana} H.,    {Voevodkin} A.,  2009, \apj, 692, 1060

\bibitem[\protect\citeauthoryear{{von der Linden}, {Best}, {Kauffmann} \&
  {White}}{{von der Linden} et~al.}{2007}]{vonderLinden2007}
{von der Linden} A.,  {Best} P.~N.,  {Kauffmann} G.,    {White} S.~D.~M.,
  2007, \mnras, 379, 867

\bibitem[\protect\citeauthoryear{{Wang}, {Li}, {Kauffmann} \& {De
  Lucia}}{{Wang} et~al.}{2006}]{Wang2006}
{Wang} L.,  {Li} C.,  {Kauffmann} G.,    {De Lucia} G.,  2006, \mnras, 371, 537

\bibitem[\protect\citeauthoryear{{Wang}, {Li}, {Kauffmann} \& {De
  Lucia}}{{Wang} et~al.}{2007}]{Wang2007}
{Wang} L.,  {Li} C.,  {Kauffmann} G.,    {De Lucia} G.,  2007, \mnras, 377,
  1419

\bibitem[\protect\citeauthoryear{{Weinmann}, {Kauffmann}, {von der Linden} \&
  {De Lucia}}{{Weinmann} et~al.}{2009}]{Weinmann2009}
{Weinmann} S.~M.,  {Kauffmann} G.,  {von der Linden} A.,    {De Lucia} G.,
  2009, arXiv:0912.2741 [astro-ph]

\bibitem[\protect\citeauthoryear{{Weinmann}, {van den Bosch}, {Yang}, {Mo},
  {Croton} \& {Moore}}{{Weinmann} et~al.}{2006}]{Weinmann2006}
{Weinmann} S.~M.,  {van den Bosch} F.~C.,  {Yang} X.,  {Mo} H.~J.,  {Croton}
  D.~J.,    {Moore} B.,  2006, \mnras, 372, 1161

\bibitem[\protect\citeauthoryear{{Wetzel}, {Cohn} \& {White}}{{Wetzel}
  et~al.}{2009}]{Wetzel2009}
{Wetzel} A.~R.,  {Cohn} J.~D.,    {White} M.,  2009, \mnras, 395, 1376

\bibitem[\protect\citeauthoryear{{White} \& {Frenk}}{{White} \&
  {Frenk}}{1991}]{White1991}
{White} S.~D.~M.,  {Frenk} C.~S.,  1991, \apj, 379, 52

\bibitem[\protect\citeauthoryear{{White} \& {Rees}}{{White} \&
  {Rees}}{1978}]{White1978}
{White} S.~D.~M.,  {Rees} M.~J.,  1978, \mnras, 183, 341

\bibitem[\protect\citeauthoryear{{Wilkins}, {Trentham} \& {Hopkins}}{{Wilkins}
  et~al.}{2008}]{Wilkins2008}
{Wilkins} S.~M.,  {Trentham} N.,    {Hopkins} A.~M.,  2008, \mnras, 385, 687

\bibitem[\protect\citeauthoryear{{Willman}, {Dalcanton}, {Martinez-Delgado},
  {West}, {Blanton}, {Hogg}, {Barentine}, {Brewington}, {Harvanek}, {Kleinman},
  {Krzesinski}, {Long}, {Neilsen} Jr., {Nitta} \& {Snedden}}{{Willman}
  et~al.}{2005}]{Willman2005}
{Willman} B.,  {Dalcanton} J.~J.,  {Martinez-Delgado} D.,  {West} A.~A.,
  {Blanton} M.~R.,  {Hogg} D.~W.,  {Barentine} J.~C.,  {Brewington} H.~J.,
  {Harvanek} M.,  {Kleinman} S.~J.,  {Krzesinski} J.,  {Long} D.,  {Neilsen}
  Jr. E.~H.,  {Nitta} A.,    {Snedden} S.~A.,  2005, \apjl, 626, L85

\bibitem[\protect\citeauthoryear{{Yoshida}, {Stoehr}, {Springel} \&
  {White}}{{Yoshida} et~al.}{2002}]{Yoshida2002}
{Yoshida} N.,  {Stoehr} F.,  {Springel} V.,    {White} S.~D.~M.,  2002, \mnras,
  335, 762

\bibitem[\protect\citeauthoryear{{Zabludoff}, {Zaritsky}, {Lin}, {Tucker},
  {Hashimoto}, {Shectman}, {Oemler} \& {Kirshner}}{{Zabludoff}
  et~al.}{1996}]{Zabludoff1996}
{Zabludoff} A.~I.,  {Zaritsky} D.,  {Lin} H.,  {Tucker} D.,  {Hashimoto} Y.,
  {Shectman} S.~A.,  {Oemler} A.,    {Kirshner} R.~P.,  1996, \apj, 466, 104

\bibitem[\protect\citeauthoryear{{Zavala}, {Jing}, {Faltenbacher}, {Yepes},
  {Hoffman}, {Gottl{\"o}ber} \& {Catinella}}{{Zavala}
  et~al.}{2009}]{Zavala2009}
{Zavala} J.,  {Jing} Y.~P.,  {Faltenbacher} A.,  {Yepes} G.,  {Hoffman} Y.,
  {Gottl{\"o}ber} S.,    {Catinella} B.,  2009, \apj, 700, 1779

\bibitem[\protect\citeauthoryear{{Zibetti}, {White}, {Schneider} \&
  {Brinkmann}}{{Zibetti} et~al.}{2005}]{Zibetti2005}
{Zibetti} S.,  {White} S.~D.~M.,  {Schneider} D.~P.,    {Brinkmann} J.,  2005,
  \mnras, 358, 949

\bibitem[\protect\citeauthoryear{{Zucker} \& et al.}{{Zucker} \&
  et~al.}{2006}]{Zucker2006}
{Zucker} D.~B.,  et al. 2006, \apjl, 643, L103

\end{thebibliography}

\section*{Appendix A    }

As discussed in Sec.~\ref{sec:stripping}, we have modified our previous
treatment of the transition between central and satellite status when galaxies
fall into larger systems. As long as the subhalo associated with a galaxy
remains outside the virial radius of its FOF group, we now continue to treat
that galaxy as an independent central object.  Thus galaxies effectively
become satellites only when they fall within $R_{\rm vir}$. This reduces the
number of satellites from the point of view of our galaxy formation modelling
and it increases the fraction of satellites which are type 2 or ``orphan''
systems with no associated subhalo. (This is because the orphans almost all
lie within $R_{\rm vir}$.) Here we illustrate the change in the effective number
of satellites in our two simulations as a function of stellar mass.

\begin{figure}
\bc
\hspace{-0.6cm}
\resizebox{8.5cm}{!}{\includegraphics{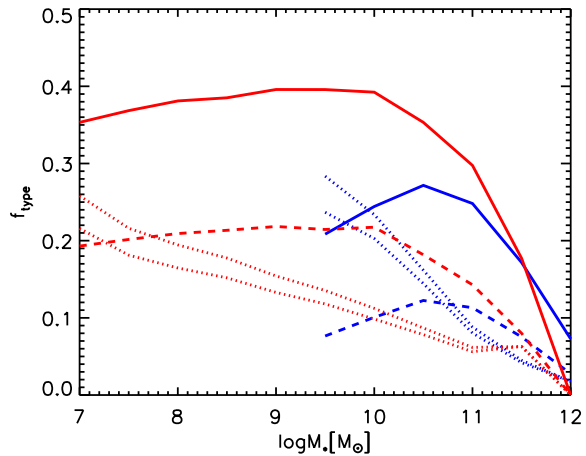}}\\%
\caption{The fraction of all galaxies which are satellites of various types
as a function of stellar mass. Blue curves are results for the MS and red
curves for the MS-II.  For each simulation and at each stellar mass, a solid
curve gives the fraction centred on a non-dominant, satellite subhalo (type 1
galaxies), a dashed curve gives the fraction which are in addition within
$R_{\rm vir}$ of halo centre, and dotted curves give the fraction with no
remaining associated subhalo (type 2 or ``orphan'' satellites; the upper curve refers to orphans within the FOF group while the lower only counts orphans within $R_{\rm vir}$). Note that in
both simulations a substantial fraction of the type 1's are actually outside
$R_{\rm vir}$ and so continue to be treated as central galaxies by our modified
prescriptions. Note also that while improved resolution reduces the number of
orphans in the MS-II, they remain a significant population even at relatively
large stellar mass.}
\label{fig:typefrac}
\ec
\end{figure}

Fig.~\ref{fig:typefrac} shows the fraction of all galaxies at each
stellar mass which are satellite systems of various types. Red curves
refer to the MS-II and are plotted down to a stellar mass of
$10^7M_\odot$, while blue curves refer to the MS and stop at its
resolution limit, $M_*\sim 10^{9.5}M_\odot$. For each simulation the
solid curve gives the fraction of galaxies which are centred on
non-dominant subhalos of their FOF groups, the dashed curve gives the
fraction which are in addition within $R_{\rm vir}$, and the upper and
lower dotted curves give the fractions which are orphans within FOF
groups and within $R_{\rm vir}$, respectively. In our previous work
(e.g. DLB07) the galaxies corresponding to the solid and upper dotted
curves were treated as satellites when modelling their evolution.  In
the current paper it is rather the galaxies corresponding to the
dashed and lower dotted curves which are treated as satellites; the
galaxies corresponding to the difference between the solid and dashed
curves continue to be treated as centrals.  Thus the effective
satellite fraction is smaller in this paper than in our previous
work. Notice that while the improved resolution of the MS-II does
decrease the number of orphan galaxies in comparison to the MS, these
remain a significant population, even at relatively high stellar
mass. Notice also that above the mass limit of the MS, the total
fraction of galaxies which are satellites agrees well between the two
simulations, demonstrating that our treatment of orphans in the MS is
indeed appropriate, as also concluded earlier when discussing
Fig.\ref{fig:clust_prof}.

\label{lastpage}
\end {document}